\documentclass[usenatbib]{mnras}
\usepackage{graphicx, amsmath, amssymb}
\usepackage{url}
\usepackage[T1]{fontenc}
\usepackage{ae,aecompl}
\usepackage{pdflscape}

\newcommand{\feh}{{\rm [Fe/H]}}
\newcommand{\myr}{\, {\rm Myr}}
\newcommand{\gyr}{{\, \rm Gyr}}
\newcommand{\kpc}{{\, \rm kpc}}
\newcommand{\pc}{{\, \rm pc}}
\newcommand{\mj}{m_j}
\newcommand{\Mj}{M_j}
\newcommand{\rh}{r_{\rm h}}
\newcommand{\rt}{r_{\rm t}}
\newcommand{\rj}{r_{\rm J}}
\newcommand{\ra}{r_{\rm a}}
\newcommand{\mtot}{\, M_{\rm tot}}
\newcommand{\msun}{\, {\rm M}_{\odot}}
\newcommand{\trhz}{\tau_{\rm {rh,0}}}
\newcommand{\ecrit}{E_{\rm crit}}

\def\camcos{Communications in Applied Mathematics and Computational science}

\title[Testing isothermal models - II. Multimass]{Testing lowered isothermal models with direct $N$-body simulations of globular clusters - II. Multimass models}
\author[M. Peuten]{M. Peuten,$^{1}$\thanks{E-mail:
m.peuten@surrey.ac.uk (MP)} A. Zocchi,$^{1,2}$ M.Gieles,$^{1}$ V. H\'enault-Brunet$^{3}$\\
$^{1}$ Department of Physics, University of Surrey, Guildford, GU2 7XH, UK\\
$^{2}$ Dipartimento di Fisica e Astronomia, Universit\`a degli Studi di Bologna, viale Berti Pichat 6/2, I40127, Bologna, Italy\\
$^{3}$ Department of Astrophysics/IMAPP, Radboud University, PO Box 9010, NL-6500 GL Nijmegen, the Netherlands}

\begin{document}

\date{}

\pagerange{\pageref{firstpage}--\pageref{lastpage}} \pubyear{2017}

\maketitle

\label{firstpage}

\begin{abstract}
Lowered isothermal models, such as the multimass Michie-King models, have been successful in describing observational data of globular clusters. In this study we assess whether such models are able to  describe the phase space properties of evolutionary $N$-body models. We compare the multimass models as implemented in \textsc{limepy} (Gieles \& Zocchi) to $N$-body models of  star clusters with different retention fractions for the black holes and neutron stars evolving in a tidal field. We find that multimass models successfully reproduce the density and velocity dispersion profiles of the different mass components  in all evolutionary phases and for different remnants retention. We further use these results to study the evolution of global model parameters. We find that over the lifetime of clusters, radial anisotropy gradually evolves from the low-mass to the high-mass components and we identify features in the properties of observable stars that are indicative of the presence of stellar-mass black holes. We find that the model velocity scale  depends on mass as $m^{-\delta}$, with $\delta\simeq0.5$ for almost all models, but the dependence of central velocity dispersion on $m$ can be shallower, depending on the dark remnant content, and agrees well with that of the $N$-body models. The reported model parameters, and correlations amongst them, can be used as theoretical priors when fitting these types of mass models to observational data. 
\end{abstract}


\section{Introduction}
\label{sec:Intro}

The amount of available data from observations of globular clusters (GCs) is steadily increasing. With the arrival of the ESA--Gaia data \citep{2016arXiv160904153G}, we are entering the era of high-precision kinematics, allowing us to study properties of GCs with unprecedented detail. This calls for adequate methods of analysing and describing them in an equally detailed way. Despite the fact that GCs are thought to be free of dark matter \citep{2010IAUS..266..365B,2013MNRAS.428.3648I,2017MNRAS.464.2174B}, and to have evolved to spherical and isotropic configurations as the result of two-body relaxation, GCs are complex systems to model. They consist of stars and stellar remnants with different masses and luminosities and primordial and dynamically processed binary stars \citep{1975MNRAS.173..729H,1989Natur.339...40G,1992PASP..104..981H,2006MNRAS.368..677H,2007MNRAS.374..344T}. The mass and luminosity functions depend on the stellar initial mass function (IMF), age and metallicity. GC stellar populations display chemical anomalies \citep{2004ARA&A..42..385G} and broadened main sequences (MS), possibly the result of variations in the helium abundance \citep{2014ApJ...785...21M}. Furthermore GCs evolve in a galactic tidal field that influences their evolution and present-day properties \citep{1990ApJ...351..121C,1999MNRAS.302..771J,2000ApJ...535..759T,2003MNRAS.340..227B, 2010MNRAS.401..105K, 2013MNRAS.436.3695R,2017MNRAS.465.3622R}. 

Modelling GCs on a star-by-star basis using direct $N$-body models has only become possible recently: \cite{2005MNRAS.363..293H} presented the first $N$-body simulation of an open cluster, \cite{2011MNRAS.411.1989Z} modelled a low-mass GC and finally \cite{2014MNRAS.445.3435H} and \cite{2016MNRAS.tmp...74W} presented the first $N$-body simulations of GCs with $N\sim10^6$. The faster Monte Carlo method allows to explore the parameter of the initial conditions to some extent \citep{2008MNRAS.389.1858H,2013MNRAS.431.2184G}. To infer properties for a large number of GCs with models with several degrees of freedom, static models that are fast to calculate are required. By using relatively simple models, that are motivated by the underlying physical processes that drive their evolution, differences between models and observations can be used to increase our understanding \citep{2011MNRAS.413.1889B}. 

In the context of GCs, the \cite{1966AJ.....71...64K} models are often compared to observations, although they cannot describe all GCs successfully. For example, \cite{2005ApJS..161..304M} find that the more extended \cite{1975AJ.....80..175W} models are better in describing the surface brightness profiles of some Galactic GCs. In addition, both King and Wilson models have isothermal cores, which are not able to describe the late stages of core collapse \citep{1980MNRAS.191..483L,1980ApJ...242..765C}. The models we are going to test and discuss in the context of GCs are multimass, anisotropic and spherical models (hereafter multimass models), which describe the properties of GCs considering their stellar mass function (MF) in the form of mass bins. This formulation allows for different behaviour of the different components. These models are defined by a distribution function (DF) which is a solution of the collisionless Boltzmann equation assuming a Maxwellian velocity distribution that is `lowered' to mimic the effect of a negative escape energy as the result of the galactic tides. The multimass formulation of a King model was first introduced by \citet[][we note that a formulation of a multimass model was already presented in \citealt{1959BAN....14..299O}]{1976ApJ...206..128D}. \cite{1979AJ.....84..752G} extended these models including radial anisotropy as formulated by \cite{1915MNRAS..75..366E} for isothermal models \citep[see also][]{1963MNRAS.125..127M}.

Since their introduction, multimass models have been successfully used in a multitude of studies such as in \cite{1977ApJ...218L.109I}, \cite{1986AJ.....91..546P}, \cite{1987AJ.....93.1114L}, \cite{1987A&A...184..144M}, \cite{1989ApJ...339..178R}, \cite{1991A&A...250..113M}, \cite{1995A&A...303..761M}, \cite{1997AJ....114.1517S},  \cite{1999A&A...345..485P} and \cite{2004AJ....127.2771R} to name a few. More recently, they have been used in observational studies, such as those by \cite{2010AJ....139..476P}, \cite{2010ApJ...713..194B}, \cite{2012ApJ...755..156S}, \cite{2015ApJ...814..144B} and \cite{2016arXiv161002300S}, as well as in theoretical studies \citep{2000MNRAS.316..671T,2015MNRAS.451.2185S}.

\cite{1977A&A....61..391D} realized that the DFs of the \cite{1954MNRAS.114..191W},  \cite{1966AJ.....71...64K} and \cite{1975AJ.....80..175W} models can be written as a single DF with one additional integer parameter. \cite{2014JSMTE..04..006G} further generalized this formulation, allowing to calculate models in between the three classical models. \cite[][hereafter GZ15]{2015MNRAS.454..576G} took up these formulation and added radial anisotropy as defined in the Michie--King models \citep{1963MNRAS.125..127M} and multiple mass components as in \cite{1979AJ.....84..752G}. GZ15 introduced a power-law dependence between mass and anisotropy radius for each mass bin, while \cite{1979AJ.....84..752G} argued that was not necessary because most events that influence the anisotropy are mass independent or not very important. GZ15 also implemented the possibility to change the degree of mass segregation with an additional parameter $\delta$ that describes the relation between the velocity scale and mass, which in most models is assumed to be equal to $1/2$.

Despite their success in describing observational data, multimass models have been criticized for several assumptions made in their construction (see \citealt{2003ASPC..296..101M} and \citealt{1997A&ARv...8....1M}). One such aspect is that multimass models have more parameters than single-mass models: in the formulation by GZ15, there are $2 N_{\rm MB} + 5$ parameters and $2$ scales, with $N_{\rm MB}$ being the number of mass bins, compared to $3$ parameters and $2$ scales for the single-mass model. It is therefore easier to fit multimass models to the data because they have more degrees of freedom \citep{2003ASPC..296..101M}. Not only the selection of the right number of mass bins, but also how they are defined is criticized as a `usual compromise between convenience and realism', as \cite{1997A&ARv...8....1M} put it. Given the numerous studies successfully using multimass models, this problem does not seem to be too much of a concern, but we nevertheless explored it in our study. Another assumption for which multimass models have been criticized is the assumption of equipartition of energy \citep{2003ASPC..296..101M,2013MNRAS.435.3272T}. Indeed the velocity scale is usually assumed to scale with the mass as $m^{-1/2}$, but we note that evolutionary multimass models only achieve partial equipartition (\citealt{1981AJ.....86..318M}; \citealt{2006MNRAS.366..227M}; GZ15; \citealt{2016MNRAS.458.3644B}) as the result of the escape velocity.

Several aspects of multimass models were already analysed with the help of Fokker--Planck \citep{2000MNRAS.316..671T} and $N$-body simulations \citep{2015MNRAS.451.2185S}. The goal of this study is to compare the multimass models in the formulation by GZ15 to a set of $N$-body models to assess the quality of the former and to analyse whether some of the above mentioned criticism is justified. In this comparison we do not include any source of uncertainties, such as observational biases, to see how good the models are under ideal conditions. Hence, we determine the MF of the multimass models directly from the $N$-body data. 

The comparison is done by fitting the multimass models to snapshots from different $N$-body models using a Markov Chain Monte Carlo (MCMC) method. Additionally, we  study the new parameters which are now available in the extended formulation of the models by GZ15. In particular, the continuous truncation parameter as introduced by \cite{2014JSMTE..04..006G} and the  parameter that controls the mass dependence of the anisotropy for each mass bin. Furthermore, we study the behaviour of the mass segregation parameter $\delta$, which in previous studies was fixed to $\delta = 1/2$. By letting this parameter free, we can test whether this assumption is justified. By varying the amount of stellar-mass black holes (BHs) and neutron stars (NSs) retained in the different $N$-body models, we also study their impact on the cluster as well as on the different parameters of the best-fitting models. 

In \cite{Zocchi2016}, a similar analysis was presented for single-mass models. This comparison showed that the single-mass models are successful in describing the different phases of the dynamical evolution. \cite{Zocchi2016} studied the development of radial anisotropy in GCs and found that the models can be used to put limits on the expected amount of radial anisotropy. 

This paper is structured as follows: in the next section, we give a brief overview of the multimass models. Then in Section~\ref{sec:N-body}, we discuss how the $N$-body models were generated and we discuss their properties. In Section~\ref{sec:Method}, we present the method used for the analysis and the challenges we encountered. The radial profiles of density, velocity dispersion and anisotropy are discussed in Section~\ref{sec:MMvsNB}. In Section~\ref{sec:csp}, we discuss the values of the best-fitting model parameters and scales, and their implications. Finally, in Section \ref{sec:Conclusion}, we discuss our results and present our conclusions.

\section{The \textsc{limepy} Models}
\label{sec:Models}

The multimass models used in this study are provided by the {\sc limepy} (Lowered Isothermal Model Explorer in {\sc PYthon})\footnote{\url{https://github.com/mgieles/limepy}} software package (GZ15). A model has different components, each representing stars in a mass range, characterized by a mean and total mass. The DF of the $j$th mass component is given by
\begin{equation}
f_{j}\left(E,J^{2}\right)=A_{j}\exp\left(-\frac{J^{2}}{2r_{{\rm a},j}^{2}s_{j}^{2}}\right)E_{\gamma}\left(g,-\frac{E-\phi\left(r_{\rm t}\right)}{s_{j}^{2}}\right),
\label{eq:LIMEPY_DF}
\end{equation}
for $E < \phi(\rt)$ and $0$ otherwise. The specific energy $E={v^{2}}/{2}+\phi(r)$ is one of the two integrals of motion, where $v$ is the velocity and $\phi(r)$ the specific potential at a distance $r$ from the centre. The energy $E$ is lowered by the potential at the truncation radius $\phi(\rt)$. The function $E_{\gamma}$ is defined as
\begin{equation}
E_{\gamma}\left(a,x\right)=\begin{cases}
\exp\left(x\right) & a=0\\
\exp\left(x\right)P\left(a,x\right) & a>0
\end{cases}
\end{equation}
with $P\left(a,x\right)\equiv \gamma\left(a,x\right) / \Gamma\left(a\right)$ the regularized lower incomplete gamma function. The other integral of motion is the specific angular momentum $J = rv_{\rm t}$, where $v_{\rm t}$ is the tangential component of the velocity vector.

The anisotropy radius $\ra$ is a parameter that controls how anisotropic the model is. The system is isotropic in the centre, radially anisotropic in the intermediate part and near $r_{\rm t}$ it is isotropic again. For small values of $\ra$, the models are strongly anisotropic and for values of $\ra$ lager than $\rt$, the models are completely isotropic. GZ15 include a power-law dependence between mass and anisotropy radius: 
\begin{equation}
r_{{\rm a},j}=r_{\rm a}\mu_{j}^{\eta}
\label{eq:raj}
\end{equation}
with $\mu_j$ the dimensionless mean mass of stars in the $j$th mass component, defined as:
\begin{equation}
\mu_{j}=\frac{m_{j}}{\bar{m}}
\label{eq:muj}
\end{equation}
where $m_j$ is the mean mass of stars in the $j$th component and $\bar m$ a reference mass which we set equal to the global mean mass. If $\eta$ is set to zero the anisotropy radius is independent of the mass as in \cite{1979AJ.....84..752G}. \textsc{limepy} expects $\ra$ to be input in units of the King radius ($r_{\rm 0}$), $\hat{r}_{\rm a} = \ra / r_{\rm 0}$, hence $\hat{r}_{\rm a}$ is the parameter we vary.

The truncation parameter $g$ was introduced by \cite{2014JSMTE..04..006G} and describes the polytropic part near the escape energy. The polytropic index $n$ relates to $g$ as $n = g+3/2$ and this formulation allows to calculate models in-between the classical models: for $g = 0$ and $\ra\gg\rt$, the DF is identical to the one from the Woolley model \citep{1954MNRAS.114..191W}. A Michie--King \citep{1963MNRAS.125..127M, 1966AJ.....71...64K} model is reproduced for $g = 1$ and for $g = 2$ one gets the non-rotating Wilson model \citep{1975AJ.....80..175W}. The range of possible values for the model parameter are $0 \leq g \leq 3.5$ because as discussed in \cite{2014JSMTE..04..006G} and GZ15, there are no finite models above $g = 3.5$. The final parameter needed to define the models is the dimensionless central potential $W_0$ \citep[][$\hat{\phi}_{0}$ in GZ15]{1966AJ.....71...64K} which specifies how centrally concentrated the model is. It is a boundary condition for solving Poisson's equation. 

Besides these parameters, there are also two constants which define the physical scales of the model: one is the global velocity scale $s$ and the other is the normalization constant $A$ which sets the phase space density. Instead of these scales, the code needs as input the total cluster mass $M_{\rm Cl}$ and a radial scale $r_{\rm scale}$ (which can be  $r_{\rm 0}$, the half-mass radius $\rh$, the viral radius $r_{\rm v}$ or $\rt$), which are internally converted to $A$ and $s$.
 
The velocity scale $s_j$ is deduced from $s$ as:
\begin{equation}
s_{j}=s\mu_{j}^{-\delta}
\end{equation}
It is usually assumed that $\delta = 1/2$, but in this study we determine the value of this parameter from the fits to the $N$-body models.

The constants $s_j$ and $A_j$ are connected to the mass in each component ($M_j$), which the user provides together with $m_j$. It must be noted that $M_{\rm Cl}$ is a required input parameter, independent from the $M_j$ parameters, because the latter are only used to compute the relative masses in each component. Only after the model is solved, $\sum_j M_j = M_{\rm Cl}$.

Given these five parameters ($g$, $W_0$, $\delta$, $\ra$ and $\eta$) and two scales ($M_{\rm Cl}$, $r_{\rm scale}$) together with the description of the mass bins ($M_j, m_j$) {\sc limepy} first calculates the density for each mass bin via:
\begin{equation}
\rho_{j}=\int f_{j}\left(E,J^{2}\right)d^{3}v
\end{equation}
Then, the dimensionless Poisson equation is solved
\begin{equation}
\nabla^{2}\hat{\phi}=-9\sum_{j}\alpha_{j}\hat{\rho}_{j}
\end{equation}
with $\alpha_j = \rho_{j,0} / \rho_0$, $\hat{\rho}_{j} = \rho_j / \rho_{j,0}$ and the dimensionless positive potential $\hat{\phi} = (\phi(\rt) - \phi)/s^2$, is iteratively solved by varying $\alpha_j$ until the calculated $M_j$ converges to the input values. After the model is solved, it is scaled to $M_{\rm Cl}$ and $r_{\rm scale}$. We can then find the likelihood for any phase space coordinate using the DF (equation~\ref{eq:LIMEPY_DF}). 

In equation (\ref{eq:muj}), we set $\bar m$ to the mean mass of the cluster. In the formulation by \cite{1976ApJ...206..128D,  1979AJ.....84..752G} and GZ15, $\bar m$ is the central density weighted mean mass. After performing several comparisons, we found that models calculated by the two different formulations give the same results within the numerical uncertainties. Furthermore we found that using the global mean mass instead speeds up the calculation, especially for models with BHs. When using the global $\bar{m}$ the meaning of two model parameters is modified compared to \cite{1976ApJ...206..128D, 1979AJ.....84..752G} and GZ15: $W_{0}$ and $\ra$ both represent their value for a hypothetical mass group with a mass of $\bar m$. Besides computational improvement, this change from the original formulation also allows us to compare the multimass $W_{0}$ value with the single-mass $W_{0}$ value, as both represent the $W_{0}$ value for the mean mass group. 

One can easily translate the values given in one $\bar m$ definition ($W_0$, $\hat{r}_{\rm a}$) to another $\bar m^*$ definition ($W_{0}^{*}$, $\hat{r}_{{\rm a}}^{*}$) by applying the following two equations:
\begin{equation}
W_{0}^{*}=W_{0}\left(\frac{\bar{m}^{*}}{\bar{m}}\right)^{2\delta}
\end{equation}
\begin{equation}
\hat{r}_{{\rm a}}^{*}=\hat{r}_{{\rm a}}\left(\frac{\bar{m}^{*}}{\bar{m}}\right)^{(\eta + \delta)}
\label{eq:rara}
\end{equation}
The $\delta$ term in equation (\ref{eq:rara}) comes from the $r_{0}$ dependence of $\hat{r}_{\rm a}$.

As further improvement to the original formulation of the \textsc{limepy} models we found that radially anisotropic models can be constructed faster if one first calculates the $M_j$ array of the corresponding isotropic model and then uses this model as starting point to solve the anisotropic model. As with the previous improvement the differences are only of numerical nature. This procedure is now implemented in the current distribution of {\sc limepy}.

\section{Description of the $N$-body models}
\label{sec:N-body}

For the computation of the $N$-body data, we use the approach presented in \cite{2010ApJ...708.1598T}: the stellar evolution is done first and separately from the dynamical evolution. We do this because Galactic GCs have different dynamical ages but have all roughly the same physical age of around $12 \gyr$. We consider models with different retention fractions of NSs and BHs and analyse them at various dynamical ages. Temporal units are always expressed in units of the initial half-mass relaxation time ($\trhz$) of the $N$-body model. 

\subsection{Set-up of the $N$-body models}

For this analysis, we run four $N$-body models, with different amounts of NSs and BHs. Each $N$-body model was set up as a cluster with $N = 10^{5}$ stars initially following the H\'{e}non isochrone model \citep{1959AnAp...22..126H} with $\rh = 2.25 \pc$. As IMF we adopted a Kroupa IMF \citep{2001MNRAS.322..231K} in the mass range between $0.1 \msun$ and $100 \msun$ without any primordial binaries. Then, by using the fitting formula by \cite{2000MNRAS.315..543H} and assuming a metallicity of $\feh = -2$, the stars were evolved to an age of about $12 \gyr$. 

We mimic the effect of supernova kick velocity by removing a certain fraction of NSs and BHs from the initial conditions described above. The retention fraction of NSs and BHs after supernova kicks is highly uncertain \citep{2012MNRAS.425.2799R,2016MNRAS.456..578M}. To bracket all possible cases, we consider four different values for the fraction of remnants that we retain in the cluster: $100\%$ (all the remnants are retained, simulation N1), $33\%$ (simulation N0.3), $10\%$ (simulation N0.1) and $0\%$ (all the remnants are removed, simulation N0). The initial half-mass relaxation time for all four clusters was $\trhz = 350 \myr$ before the stellar evolution and the removal of the dark remnants, after these steps the $\trhz$ values are $412 \myr$ for N1, $426 \myr$ for N0.3, $427 \myr$ for N0.1 and $428 \myr$ for N0.

The clusters are evolved on a circular orbit with a circular velocity of $V_{\rm c} = 220 \, \rm km/s$ at a distance of $R_{\rm G} = 4 \kpc$, in a singular isothermal galactic potential to mimic a galaxy. The equation of motion is solved in an inertial reference frame centred on the cluster. 

These four stellar systems were then dynamically evolved with the state-of-the art $N$-body integrator \textsc{nbody6} \citep{2003gnbs.book.....A}, in the variant with GPU support \citep{2012MNRAS.424..545N}, until total dissolution of each cluster, i.e. until less than 100 objects are left in the cluster. Every object reaching a distance greater than twice the Jacobi radius ($\rj$) is considered lost and is removed from the $N$-body model. As the stellar evolution is done before the actual $N$-body simulation, binaries which formed in the course of the simulations were also only evolved dynamically.  

A snapshot of each cluster is taken every $\gyr$, resulting in 48 snapshots\footnote{The snapshots can be retrieved from \url{http://astrowiki.ph.surrey.ac.uk/dokuwiki/doku.php?id=tests:collision:mock_data:challenge_2}}
 (11 for model N1, 13 for model N0.3, 12 for model N0.1 and 12 for model N0) which we fit the multimass models to (Section \ref{sec:Method}).

\subsection{Selecting bound objects}
\label{subsec:BoundSel}

Because multimass models describe bound objects in a cluster, we removed any unbound object from the $N$-body models. We discuss here how we selected the unbound objects for each $N$-body snapshot.

First, we determine the Jacobi radius
\begin{equation}
r_{\rm J}=\left(\frac{G M_{\rm Cl}}{2\Omega^{2}}\right)^{1/3}
\label{eq:r_J}
\end{equation}
in which $M_{\rm Cl}$ is the total mass within $\rj$ and $\Omega={V_{\rm c}}/{R_{\rm G}}$ is the orbital angular velocity. As a first guess, we set $M_{\rm Cl}$ equal to the total mass of all stars in the snapshot and then determine $\rj$ through an iterative approach. 

With $\rj$ determined we are now able to calculate the specific critical energy which is equal to the potential at $\rj$
\begin{equation}
\ecrit=\phi\left(\rj\right)=-\frac{G M_{\rm Cl}}{\rj}\label{eq:E_crit}
\end{equation}
The true critical energy is different as equation~({\ref{eq:E_crit}) neglects the tides. We adopted this definition nevertheless to be consistent with the multimass models, which also do not account for the changed potential due to tidal effects. We considered an object bound if it is within $\rj$ and for its energy it holds $E_i < \ecrit$, and we only used these bound objects in the rest of this analysis.

\subsection{Properties of the $N$-body models}
\label{subsec:NBodyDisc}

\begin{figure}
\includegraphics[width=1\columnwidth]{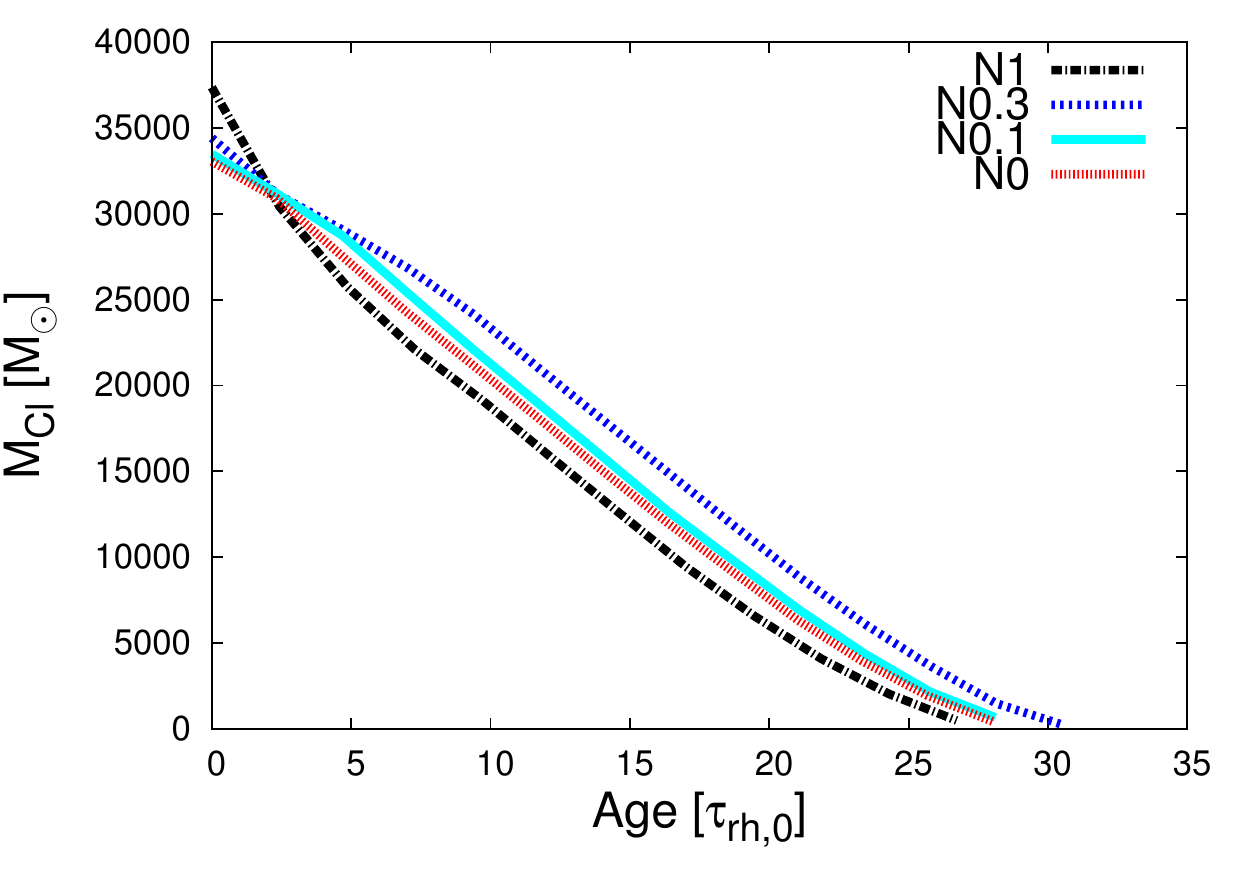} 
\caption{Evolution of the cluster mass $M_{\rm Cl}$ for the four $N$-body models. Age is given in units of their $\trhz$. \label{fig:N-Body-Mass}}
\end{figure}

\begin{figure}
\includegraphics[width=1\columnwidth]{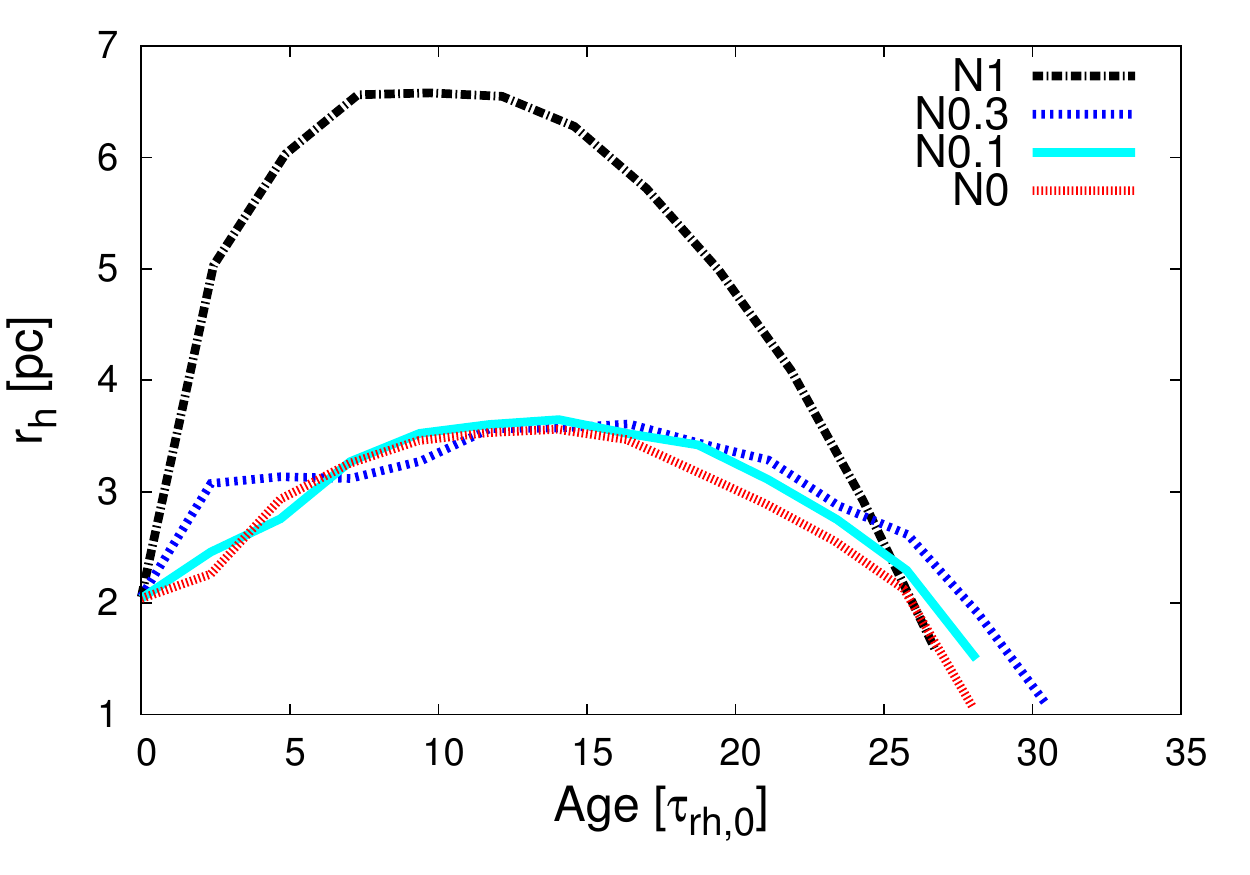} 
\caption{Evolution of the cluster half-mass radius $\rh$ for the four $N$-body models. Age is given in units of their $\trhz$. \label{fig:N-Body-Rh}}
\end{figure}

\begin{figure}
\includegraphics[width=1\columnwidth]{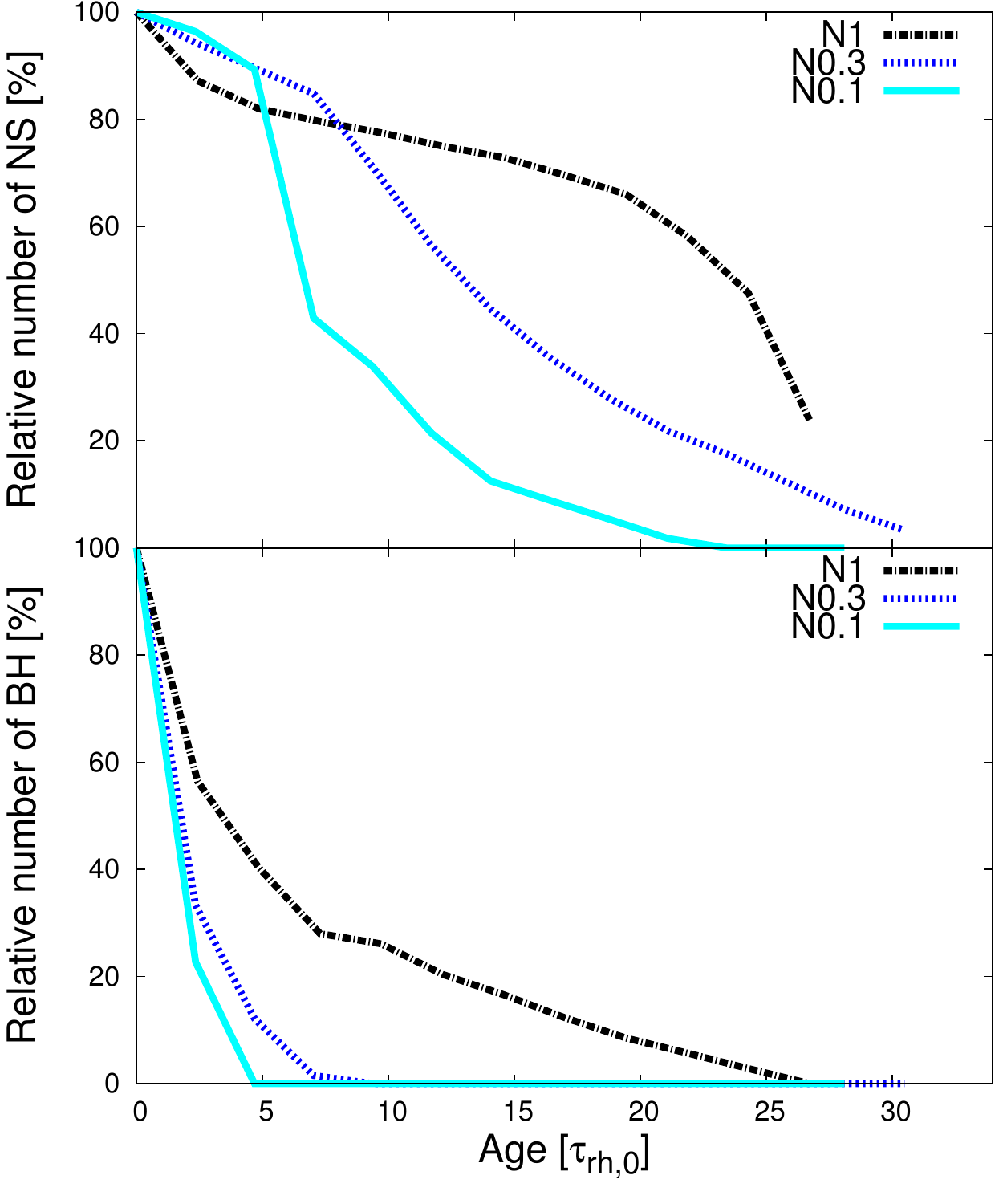} 
\caption{Evolution of the relative number of NS (top) and BH (bottom) as function of time in units of $\trhz$ for the three models which initially retained BHs and NSs. \label{fig:N-Body-BHNS}}
\end{figure}

\begin{figure}
\includegraphics[width=0.95\columnwidth]{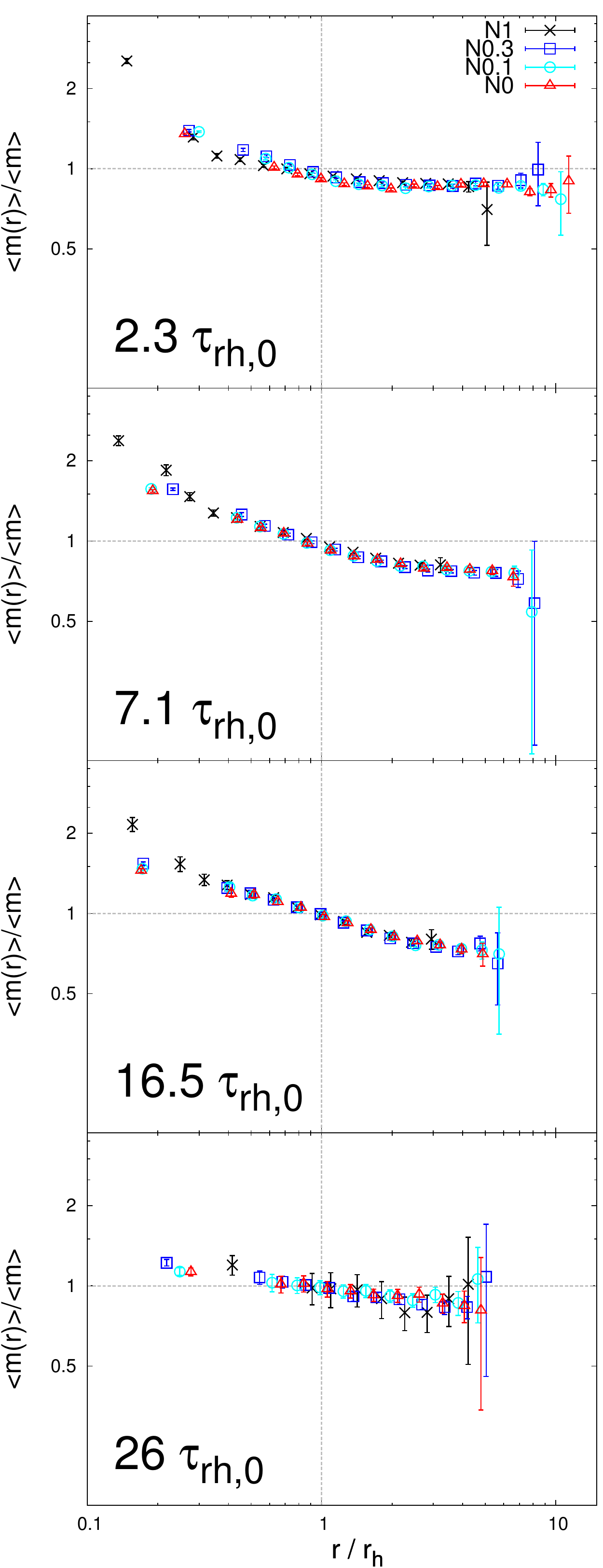}
\caption{Relative mean mass (i.e. mean mass of stars in radial bins divided by the total mean mass) as a function of the distance from the cluster centre in units of $\rh$, for all simulations at four different times: $2.3 \trhz$, $7.1 \trhz$, $16.5 \trhz$ and $26 \trhz$. Triangles (red), circles (cyan), boxes (blue) and stars (black) refer to simulation N0, N0.1, N0.3 and N1, respectively. Error bars denote $1\sigma$ uncertainties. \label{fig:AVG-Plot-Selection}}
\end{figure}

Fig.~\ref{fig:N-Body-Mass} shows how $M_{\rm Cl}$ for the four different $N$-body models evolves over the course of the simulation. It is apparent that the cluster with 100\% initial BH and NS retention (simulation N1) has the highest initial mass loss, but there seems to be no direct correlation between the number of BHs and NSs and initial mass loss as can be seen from the other three $N$-body models. Over the course of evolution the four models seem to have aligned their mass-loss rate which is in accordance with the findings of \cite{1987ApJ...322..123L} and \cite{2011MNRAS.413.2509G} that the escape rate of clusters with the same mass mainly depends on the tidal field, which is the same for all four models. 

In Fig.~\ref{fig:N-Body-Rh}, we have plotted the evolution of $\rh$ for the four different $N$-body models. As can be seen in the figure, increasing the retention fraction of the BHs leads to an expansion of the cluster: simulation N1, with $100\%$ NS and BH retention has an $\rh$ which is on average twice as large as the $\rh$ from simulation N0 with no NS and BH retention. The cluster in simulation N0.3 loses all of its BHs at around $7 \trhz$ and for the rest of the simulation its $\rh$ resembles that of N0. The effect of stellar-mass BHs on the radius evolution was also described by \cite{2008MNRAS.386...65M}. The global evolution of $\rh$ however is essentially the same, independent of BHs and NSs retained, and follows the description in \cite{2011MNRAS.413.2509G}. In the first half of their lifetime, the clusters are in the expansion-dominated phase while in the second half the clusters are in the evaporation-dominated phase during which $\rh$ decreases again until total dissolution. 

Fig.~\ref{fig:N-Body-BHNS} shows the relative number evolution for the BHs and NSs in the simulation N1, N0.3 and N0.1. As can be seen in the bottom figure, the cluster from $N$-body simulation N1 with $100\%$ initial BHs and NSs retention is the only cluster that retains its BHs almost until to the end of its lifetime, while the cluster from $N$-body simulation N0.3 loses all its BHs at around $7 \trhz$ and the one from simulation N0.1 already at $2 \trhz$. Looking at the NSs in the top of Fig.~\ref{fig:N-Body-BHNS}, we see their initial loss is not as strong as for the BHs. But as soon as all BHs have left the cluster, the NSs escape rate increases such that the cluster N0.1 loses all its NS at around $\sim 21 \trhz$. Only the cluster from $N$-body simulation N1 has a population of NS left at the end of its lifetime. After all BHs are lost, NSs, then being the most massive objects in the cluster, segregate to the centre and are then ejected from the cluster due to interactions they experience with each other.

Tables \ref{tab:NBodyN1}--\ref{tab:NBodyN0} list various properties of the different snapshots, such as the dynamical age, the bound mass and the number of NSs and BHs.

\subsection{Mean mass at different radii}

As in \cite{P2016}, we find that the mean mass profile is independent of  the remnant retention fraction. In Fig.~\ref{fig:AVG-Plot-Selection}, we plot this  for all four $N$-body models at four different times in their evolution: $2.3 \trhz$, $7.1 \trhz$, $16.5 \trhz$ and $26 \trhz$. Looking at the different times we see that the overall behaviour is the same for all models independent of their dark remnant population. Some divergences between the different $N$-body models can be seen in the first snapshot at $2.3 \trhz$ but over the course of evolution these differences diminish. The profiles get flatter over time. This is comparable to the behaviour found for a set of $N$-body models where the dynamical and stellar evolution were done concurrently in \cite{P2016}. Here, the evolution over time is less strong because the stellar evolution was done before the dynamical evolution. We are not aware of a theory providing an explanation for this attractor solution of $\bar{m}(r)$, but for single-mass system it is known that after several relaxation times the evolution becomes self-similar \citep{1961AnAp...24..369H,1965AnAp...28...62H}. Also it had been shown that the evolution of radii and mass of the multimass systems is comparable to those of single-mass systems \citep{1995ApJ...443..109L,2010MNRAS.408L..16G} but faster. Furthermore \cite{1996MNRAS.279.1037G,1997MNRAS.286..709G} showed in multimass $N$-body models that after some time the mean mass in Lagrangian shells stops evolving and, to a first approximation, stays constant. Although we do not explore this here, this result could be used as a theoretical prior when comparing multimass models to data.

\section{Method}
\label{sec:Method}

To determine the best-fitting multimass models for each snapshot we use the MCMC software package \textsc{emcee} \citep{2013PASP..125..306F}, which is a pure-\textsc{python} implementation of the Goodman \& Weare's Affine Invariant MCMC Ensemble sampler \citep{10.2140/camcos.2010.5.65}. The \textsc{python} implementation makes it straightforward to couple it with {\sc limepy}. Furthermore, we benefit from the fact that we created a distributed grid computing version of \textsc{python}'s \textit{map()} function, thereby dynamically distributing efficiently the workload from several MCMC runs over all available CPU cores at the University of Surrey Astrophysics computing facilities. 

The fitting process consists of computing a multimass model based on the input parameters provided by the MCMC walker position in parameter space and the current mass bin description (see the next section). Then the likelihood for each star in all mass bins is calculated using the DF (see equation~\ref{total-likelihood}) and the phase space position of the star from the $N$-body snapshot. By randomly varying the walker positions in parameter space, the MCMC algorithm tries to find those parameters which maximize the product of these individual likelihoods. The best-fitting value of each parameter is estimated as the median of the marginalized posterior distribution using all walker positions from all chains after removing the initial burn-in phase. This generally coincides with the value of the parameter providing the largest likelihood. For the $1\sigma$ errors, we use the values from the 16th and 84th percentiles.

\subsection{Determining the mass bins}
\label{sec:MassBins}

As  mentioned in Section~\ref{sec:Intro}, the mass bin selection for multimass models is in most cases a choice of convenience \citep{1997A&ARv...8....1M} as throughout the literature there is no general rule on how to select the best. This can be partially explained by the fact that most publications consider different data for their analyses, and have different research targets, leading to different approaches on how to set up the MF. However, we do know everything about our $N$-body models, and this allows us to test the mass bin selection for multimass models. In particular, we want to understand what the minimum number of mass bins is to get a stable result and how to choose the bins.

For this analysis we use the $N$-body snapshot of simulation N1 at the time of $2.9 \, \trhz$. As we wanted to trace the overall evolution of the different star types, we opted against mixing them and therefore we give every star type at least one bin. This means that we  have at least five mass bins, one each for the MS stars, the evolved stars\footnote{In this work, every post MS star which is not a remnant is called an ES.} (ES), the white dwarfs (WD), NSs and BHs. Looking at the BHs, NSs and ESs we decided to not further split them into several bins given the fact that they have either a rather small range of possible masses and/or are to low in number to justify the split. This leaves us with MS stars and WDs which are both numerous and do have a large mass range: $0.1 - 0.83 \msun$ for the MS stars and $0.55 - 1.44 \msun$ for the WDs in our $N$-body snapshots. 

First, we determine how the bin selection influences the results of the analysis. For this, we choose four different binning methods: a logarithmic binning, a linear binning, a binning where in each mass bin there is an equal number of stars and a binning where there is the same amount of mass in each mass bin. We fixed the number of WD mass bins to one and then for each bin type we calculated the multimass model repeatedly with increasing number of MS star bins. The general idea here is that with increasing number of mass bins, the overall parameters like $M_{\rm Cl}$, $\rh$, etc., should converge to the value one would get for the ideal case, where each star has its own mass bin. We find that the results are almost independent of the way one chooses the binning: the number of MS mass bins needed to converge is the same and the difference between the different models are for all properties generally less than $5\%$. We therefore choose for the further analysis the logarithmic binning.

Then we determine the minimum number of bins needed to get stable results, as increasing the number of bins also increases the computation time of the models. We varied the number of mass bins of the MS stars and the WDs independently from each other. Here again, we see that with increasing number of mass bins the different quantities converge. We find that we need at least two WDs mass bins and at least four MS stars mass bins for the different quantities to converge. Increasing the number of bins any further does not improve the results (values are comparable within $5\%$). For our further analysis, we opt to use five MS stars mass bins and three WDs mass bins. Therefore, in total we consider eleven mass bins (MS: 5; ES: 1; WD: 3; NS: 1; BH: 1) to set up our multimass models. Tables \ref{tab:MFN1}--\ref{tab:MFN0} list the mass bins for all $N$-body snapshots used.

\subsection{Artificial background population}
\label{subsec:ABP}

Before we present the results, we discuss a particularity which we encountered in our analysis. The potential of \textsc{limepy} models is spherical, however, the true potential of the cluster is triaxial because of the effect of tides. Also the Lagrange points of the cluster, through which stars can escape \citep{2000MNRAS.318..753F,2001MNRAS.325.1323B,2010MNRAS.401..105K,2016arXiv161202253C}, are not accounted for in the multimass model. Therefore, the models are not able to describe the objects near the critical energy correctly. From this, it follows that some of the objects which are unbound in the true potential are still found bound in our definition and cannot be described by the model correctly. These objects pose a problem because they drive the fit to unrealistic parameter values.

To cope with this problem, we introduced an artificial background population with a constant likelihood (i.e. uniform distribution) in phase space. We added the artificial background population with a total mass of around $1\%$ of the original cluster mass to the $N$-body snapshot. This background population has the same MF as the cluster. The upper limit for the maximal distance and velocity are chosen to be twice the maximal values from the original snapshot ($r_{\rm max}$ and $v_{\rm max}$).

We describe the likelihood function of the background model as 
\begin{equation}
\mathfrak{L}_{B}=\frac{M_{\rm Back}}{V \mtot}
\end{equation}
where $\mtot$ is the total mass of the snapshot including the artificial background and $M_{\rm Back}$ is the mass of the background only. The phase space volume $V$ is defined as
\begin{equation}
V=\frac{4}{3}\mathrm{\pi}\left(2r_{\rm max}\right)^{3} \times \frac{4}{3}\mathrm{\pi}\left(2v_{\rm max}\right)^{3}
\end{equation}

The total likelihood of an object for a given model is calculated as
\begin{equation}
\mathfrak{L}=\frac{f(E,J^{2})}{\mtot}+\frac{M_{\rm Back}}{V \mtot}
\label{total-likelihood}
\end{equation}

When integrated over the whole phase space volume within $2 v_{\rm max}$ and $2 r_{\rm max}$ the first term equals to $M_{\rm Cl}/\mtot$ and the second to $M_{\rm Back}/\mtot$, giving a total likelihood of unity, as required.

\subsection{MCMC results}

\begin{figure*}
\includegraphics[width=1\textwidth]{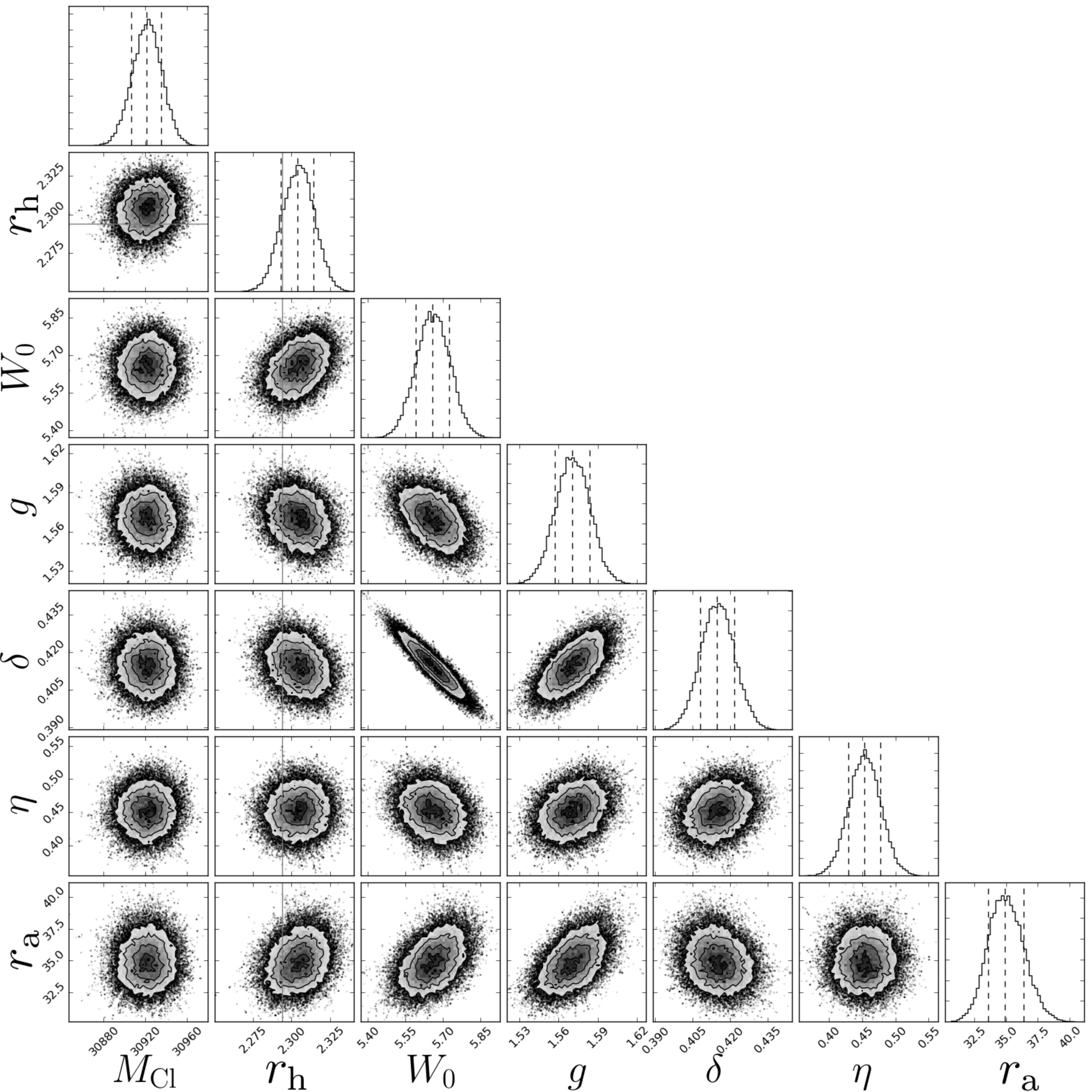}
\caption{Marginalized posterior probability distribution and 2D projections of the posterior probability distribution for the model parameters and scales. This figure shows the results of the MCMC fitting to the $N$-body model N0 at $2.4 \trhz$. The dashed lines in the marginalized posterior probability distribution indicate the 16th, 50th and 84th percentiles.\label{fig:MCMC-GC2-1}}
\end{figure*}

\begin{figure*}
\includegraphics[width=1\textwidth]{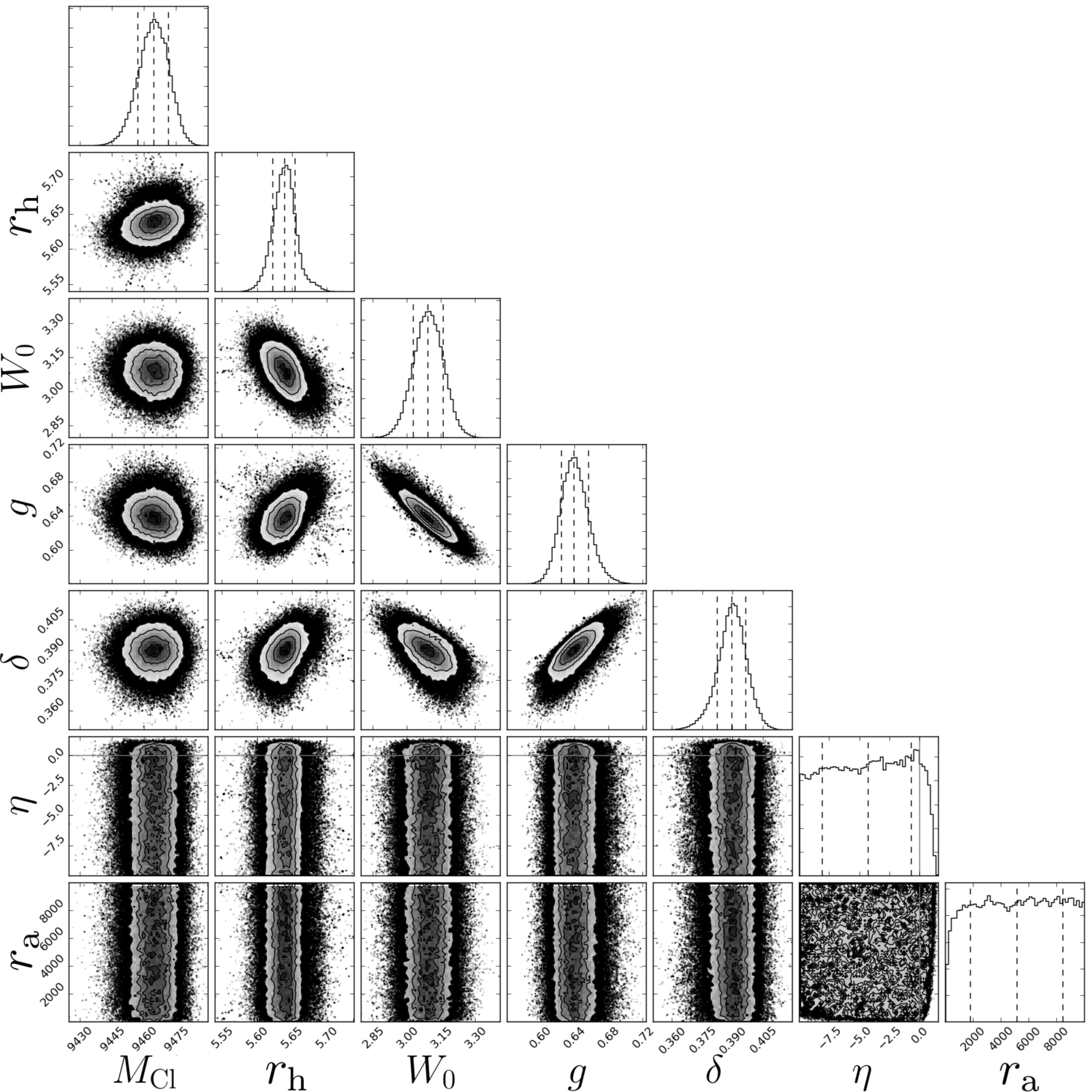}
\caption{Marginalized posterior probability distribution and 2D projections of the posterior probability distribution for the model parameters and scales. This figure shows the results of the MCMC fitting to the $N$-body model N1 at $17.0 \trhz$. The dashed lines in the marginalized posterior probability distribution indicate the 16th, 50th and 84th percentiles. The best-fitting values of $\eta$ and $\ra$ are unconstrained because this stellar system is not radially anisotropic.\label{fig:MCMC-GC1-7}}
\end{figure*}

We initiate the MCMC walkers in a randomly chosen sphere in parameter space. For some snapshots, we run several fits with different initial conditions to test for any divergence. We chose flat priors restricted mainly by currently observed values for the parameters and/or by the range in which they are considered physically valid. For the MCMC fitting, we started out with around $500$ walkers and found good fits for the $N$-body model N0 without BHs and NSs. For the other models, prominently those with BHs, converging fits were only achieved with at least $2000$ walkers. On average, each MCMC chain was run for $1000$ iterations and convergence was reached after around $300$ iterations, which we trimmed from the MCMC chains for the calculation of the best-fitting parameters. The MCMC chain took on average longer to converge in snapshots with BHs than in snapshots without. In some cases, we also had to adjust the \textsc{emcee} scale parameter $a$, which is generally set to $2$, to increase the acceptance rates (for details on how this affects the MCMC algorithm see the discussion in \citealt{2013PASP..125..306F}, their Eq.~2). 

In Figs.~\ref{fig:MCMC-GC2-1} and \ref{fig:MCMC-GC1-7}, we show the marginalized posterior probability distribution for each parameter as well as the 2D projections of the posterior probability distribution representing the covariance between the different fitting parameters for two MCMC runs. Figs.~\ref{fig:MCMC-GC2-1} and \ref{fig:MCMC-GC1-7} show the results of the fitting to the $N$-body model N0 at $2.3 \trhz$, and to the $N$-body model N1 at $17.0 \trhz$, respectively. The obvious difference between the two models is that for model N1 there are two parameters, namely $\ra$ and $\eta$ that do not converge to a single value. The stellar system in this particular $N$-body snapshot is isotropic: values of $\ra$ larger than $\rt$ generate isotropic models, equally likely to reproduce the data, and for this reason, the values of $\ra$, and consequently of $\eta$, cannot be constrained. 

Looking at the 2D projections of the posterior probability distribution in Figs.~\ref{fig:MCMC-GC2-1} and \ref{fig:MCMC-GC1-7}, one can see that they are nearly circular for most of the parameter pairs. This shows that when using the full phase space information of each star, degeneracies between the different parameters can be alleviated. 

Tables \ref{tab:MCMCN1}--\ref{tab:MCMCN0} list the best-fitting parameters for all the $N$-body snapshots we considered.

\section{Comparison of multimass models and $N$-body models}
\label{sec:MMvsNB}

In the first part of our analysis, we compare the best-fitting multimass models with the results directly computed from the $N$-body snapshots for each model at all times. 
 
\subsection{Mass density profile}
\label{subsec:MDP}

\begin{figure*}
\includegraphics[width=1\textwidth]{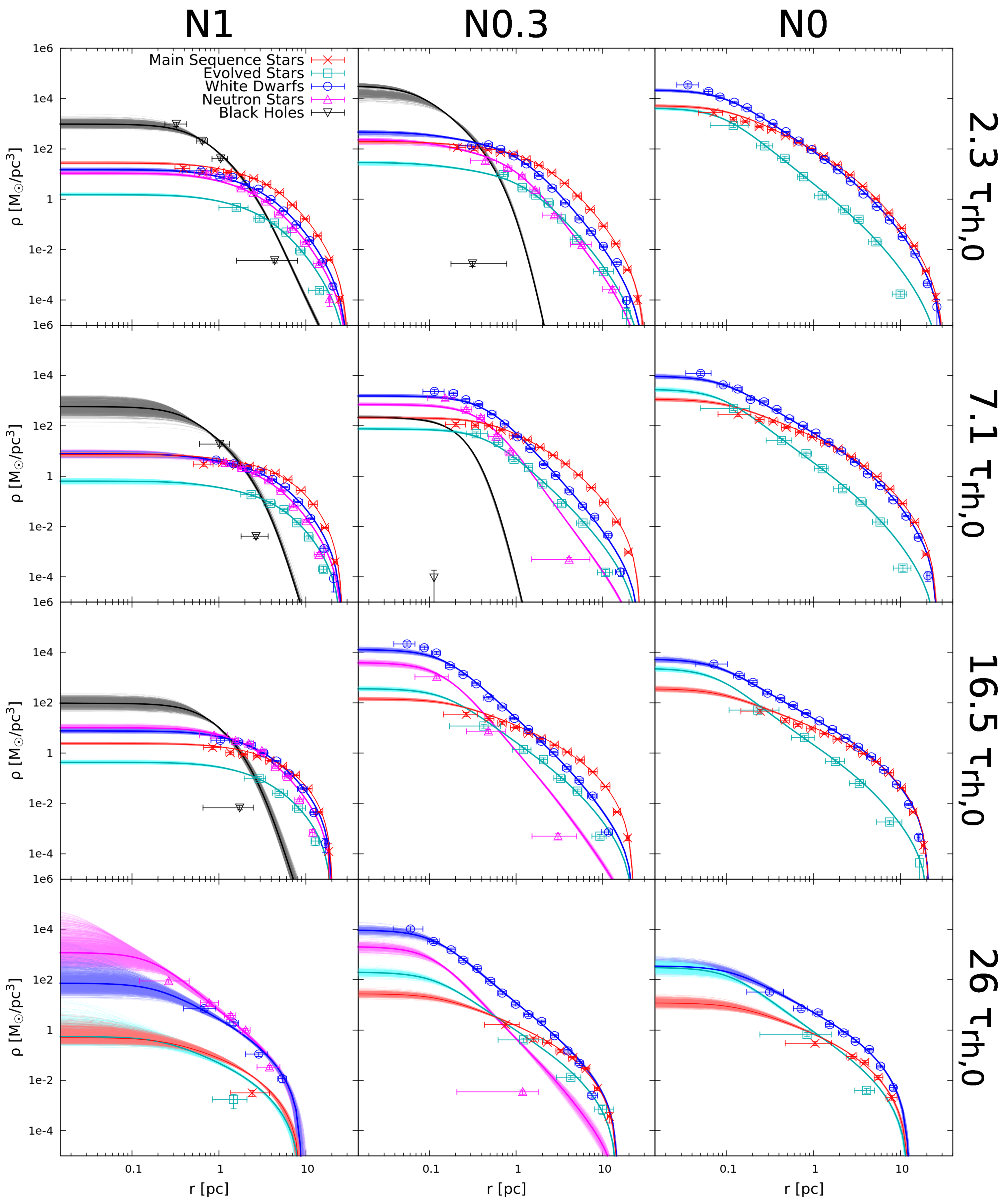} 
\caption{Comparison of the mass density profiles for models N1, N0.3 and N0 at four different ages: $2.3 \trhz$, $7.1 \trhz$, $16.5 \trhz$ and $26 \trhz$. The points represent the binned $N$-body data, the thick lines represent the best-fitting multimass models profiles and the thin lines represent the results from the walker positions at the last iteration. Red represents the MS stars, cyan -- ESs, blue -- WDs, pink -- NSs and black -- BHs. Error bars denote $1\sigma$ uncertainties. \label{fig:MDP3x4}}
\end{figure*}

First, we compare the mass density profiles of the best-fitting multimass models and the $N$-body models. For this, we binned each mass bin of the $N$-body data such that in each radial bin there are at least $30$ objects and each radial bin has a minimal radial width of $0.15 \pc$. We assumed Poisson errors for the uncertainties of the binned data and define the position uncertainty by the 16th and 84th percentiles of the distribution of the positions of the objects in each bin. In Fig.~\ref{fig:MDP3x4}, we compare the mass density profiles for the three models N1, N0.3 and N0 at four different times: $2.3 \trhz$ which is the first snapshot for each $N$-body model, $7.1 \trhz$, $16.5 \trhz$ and $26 \trhz$ the snapshot at the end of the clusters lifetime. For clarity  we only show one mass bin per stellar type. We did not include the results of model N0.1, since they are similar to the results of model N0. Together with the best-fitting result from the multimass models, we also plot the results from the walker positions at the last iteration of the MCMC routine, reflecting the uncertainties of the results.

As can be seen from Fig.~\ref{fig:MDP3x4}, the best-fitting multimass models reproduce the mass density profiles of the different mass components. Differences are only found in the outermost regions and innermost regions as well as for cases where the number of objects in a mass bin is low. For the outer regions, a difference is expected: as already discussed in Section~\ref{subsec:ABP}, the models assume that the cluster is spherically symmetric, which is not the case in the $N$-body models as the clusters are slightly elongated due to tidal forces. 

For the differences in the most central parts one sees that the mass density is underestimated for the heavier mass bins while for the lighter mass bins it is overestimated. This could be explained by the fact that these models are post core collapse, therefore their density profiles are slightly different from isothermal models \citep{1980MNRAS.191..483L}. Given that the differences in the centre are small, one can see that multimass models are able to describe post-collapse models. 

The overall agreement between multimass and $N$-body models for the density profiles is consistent with the findings of \cite{2016arXiv161002300S} who fitted Michie--King models to observational data of NGC 5466, NGC 6218 and NGC 6981: for all three GCs, the multimass models reproduce the observed mass density profiles (see their fig.~6). Differences are only found in the outer regions, most likely for the same reasons as discussed above.

When one compares the different models in Fig.~\ref{fig:MDP3x4} at the same dynamical ages it can be seen that models without BHs are denser and the stars are found far more concentrated than in models with BHs. The BHs in the centre `push' the lower mass stars out of the core, which results in a large core radius ($r_{\rm c}$) as well as a larger $\rh$, an effect already studied by \cite{P2016} for the cluster NGC 6101. In the evolution of model N0.3, one can see how the cluster changes when all BHs have been lost: the central regions get efficiently populated by the next lighter objects and the resulting mass density profile of the cluster looks as concentrated as the one from model N0 at that same dynamical age, leaving no clue about its diminished BH population.
   
\subsection{Velocity dispersion}
\label{subsec:VDP}

\begin{figure*}
\includegraphics[width=1\textwidth]{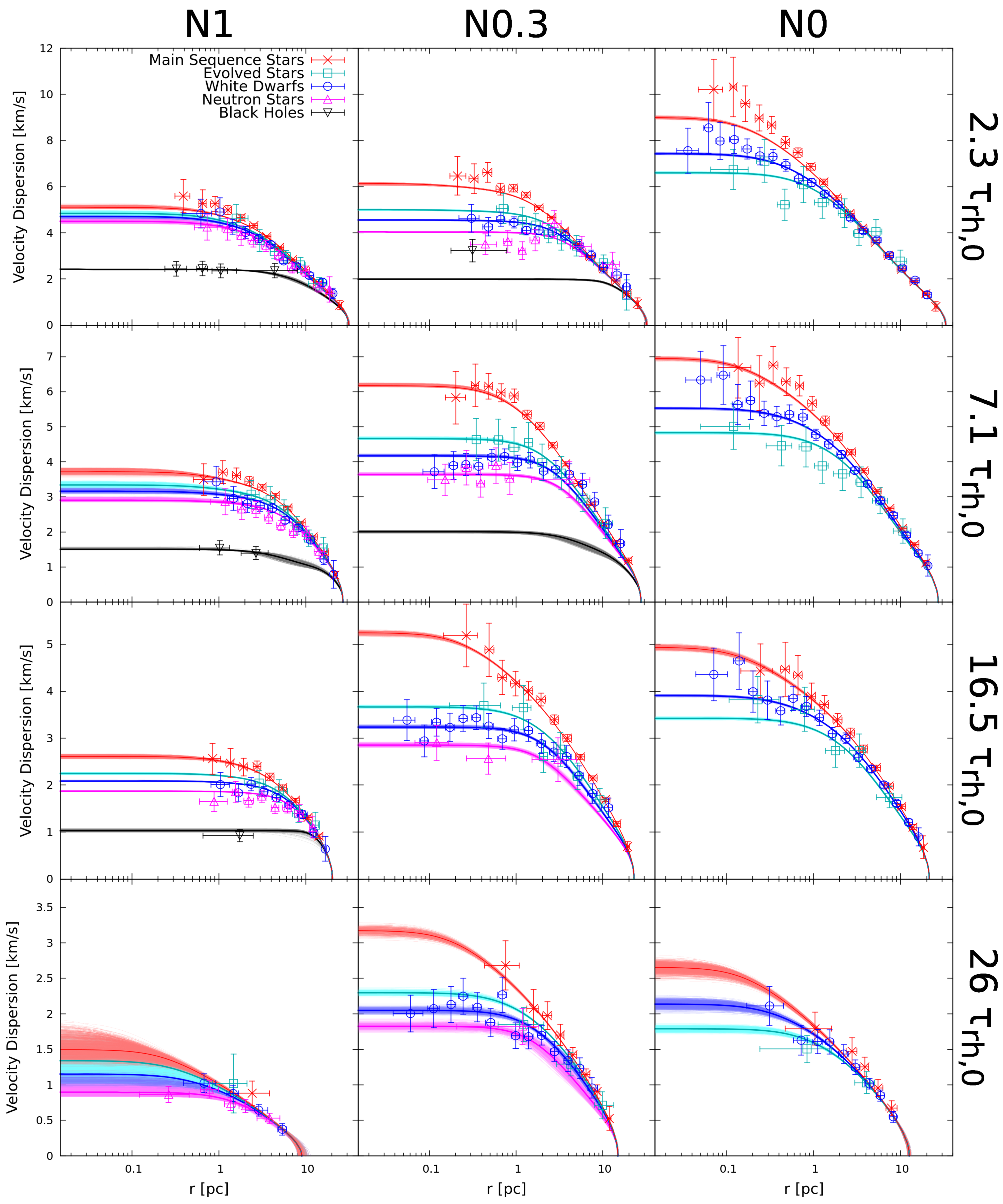} 
\caption{Comparison of the velocity dispersion profiles for models N1, N0.3 and N0 at four different ages: $2.3 \trhz$, $7.1 \trhz$, $16.5 \trhz$ and $26 \trhz$. The points represent the binned $N$-body data, the thick lines represent the best-fitting multimass models profiles and the thin lines represent the results from the walker positions at the last iteration. Red represents the MS stars, cyan -- ESs, blue -- WDs, pink -- NSs and black -- BHs. Error bars denote $1\sigma$ uncertainties. \label{fig:VDP3x4}}
\end{figure*}

For the comparison of the velocity dispersion profiles in Fig.~\ref{fig:VDP3x4}, we used the same snapshots as for the mass density profiles. For the calculations of the velocity dispersions, we are using a mass-weighted approach to make the values comparable to the values from the multimass model as they are calculated for the mean mass of each mass bin. The velocity dispersion is therefore calculated as:  
\begin{equation}
\sigma_{k}^{2}=\frac{\sum_{i}^{N}m_{i}\left[v_{k,i}-\left\langle v_{k}\right\rangle \right]^{2}}{\sum_{i}^{N}m_{i}}\qquad k=r,\theta,\phi
\end{equation}
with 
\begin{equation}
\left\langle v_{k}\right\rangle =\frac{\sum_{i}^{N}m_{i} v_{k,i}}{\sum_{i}^{N}m_{i}}\qquad k=r,\theta,\phi
\end{equation}
the mass-weighted mean velocity for each component. The calculation of the uncertainties of the binned $N$-body data was done using the description from \cite{1993ASPC...50..357P} using their equation (12). Again, the results from the best-fitting multimass models are in agreement with the data from the $N$-body models. As with the mass density profiles, small difference can be seen in the outermost regions. In the plot of model N0.3 at $7.1 \trhz$, there is no value from the $N$-body snapshot for the BHs as there is only one BH left, in which case $\sigma$ is undefined.

When comparing the different models at the same dynamical age we find that in clusters with BHs, the velocity dispersions for the different mass bins are smaller than in clusters without BHs (see discussion in Section~\ref{subsec:delta}). As the cluster is losing its BH population (see for example the evolution of model N0.3), the velocity dispersions of the different mass bins increase to the values seen in model N0 which had all its BHs removed before the actual $N$-body evolution, again leaving no hint of the lost BH population. In Section~\ref{subsec:delta}, we will look again at this relation and discuss an explanation for this behaviour. 

\subsection{Radial anisotropy}
\label{subsec:anisotropy}

In this section, we consider the anisotropy of the velocities. In Section~\ref{subsec:Beta}, we consider the anisotropy profile within the cluster and in Section~\ref{subsubsec:Kappa} we consider the global anisotropy of the cluster as a whole.

\subsubsection{Anisotropy profiles}
\label{subsec:Beta}

\begin{figure*}
\includegraphics[width=1\textwidth]{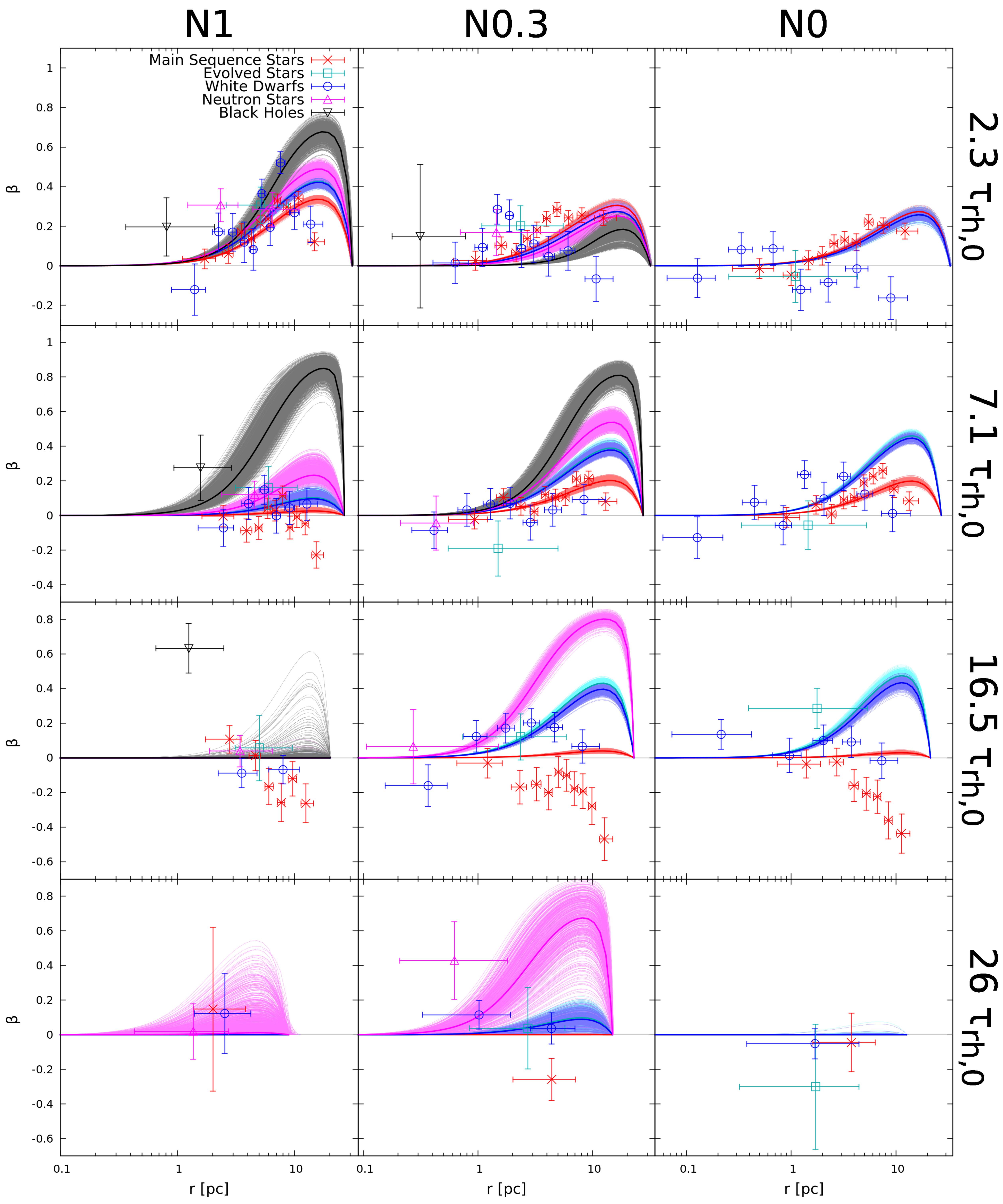} 
\caption{Comparison of the anisotropy profiles for models N1, N0.3 and N0 at four different ages: $2.3 \trhz$, $7.1 \trhz$, $16.5 \trhz$ and $26 \trhz$. The points represent the binned $N$-body data, the thick lines represent the best-fitting multimass models profiles and the thin lines represent the results from the walker positions at the last iteration. Red represents the MS stars, cyan -- ESs, blue -- WDs, pink -- NSs and black -- BHs. Error bars denote $1\sigma$ uncertainties. \label{fig:Beta4}}
\end{figure*} 

The anisotropy parameter $\beta$ is defined as \citep{1987gady.book.....B}:
\begin{equation}
\beta \equiv 1-\frac{\sigma_{\rm t}^{2}}{2\sigma_{\rm r}^{2}}
\end{equation}
with $\sigma_{\rm r} $ the radial velocity dispersion and $\sigma_{\rm t}$ the tangential velocity dispersion. For $\beta < 0$, the orbits are tangentially biased, for $\beta = 0$ they are isotropic, for $0 < \beta < 1$ they are radially biased and for $\beta = 1$ they are radial. 

In Fig.~\ref{fig:Beta4}, we compare the anisotropy profiles from the best-fitting multimass models for a selection of mass bins to the anisotropy profiles from the $N$-body snapshots. As the $\beta$ parameter is more affected by random scatter, we had to bin the data from the $N$-body snapshots differently than in the previous two plots. We varied the number of objects per bin, such that the average uncertainty in $\beta$ is $\leq 0.1$ and there are no more than $10$ radial bins per mass bin to not overcrowd the plots. For the ESs and the BHs, the average $\beta$ uncertainty was always well above $0.1$ which is why we only show one radial bin for each in all snapshots. 

Comparing the predictions from the best-fitting multimass model with the results from the $N$-body data, we find that when the snapshot has some degree of radial anisotropy the multimass models qualitatively reproduce them. This can be seen best with the mass bins from the MS stars. Also differences between the best-fitting multimass prediction and the binned data can be seen at the outer regions of the cluster. When some of the mass bins are tangentially anisotropic the best-fitting model is isotropic as our multimass models cannot describe any other kind of anisotropy.

Looking at the data from the snapshots itself, we see that the heaviest mass bins become more radially anisotropic, while the low-mass bins become first isotropic and then tangential anisotropic. We will discuss the evolution of the anisotropy further in Section~\ref{subsec:eta} when we analyse the best-fitting $\eta$ parameter. 

\subsubsection{Global anisotropy}
\label{subsubsec:Kappa}

\begin{figure*}
\includegraphics[width=1\textwidth]{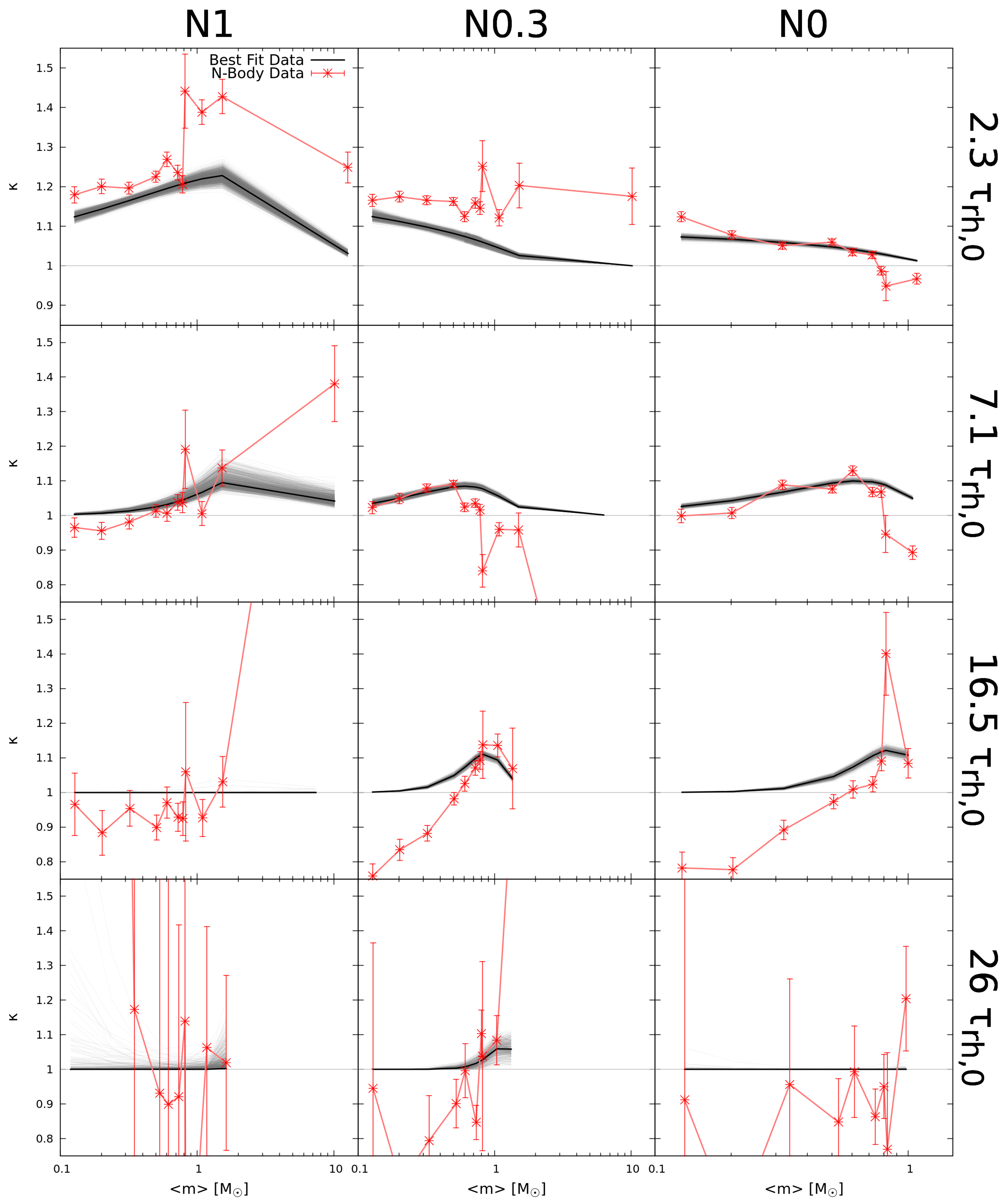} 
\caption{Comparison of the values of the global anisotropy parameter $\kappa$ as a function of mass of the different components for models N1, N0.3 and N0 at four different ages: $2.3 \trhz$, $7.1 \trhz$, $16.5 \trhz$ and $26 \trhz$. The red points represent the $N$-body data, the thick black lines represent the best-fitting multimass models values and the thin grey lines represent the results from the walker positions at the last iteration. Error bars denote $1\sigma$ uncertainties.  \label{fig:Kappa4}}
\end{figure*}

To quantify the global anisotropy we use the parameter $\kappa$, that was introduced by \cite{1981SvA....25..533P} and is defined as
\begin{equation}
\kappa=\frac{2  K_{\rm r}}{K_{\rm t}} 
\label{eq:kappa}
\end{equation}
where $K_{\rm r} = 0.5 \sum_{i} m_{i} v_{{\rm r},i}^{2}$ is the radial component of the kinetic energy and $K_{\rm t} = 0.5 \sum_{i} m_{i} v_{{\rm t},i}^{2}$ the tangential component. For $\kappa = 1$, the models are isotropic, for $\kappa > 1$ they are radially anisotropic and for $\kappa < 1$ they are tangentially anisotropic.

In Fig.~\ref{fig:Kappa4}, we compare the values of $\kappa$ obtained for the best-fitting multimass models and for the $N$-body snapshots. For the uncertainties of the $N$-body data, we used Poisson statistics. The best-fitting multimass models are able to qualitatively reproduce the overall behaviour of the $\kappa$ parameter. It can be seen that at later times, the low-mass stars are tangentially anisotropic, which cannot be reproduced by {\sc limepy}. This also explains why the best-fitting value of $\kappa$ (and $\beta$) from the multimass models does not converge to unity immediately (or $\beta\simeq0$): there are still  radial orbits left and the tangential orbits are treated as isotropic, so the best-fitting results for $\kappa$ (and $\beta$) still indicate some radial anisotropy, resulting in a smoother transition from radial anisotropy to isotropy with respect to what is observed in the $N$-body data. Clusters that are dominated by tangential orbits are therefore, by construction, not well reproduced by the multimass models used in our study (see also \citealt{2015MNRAS.451.2185S}).

Looking at the $N$-body data, we see that $\kappa$ of the lowest mass bins typically goes down with time, while $\kappa$ of the heaviest mass bins goes up. For model N1 with BHs, this change is faster than for model N0 with no BHs and NSs. We refer the reader to Section~\ref{subsec:eta} for a further analysis. 

\section{Analysis of model parameters}
\label{sec:csp}

We focus here on the best-fit ting parameters resulting from our fitting procedure. The two model scale parameters can be computed directly from the $N$-body data, therefore we use them to assess the quality of the multimass models. For the other five model parameters, we are analysing their evolution to see whether they can give us some further insights into the clusters. Furthermore, we also discuss the evolution of two additional quantities ($\rt$ and $\kappa$) to assess the quality of the models.

\subsection{Total cluster mass}
\label{subsec:MCL}

\begin{figure}
\includegraphics[width=1\columnwidth]{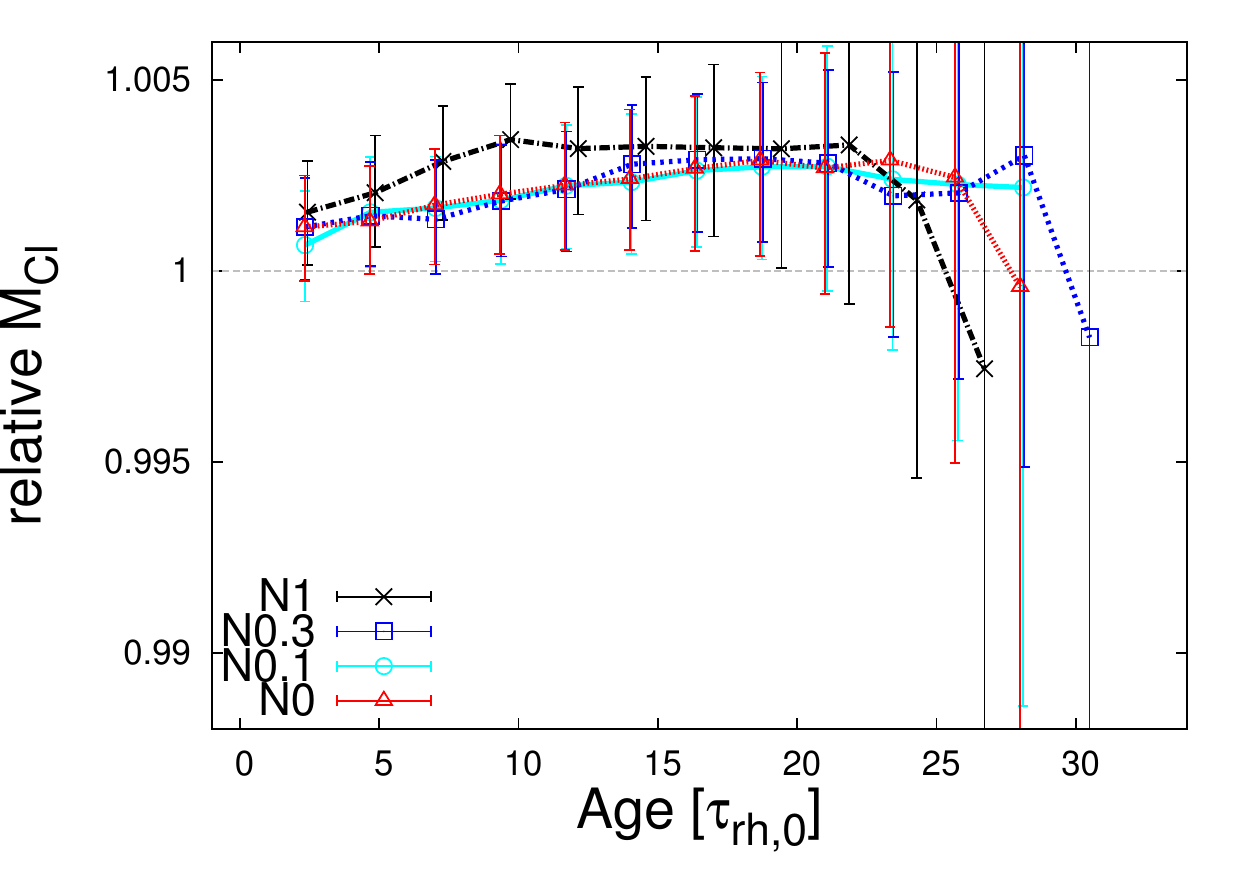} 
\caption{Best-fitting value of the total mass of the cluster, $M_{\rm Cl}$, divided by the true mass calculated from the $N$-body snapshots as a function of time in units of $\trhz$. The multimass models reproduce the true masses within $1\%$. Error bars denote $1\sigma$ uncertainties. \label{fig:RMC}}
\end{figure}

In Fig.~\ref{fig:RMC}, we plot the $M_{\rm CL}$ from the best-fitting multimass model divided by the true cluster mass as measured in the $N$-body model for all four $N$-body models throughout their cluster lifetime. As can be seen in this figure, the best-fitting value is always within $1\%$ of the $N$-body value but almost none of these are consistent within the $1\sigma$ uncertainties: there is some systematic error in the multimass models which is not accounted for yet. However, given that the difference is always smaller than $1\%$ this effect is negligible. The results are comparable to the single-mass results from \cite{Zocchi2016} where an accordance within $5\%$ of the true values was found. 

\cite{2015MNRAS.451.2185S} found that single-mass models underestimated the mass of an $N$-body system by $50\%$, depending on its dynamical state. There are several differences between their study and ours which could lead to such different results. They are using simulated observations as input for their analysis, which affects the recovery of the MF. \cite{2015MNRAS.448L..94S} found that approximating mass-segregated clusters by single-component models leads to an underestimation of the mass by a factor of two or three, especially for metal-rich GCs. Also the models used in \cite{2015MNRAS.451.2185S} have less parameters than the ones used here, as for example they do not incorporate the variable truncation parameter $g$. \cite{Zocchi2016} showed for single-mass models that the total mass is better recovered when allowing $g$ to be free. We discuss the effect of $g$ in Section~\ref{subsec:g}. 

\subsection{Half-mass radius}

\begin{figure}
\includegraphics[width=1\columnwidth]{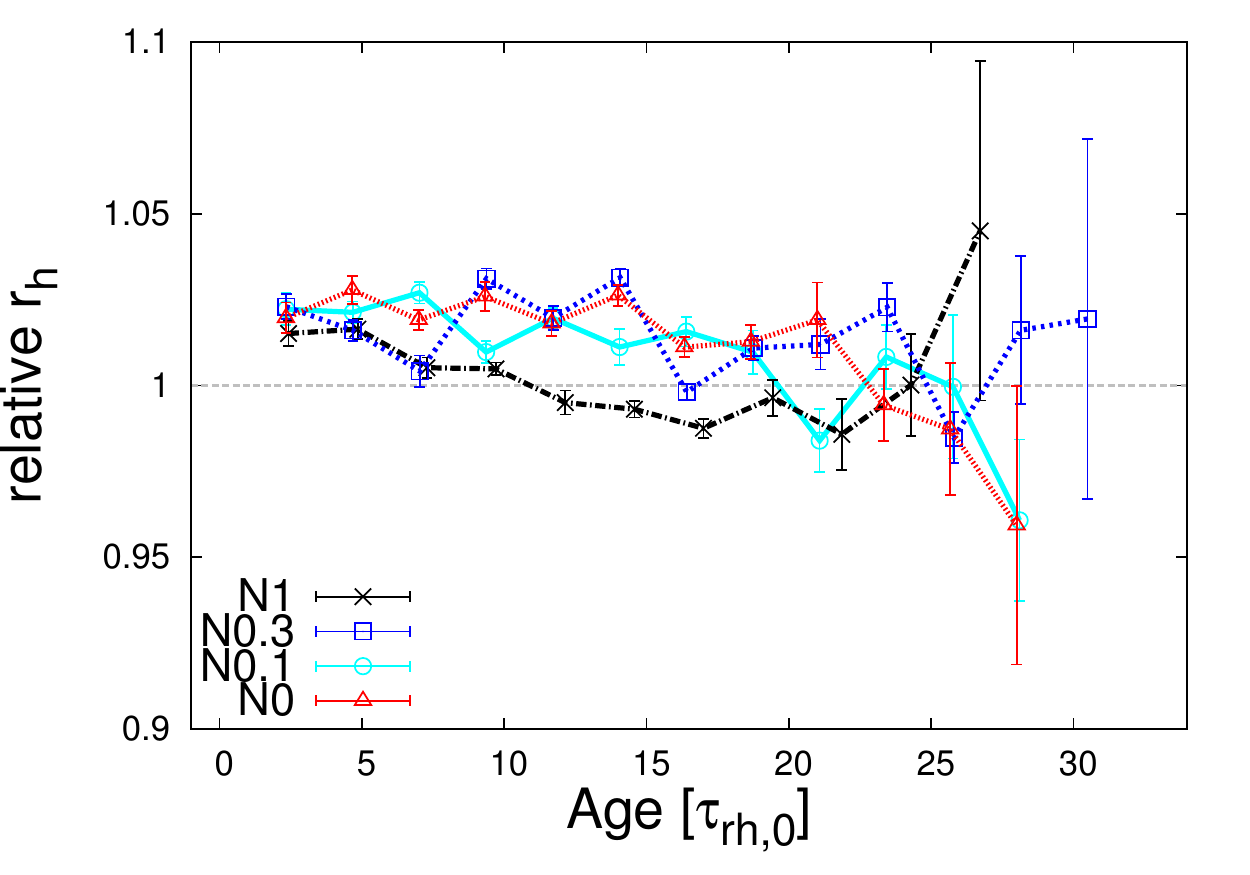} 
\caption{Best-fitting half-mass radius $\rh$ divided by the true value calculated from the $N$-body snapshots as a function of time in units of $\trhz$. The multimass models reproduce the true half-mass radii within $5\%$. Error bars denote $1\sigma$ uncertainties. \label{fig:Rh_rel}}
\end{figure}

The second scale parameter that can be computed from the $N$-body models is $\rh$: In Fig.~\ref{fig:Rh_rel}, we plot $\rh$ from the best-fitting multimass models divided by the true value as computed from the $N$-body data, for all four clusters throughout their lifetime. Again we find good agreement, within a few percent. As with $M_{\rm Cl}$, only a few of the data points show agreement within $1\sigma$ uncertainties. These results are comparable to the result for the single-mass case by \cite{Zocchi2016} who found an agreement within $7\%$. 

\subsection{Dimensionless central potential}

\begin{figure}
\includegraphics[width=1\columnwidth]{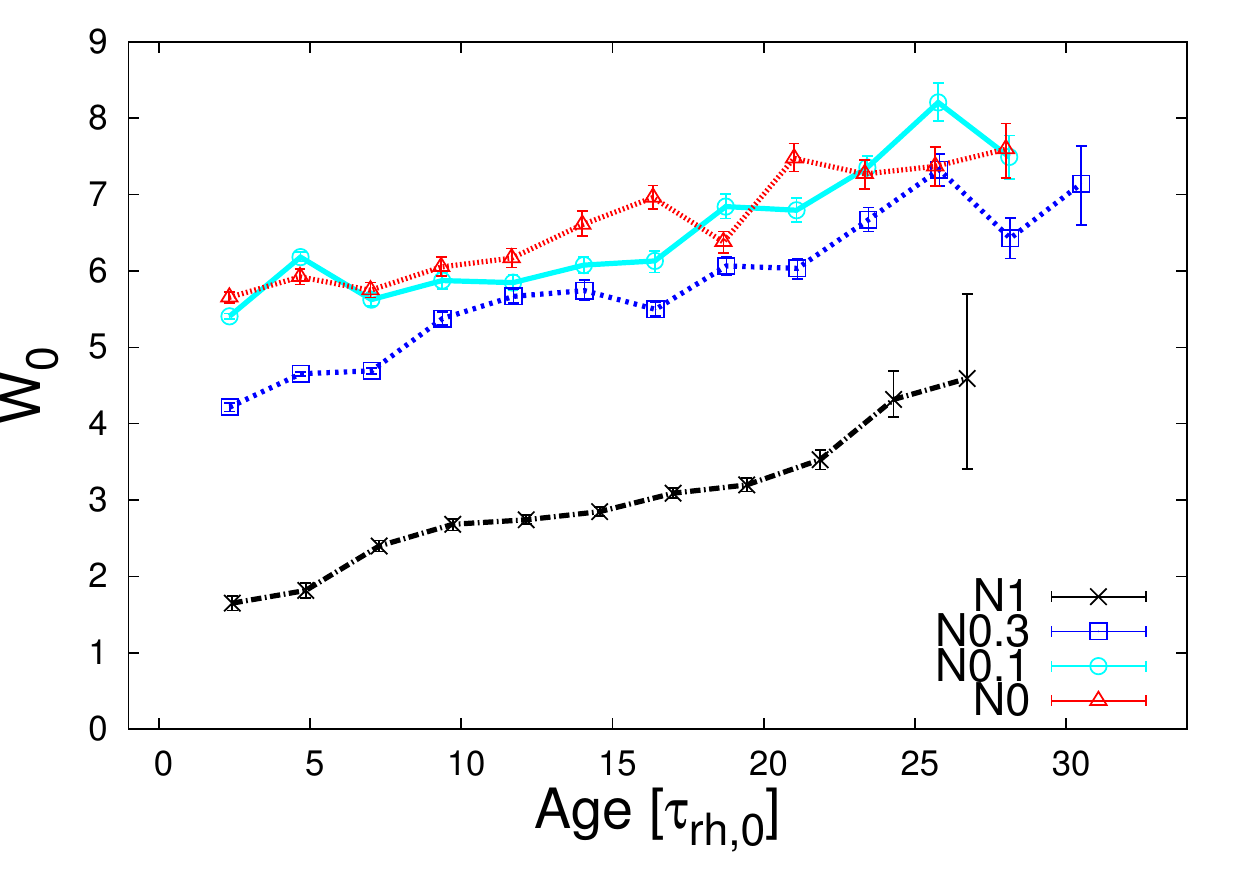} 
\caption{Best-fitting value of the central dimensionless potential, $W_0$, obtained for all four $N$-body models, as a function of time in units of $\trhz$. Error bars denote $1\sigma$ uncertainties. \label{fig:W0MM}}
\end{figure}

\begin{figure}
\includegraphics[width=1\columnwidth]{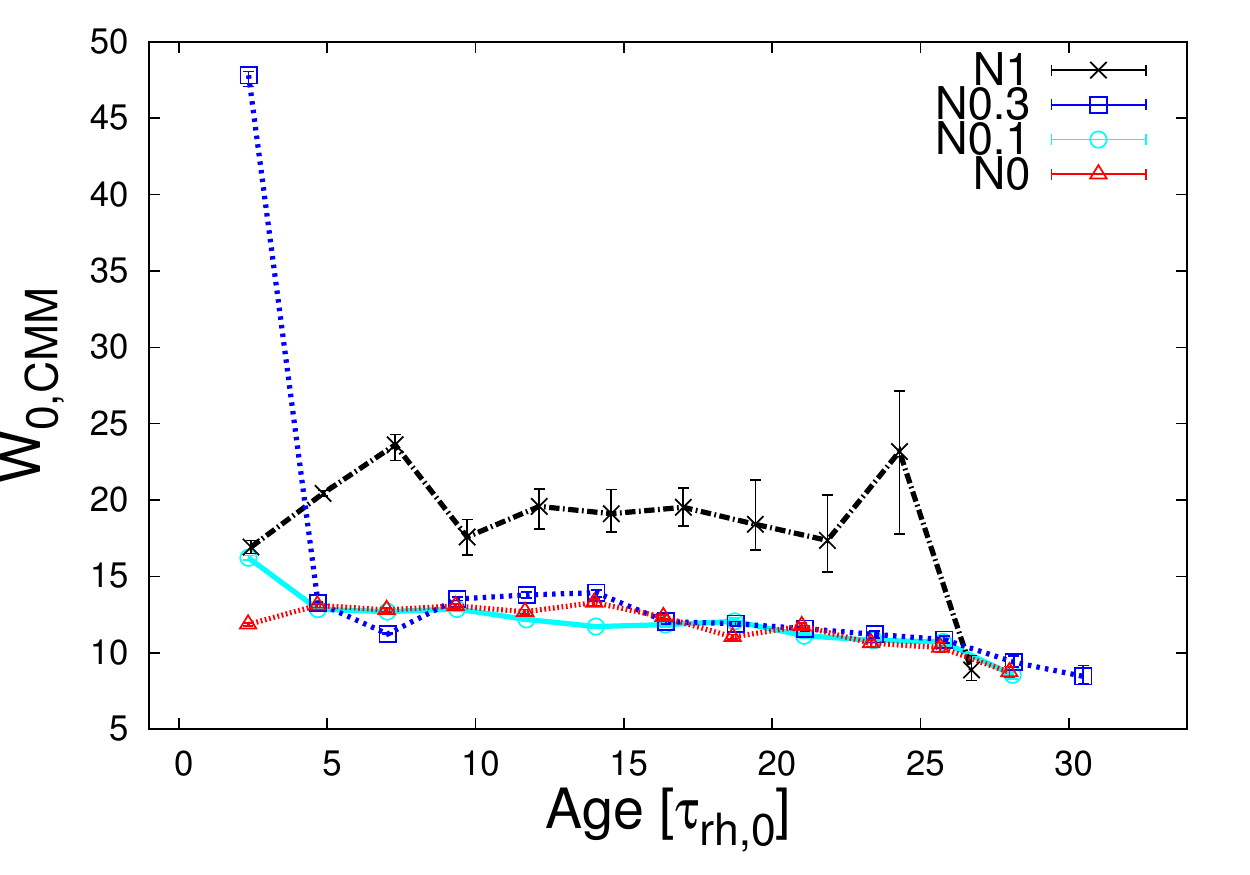} 
\caption{Central dimensionless potential obtained for the best-fitting multimass models when considering the central density weighted mean mass as the reference mass $\bar m$ for the four $N$-body models over time in units of $\trhz$. Error bars denote $1\sigma$ uncertainties. \label{fig:W0}}
\end{figure}

\begin{figure}
\includegraphics[width=1\columnwidth]{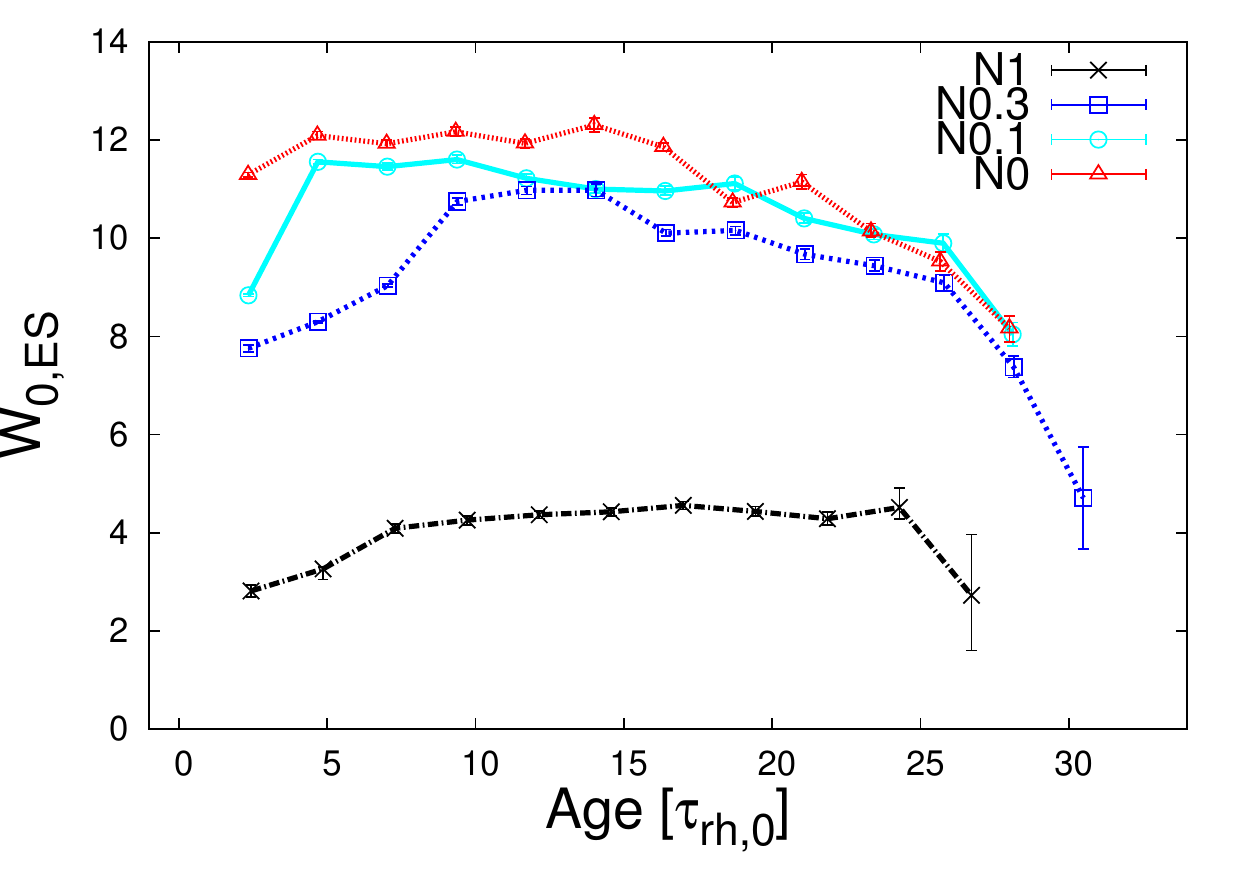} 
\caption{Central dimensionless potential of the ESs for the four $N$-body models over time in units of $\trhz$. Error bars denote $1\sigma$ uncertainties. \label{fig:W0_ES}}
\end{figure}

Now that we have shown that multimass models can reproduce the most important cluster properties, we focus on analysing the other fitting parameters. First, we look at the dimensionless central potential $W_0$ for which the evolution over the whole lifetime for the four $N$-body models is plotted in Fig.~\ref{fig:W0MM}. As discussed in Section~\ref{sec:Models}, the $W_0$ value in multimass models represents the dimensionless central potential of a hypothetical mass group with a mass equal to the global mean mass. As the global mean mass  increases from $(0.36 \pm 0.01) \msun $ to $(0.8 \pm 0.2) \msun$ during the evolution of the four $N$-body models, the $W_0$ values do not refer to the same stellar population and this needs to be kept in mind when comparing $W_0$ values of different $N$-body models and/or at different times. Using the central density weighted mean mass instead of the global mean mass has other issues, as can be seen for example in the snapshots of model N0.3 at $4.7 \trhz$ or in model N1 at $26.7 \trhz$ in Fig.~\ref{fig:W0}. In both cases, the number of BHs decreases to (almost) zero reducing the central density weighted mean mass more than the global mean mass, explaining the more significant change of the $W_0$ values in this figure.

It is also possible to use the values of $W_0$ obtained for different mass bins for a comparison. As an example we consider the ESs mass bin, because not only does the average mass of the ESs not vary in our models [$\bar{m}_{\rm ES} = (0.821 \pm 0.006) \msun$], but it also represents the objects that are easiest to observe. In Fig.~\ref{fig:W0_ES}, we plot the evolution of the $W_0$ values of the ESs for all four $N$-body models over their entire lifetime. The uncertainties are estimated by calculating the $W_0$ values of the ESs for the last 10 iterations of all the walkers and then using the values from the 16th and 84th percentiles as $1 \sigma$ uncertainties.

Looking at Figs~\ref{fig:W0MM} and \ref{fig:W0_ES}, one can see the effect of BHs as discussed in Section~\ref{subsec:MDP}: in clusters with BHs, the other stars are pushed outwards and the cluster appears less concentrated with a low value of $W_0$ for the mean mass stars and ESs. While clusters without BHs instead appear much more concentrated as the normal stars can occupy the centre and therefore have a higher $W_0$ value for the mean mass stars and for the ESs. Therefore, clusters with low $W_0$ value for the observable stars are much more likely to be hosting a BH population than clusters with a high $W_0$ value \citep{2004ApJ...608L..25M,2008MNRAS.386...65M,P2016}.

\subsection{Truncation parameter}
\label{subsec:g}

\begin{figure}
\includegraphics[width=1\columnwidth]{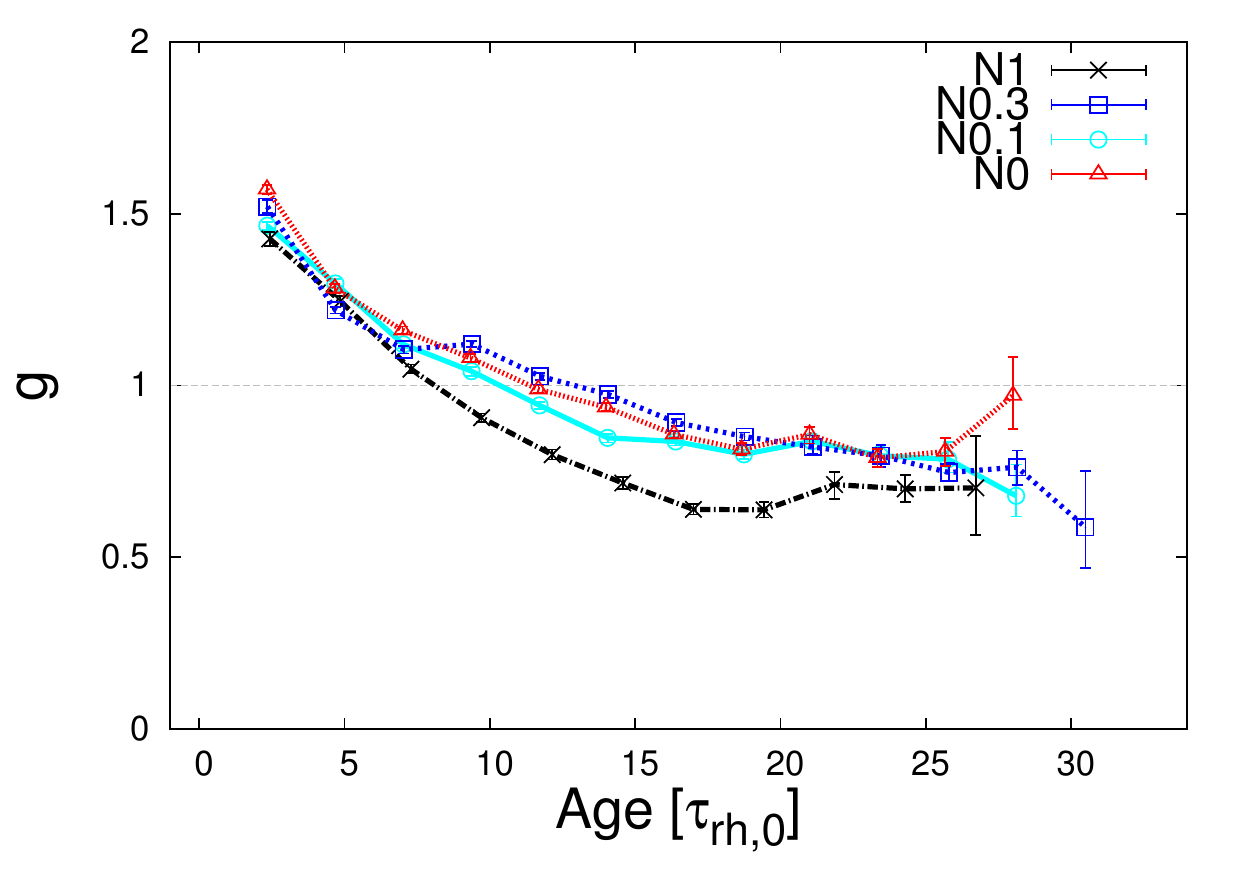} 
\caption{Best-fitting values of the truncation parameter $g$ obtained for the four $N$-body models over time in units of $\trhz$. Error bars denote $1\sigma$ uncertainties.\label{fig:G}}
\end{figure}

The truncation parameter $g$ provides an indication of the effect of external tides on the stellar system. In Fig.~\ref{fig:G}, we plot the evolution of $g$ for the four clusters over their  lifetime. The evolution is similar for all four $N$-body models: at the beginning $g$ is around $1.5$ which represents a model in between a \cite{1975AJ.....80..175W} model ($g=2$) and a \cite{1966AJ.....71...64K}  model ($g=1$). As the clusters fill their Roche volume, the tides interact with the clusters, stripping their outermost stars and thereby making the truncation in energy space steeper. This evolution is reflected in the truncation parameter $g$ decreasing as the cluster evolves, converging at the end of the lifetime to a value of around $g \approx 0.73$ which represent a model between a \cite{1966AJ.....71...64K} and a \cite{1954MNRAS.114..191W} model ($g=0$). 

The results are comparable to the single-mass model findings of \cite{Zocchi2016}, though they start with a higher truncation parameter, due to the fact that they start with a smaller initial $\rh / \rj$ ratio ($=0.01$) than we do ($\rh / \rj = 0.7$). We must note that we only use bound objects in our analysis, which might lead to smaller values of the truncation parameter as in the outer regions of the clusters ($0.8\rj - \rj$) most objects are energetically unbound \citep{2016arXiv161202253C}.

As before, we see a difference between the model with BHs (N1) and the models which are mostly BH free (N0, N0.1 and N0.3): at the beginning the value of $g$ decreases faster for model N1 than for the other three models and therefore also converges quicker to its final value. This behaviour in the first half of evolution is not too surprising given that $\rh$ of that model is roughly twice as large as the others, thereby the impact on the steepness of the truncation is stronger (see Section~\ref{subsec:NBodyDisc}).

\cite{2005ApJS..161..304M} found that Wilson models are equally good or  better in describing a sample of Galactic and extragalactic clusters than King models. With our results of the evolution of $g$, one can interpret \cite{2005ApJS..161..304M} findings that clusters with large $g$ are still dynamically expanding towards filling the Roche volume (see also \citealt{2012MNRAS.419...14C}).

\subsection{Mass segregation}
\label{subsec:delta}

\begin{figure}
\includegraphics[width=1\columnwidth]{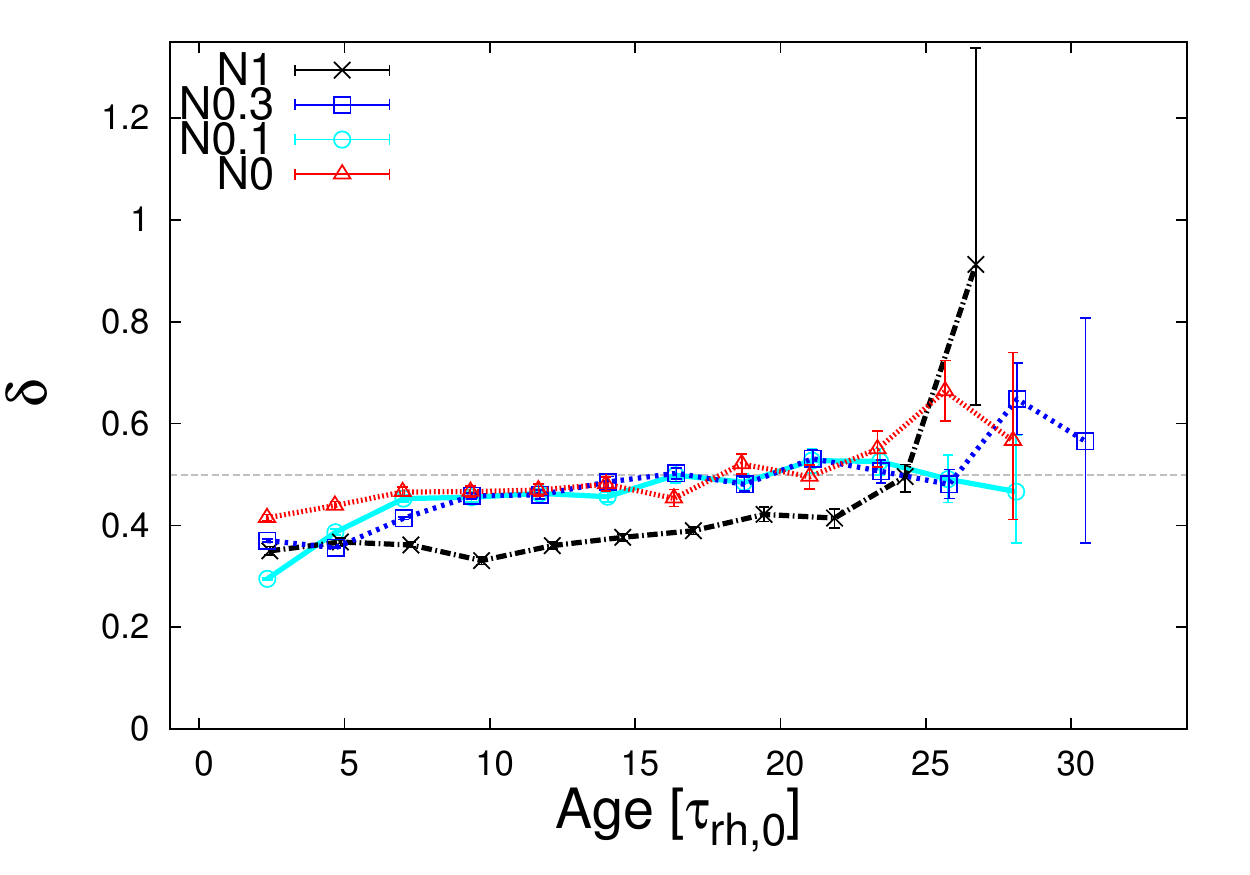} 
\caption{Best-fitting values of mass segregation parameter $\delta$ for all four $N$-body models over time in units of $\trhz$. Error bars denote $1\sigma$ uncertainties.\label{fig:Delta}}
\end{figure}

\begin{figure*}
\includegraphics[width=1\textwidth]{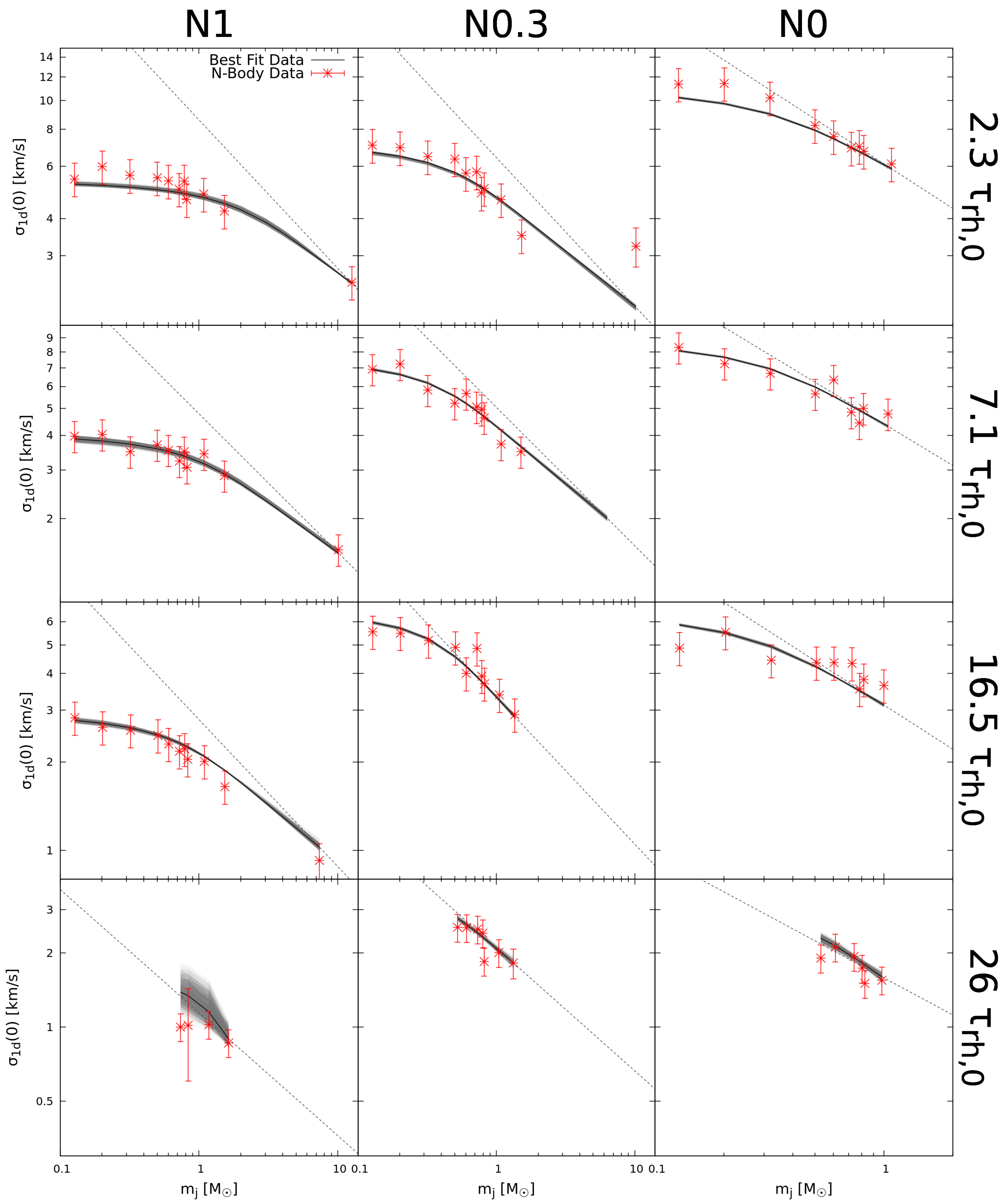} 
\caption{Comparison of the central velocity dispersion for the different mass bins for Models N1, N0.3 and N0 at four different dynamical ages: $2.3 \trhz$, $7.1 \trhz$, $16.5 \trhz$ and $26 \trhz$. The red points represent the binned $N$-body data, the black lines represent the best-fitting multimass models central velocity dispersion and the thin grey lines represent the results from the walker positions at the last iteration. The dashed black line shows a $\sigma_{1d}(0) \propto m_{j}^{-1/2}$ reference line. For the snapshots with a BH population several additional mass bins in the high-mass end were included to better show the relation. Error bars denote $1\sigma$ uncertainties.\label{fig:vd-vs-m4}}
\end{figure*}

In Fig.~\ref{fig:Delta}, we plot the evolution of the best-fitting value for $\delta$ during the whole lifetime of the four $N$-body models. In the initial stages of their evolution, all four $N$-body models are still in the process of segregation, as they were set up without any primordial mass segregation. Over the course of evolution of the clusters, the value converges to around $\delta = 0.5$. This is in accordance with findings of \cite{2016arXiv161002300S}, who study mass segregation in observations of GCs. At late stages, there are some snapshots for which the best-fitting value is $\delta\gtrsim0.5$, however the results are compatible with $0.5$ within $3\sigma$. \cite{2015MNRAS.451.2185S} found that for some of their late $N$-body snapshots, the multimass models underestimate the amount mass segregation. The same is also found for the best-fitting multimass model to the observations of NGC 6218 in \cite{2016arXiv161002300S}. However, we do not find this in these models.

To further analyse the behaviour of $\delta$ over the course of the cluster evolution, we additionally plot in Fig.~\ref{fig:vd-vs-m4} the central velocity dispersion for the different mass bins from the $N$-body data together with the predicted central velocity dispersion from the best-fitting multimass model (see also Fig.~\ref{fig:VDP3x4}). Given the results from Section~\ref{subsec:VDP}, it is no surprise that the best-fitting multimass models are able to reproduce the true values.
 
Despite the fact that the best-fitting models have $\delta \approx 0.5$, this does not mean that the multimass models are in a state of energy equipartition, as can be seen in Fig.~\ref{fig:vd-vs-m4}. This was already pointed out by \cite{1981AJ.....86..318M} and \cite{2006MNRAS.366..227M} as well as by GZ15. In a mass-segregated multimass model the following relation
\begin{equation}
m_{j}s_{j}^{2}=m_{i}s_{i}^{2}\ \forall j,i\ j\neq i
\end{equation}
holds true with $s_j$ and $s_i$ being the velocity scale of two different mass bins. For the 1D velocity dispersion at the centre the relation:
\begin{equation}
m_j s^2_j = m_j \sigma^2_{1d,j0}\ \forall j
\end{equation}  
only holds true for $W_{0}\rightarrow\infty$. In the $N$-body models, we are studying here $W_{0}\ll\infty$ and therefore $\sigma_{1d,j0} < s_j$, which means that our multimass models are never in a state of energy equipartition despite having $\delta = 0.5$. 

Furthermore \cite{2016MNRAS.458.3644B} showed, using Monte Carlo simulations, that in clusters only objects above a certain critical mass $m_{\rm eq}$ can be in energy equipartition. The value of $m_{\rm eq}$ depends on the mass spectrum of the cluster, and is larger for cluster with a wider mass spectrum, i.e. when BHs are retained. This trend can also be seen in Fig.~\ref{fig:vd-vs-m4}, where the number of mass bins following the $\sigma_{1d}(0) \propto m_{j}^{-1/2}$ relation, which are therefore in energy equipartition, is only greater than one for the models without BHs. For models with BHs, only the BHs can be in energy equipartition as they are the only ones which have a mass greater than $m_{\rm eq}$. This leads to the largest part of the other objects in these clusters having a smaller spread in the velocity dispersions, which is the reason for the reduced mass segregation in the observable stars in clusters with BHs.

Looking at the results in Fig.~\ref{fig:Delta}, we can conclude that setting the mass segregation parameter to a fixed value of $\delta = 0.5$ as in its initial formulation of the multimass models by \cite{1976ApJ...206..128D} is indeed justified to model all but the youngest clusters. As those are not yet fully mass segregated, the value of $\delta$ must therefore be smaller, something also found by \cite{2015MNRAS.451.2185S, 2016arXiv161002300S} for young clusters.

\subsection{Anisotropy radius }

\begin{figure}
\includegraphics[width=1\columnwidth]{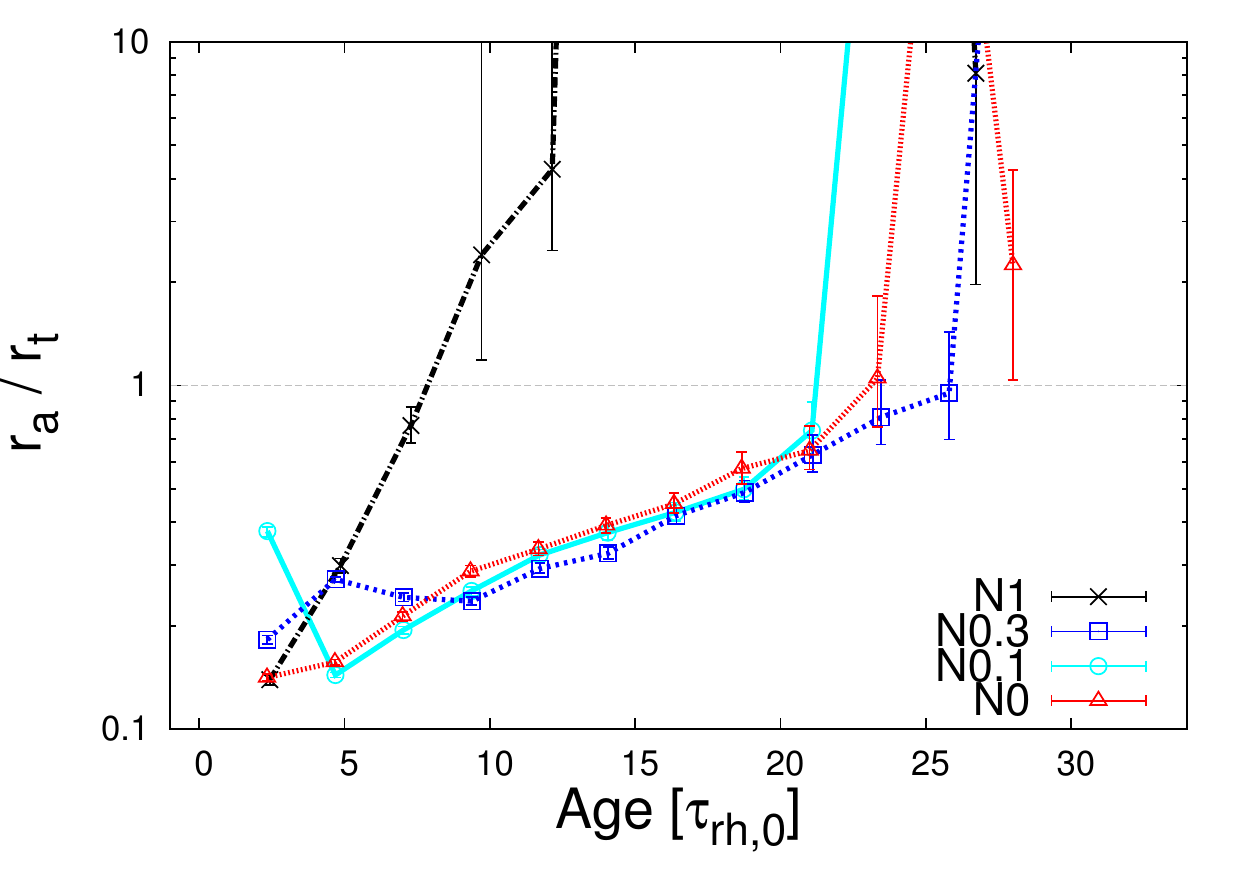} 
\caption{Anisotropy radius $\ra$ in units of the truncation radius $\rt$ as obtained for the best-fitting multimass models for all four $N$-body models over time in units of $\trhz$. Error bars denote $1\sigma$ uncertainties.\label{fig:RaRt}}
\end{figure}

The last two fitting parameters are coupled together as they both determine $r_{{\rm a},j}$ for each mass bin as can be seen in equation~(\ref{eq:raj}). 

First we focus on $\ra$. In Fig.~\ref{fig:RaRt} we plot $\ra/\rt$ of the best-fitting model, for all four $N$-body models during their lifetime. If $\ra/\rt\gtrsim1$, the cluster is isotropic (see Section~\ref{sec:Models}). Considering the definition of $r_{{\rm a},j}$ (equation~\ref{eq:raj}), it follows that even if $\ra$ is well above $\rt$ for some mass bins $r_{{\rm a},j}$ can still be below $\rt$ and therefore these mass bins still show some degree of radial anisotropy. 

Looking at Fig.~\ref{fig:RaRt}, one can see that the model which retained all its BHs (N1) behaves differently from the other models. Model N1 is only radially anisotropic at the beginning of the lifetime and quickly becomes isotropic. The model without  BHs (N0) loses its radial anisotropy more slowly and only at the end of its lifetime it becomes isotropic/tangentially anisotropic. The two models in between (N0.3 and N0.1) behave at the beginning differently: as long as they still have some BHs left their $\ra$ value evolves independently, but after all BHs are lost their $\ra$ value drops to the $\ra$ value of N0 at the same dynamical age and from then on essentially follows the $\ra$ evolution of model N0.

These results obtained for the BH-free clusters are comparable, albeit with smaller magnitude, to the results found for the single-mass case by \cite{Zocchi2016}. They showed that the anisotropy radius is monotonically decreasing till the cluster reaches core collapse after which it is monotonically increasing, and it eventually becomes so large that the corresponding model is isotropic.

Before we discuss the possible physical reasons behind the evolution of $\ra$, we first have a look at the fitting parameter $\eta$, which is needed to include a dependence of the anisotropy radius on the mass.

\subsection{Mass-dependent anisotropy}
\label{subsec:eta}

\begin{figure}
\includegraphics[width=1\columnwidth]{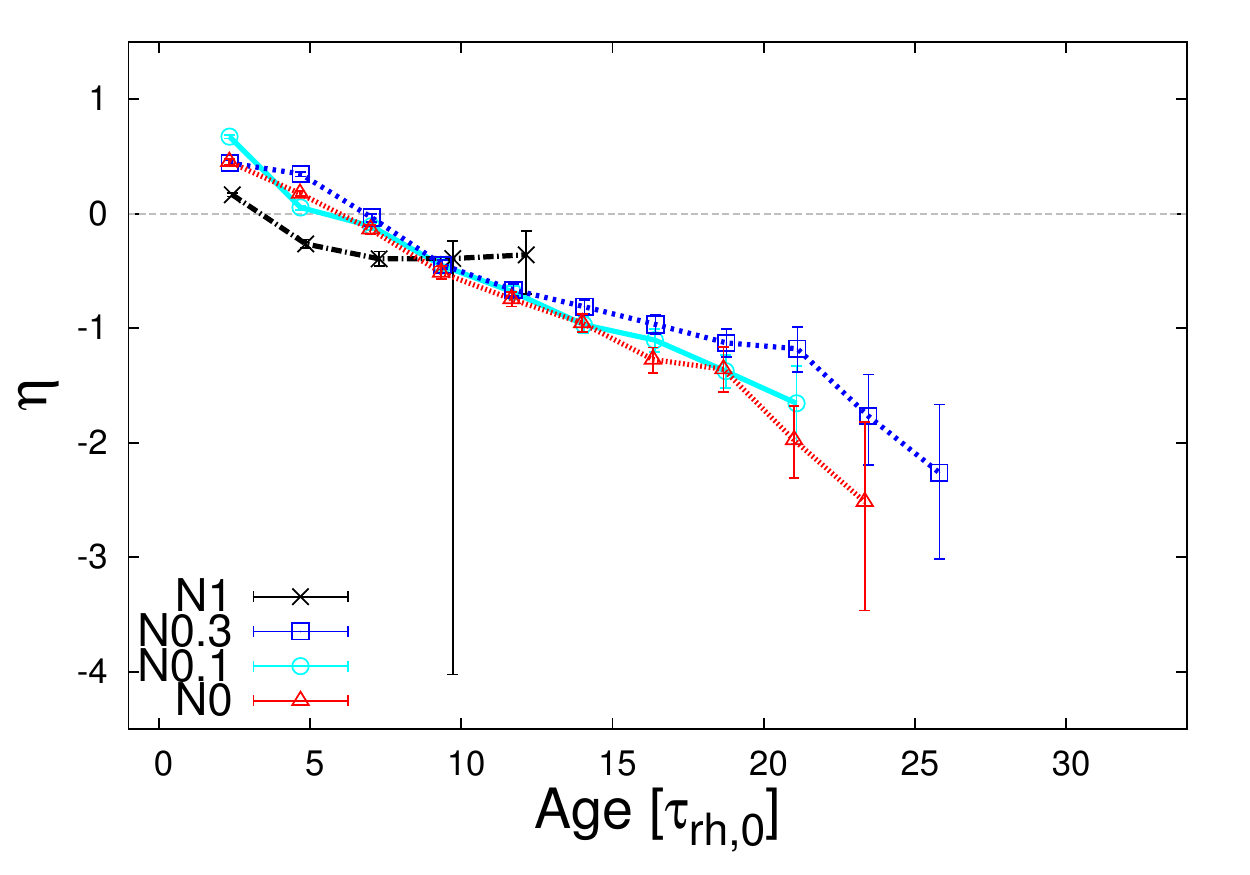} 
\caption{Best-fitting anisotropy parameter $\eta$ for all four $N$-body models over time in units of $\trhz$. Error bars denote $1\sigma$ uncertainties.\label{fig:Eta}}
\end{figure}

The anisotropy parameter $\eta$ is a novel fitting parameter in multimass models. In Fig.~\ref{fig:Eta}, we plot the values of $\eta$ for the four clusters over time. We only plot this for the snapshots showing some degree of anisotropy. The most important feature is that $\eta$ evolves for all clusters from a value of $\eta \approx 0.5$ at the beginning of the clusters lifetime to a value of $\approx -2.5$ at the end of their lifetime. The model which retains BHs throughout its lifetime (N1) stops the evolution earlier with a value of $\eta \approx -0.5$. 

The evolution of $\eta$ is not surprising given what we already saw in Section \ref{subsec:anisotropy}, where we showed that the amount of radial anisotropy is decreasing in the low-mass bins and is increasing in the high-mass bins. This is reflected in the development of $\eta$ changing from a positive value to a negative one over time. This trend is comparable to what \cite{2015MNRAS.451.2185S} found in their analysis of the \textit{W5rh1R8.5} $N$-body model from \cite{2003MNRAS.340..227B}: they found that the low-mass stars, which are preferentially located in the cluster outer regions due to mass segregation, become tangentially anisotropic. The reason for this behaviour is that interactions occurring in the cluster centre kick stars into the cluster halo on to radial orbits \citep{1968MNRAS.138..495L,1975ApJ...201..773S}. As stars on radial orbits reach the cluster boundary with positive velocity, they can escape the cluster more efficiently, thereby depleting the low-mass population from stars with radial orbits, leaving only the stars with tangential orbits in the cluster.

To test the relevance of $\eta$, we rerun fits to model N0 but this time fixing $\eta = 0$ comparable to the original formulation by \cite{1979AJ.....84..752G}. In these fits, the most obvious difference is that with increasing time, and therefore also with a higher  absolute value of $\eta$, the uncertainties of $\ra$ increase up to five times the value recovered in the fits with a non-fixed $\eta$ value. Therefore, the introduction of $\eta$ improves the ability to describe the data.

\subsection{Truncation radius}

\begin{figure}
\includegraphics[width=1\columnwidth]{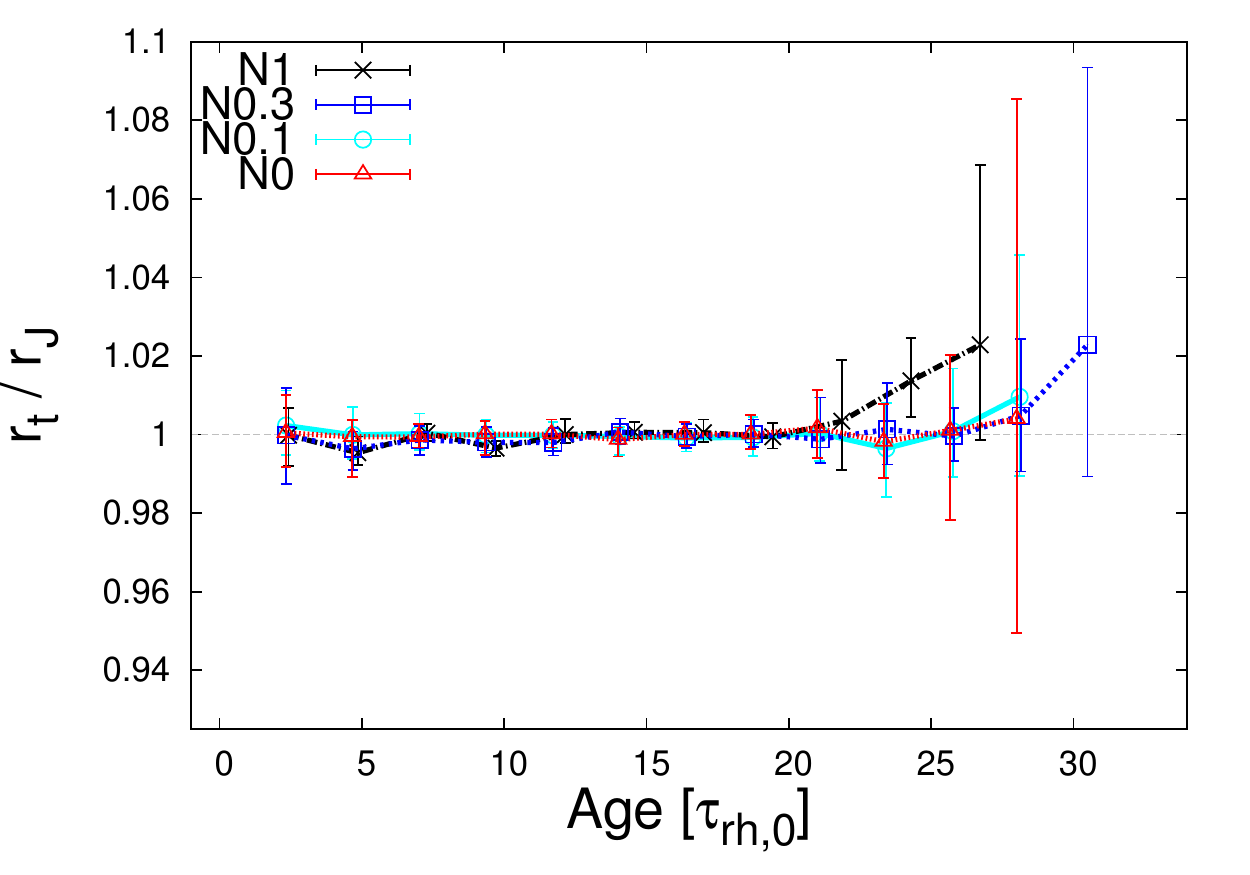} 
\caption{Ratio of the truncation radius $\rt$ obtained for the best-fitting models to the Jacobi radius $\rj$ determined from the $N$-body snapshots for all four models as a function of time in units of $\trhz$. Error bars denote $1\sigma$ uncertainties.
\label{fig:RtRj}}
\end{figure}

At the end of our comparison, we look at two quantities on which we do not fit but which get computed by the multimass models and which can be calculated for the $N$-body models. In Fig.~\ref{fig:RtRj}, we plot $\rt$ divided by $\rj$ as determined in Section~\ref{subsec:BoundSel} for the four $N$-body models over their whole lifetime. The values of $\rt$ and their uncertainties were computed using all the walker positions of the last ten iterations of the MCMC runs. This figure shows that $\rt$ stays within $3\%$ of the computed $\rj$ and in all but two cases the results are consistent within their $1\sigma$ uncertainties. The largest discrepancies can be seen at the end of the lifetime of the clusters. Compared to the single-mass models in \cite{Zocchi2016} which showed divergence of a factor of two, the multimass models are able to reproduce $\rt$ accurately.

\subsection{Global anisotropy parameter}

\begin{figure}
\includegraphics[width=1\columnwidth]{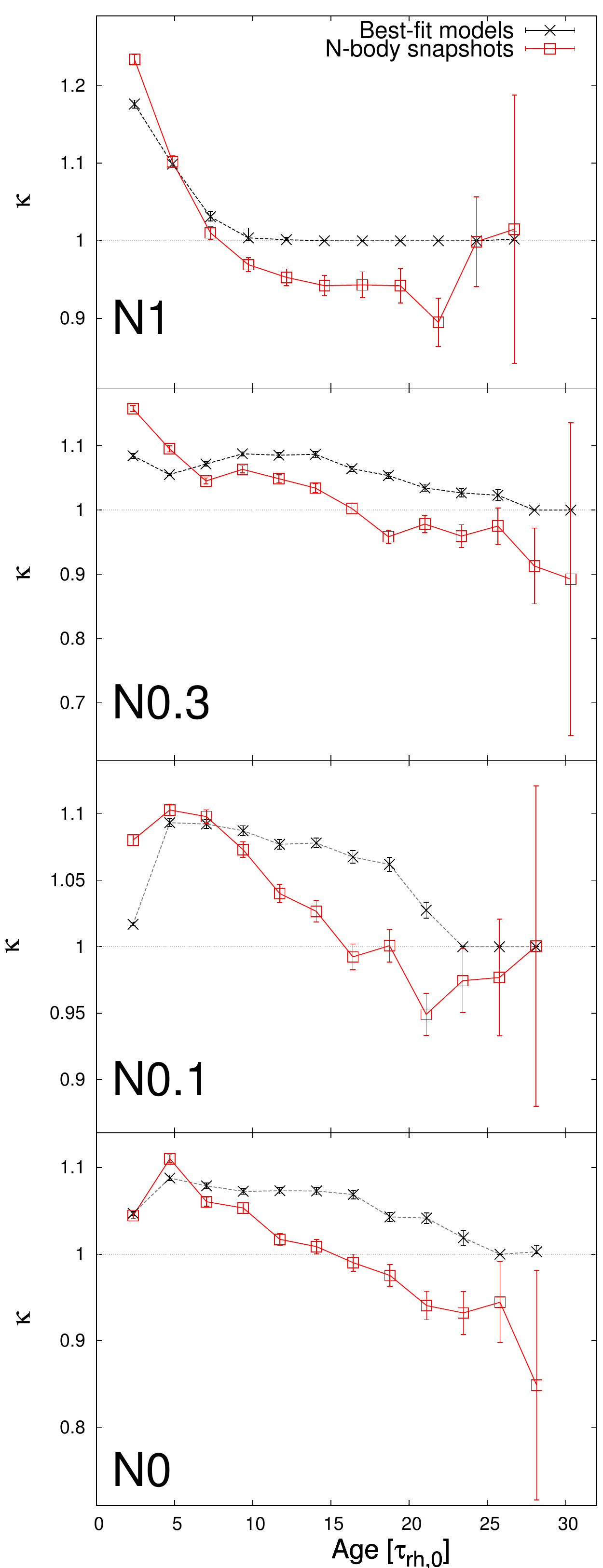} 
\caption{Comparison of the values of the global anisotropy parameter $\kappa$ for the four different $N$-body models over their whole lifetime. The black dashed lines represent the best-fitting multimass estimate, and the red solid lines represent the true values directly calculated from the $N$-body snapshots. \label{fig:GKappa}}
\end{figure}

In Fig.~\ref{fig:GKappa}, we look at the evolution of the global value of $\kappa$. Here, we have plotted the comparison between the best-fitting value inferred using the walker positions of the last 10 iterations of the MCMC runs from each snapshots to the one calculated from the $N$-body snapshots directly. These are calculated applying equation (\ref{eq:kappa}) to all objects in the $N$-body model. For the uncertainties of the $N$-body data, we used Poisson statistics. As before the best-fitting multimass models qualitatively reproduce the overall trend as long as the $N$-body model is radially anisotropic. When the $N$-body snapshots becomes tangentially anisotropic, the best-fitting multimass models are isotropic. Hence, the multimass $\kappa$ values still shows some radial anisotropy where the true cluster is already dominated by tangentially anisotropic orbits. For all models $\kappa < 1.7 \pm 0.25$, from which it follows that all models are stable against radial orbit instability as discussed in  \cite{1981SvA....25..533P}.

\section{Discussion and conclusion}
\label{sec:Conclusion}

In this study, we assessed the validity of the multimass anisotropic models provided by the \textsc{limepy} software (GZ15) fitting them to $N$-body models. We find that the $N$-body models are well described  by multimass models, a result which is fortunate given the long list of observational studies using multimass models to analyse GCs (see Section~\ref{sec:Intro}). \cite{Zocchi2016} showed for the single-mass case that the \textsc{limepy} models are able to describe clusters at all evolutionary phases. Although the agreement is not perfect, the systematic differences are negligible for most applications and parameters of interest (see also the discussion in Section~\ref{subsec:MCL}).

Our comparison shows that the best-fitting total cluster masses are off by no more than $1\%$ from the true value as computed from the $N$-body snapshots. The best-fitting cluster half-mass radius is reproduced within $5\%$  and the truncation radius is reproduced within $3\%$.  

We find that the mass density and velocity dispersion profiles of the different mass bins are well reproduced by the multimass models. If the $N$-body snapshot is radially anisotropic then the multimass models are generally able to reproduce it. 

We show that in the $N$-body models, regardless of initial BHs and NSs retention, the truncation parameter $g$ evolves from roughly $1.5$ to about $0.7$. The general trend can be explained by the tidal effects stripping the loosely bound stars. We find that the best-fitting mass segregation parameter $\delta$ converges to a value close to $0.5$ for our $N$-body models, which is the value used in the original formulation by \cite{1976ApJ...206..128D}. Only for young clusters which are not yet mass segregated is the best-fitting value smaller and for models with BHs it is $\sim 0.4$.

The newly introduced $\eta$ parameter shows that the anisotropy radius is mass dependent and that this mass dependence changes in our $N$-body models over time from $\eta = 0.5$ where the lighter stars are more radially anisotropic to $\eta = -2.5$ ($\eta = -0.5$ for the model which initially retained all its BHs) where the heavy objects are more radially anisotropic. We find in this study that the effects which influence the anisotropy radius are more mass dependent than initially thought and therefore $\eta$ is another relevant parameter when analysing radial anisotropy with multimass models. Furthermore, we find that clusters with a BH population can be tangentially anisotropic for most of their lifetime.

The $W_0$ parameter for the observable ESs is lower for the clusters with BHs than for the clusters without BHs. Therefore, clusters which still harbour a stellar-mass BH population should appear less dense when looking at the observable stars. $N$-body simulation N0.1, which loses its BHs within its first $\trhz$, does not show any strong differences to simulation N0, despite having a population of NSs. The influence of the NSs on a cluster is therefore negligible, compared to the impact stellar-mass BHs have on a cluster.

We conclude that the \textsc{limepy} multimass models are an adequate tool to study the global properties of GCs, as the results from the comparison with $N$-body models show a good agreement with their properties inferred from multimass models.

\section*{Acknowledgements}

This research was done as part of the \textit{Gaia} Challenge\footnote{\url{http://astrowiki.ph.surrey.ac.uk/dokuwiki/doku.php?id=start}}, whose goal is to compare and improve mass modelling techniques in preparation of data releases of the ESA \textit{Gaia} mission. We thank the organizers of the Third Gaia Challenge Workshop (Barcelona, 2015) for a very productive meeting. We thank the anonymous referee for constructive suggestions. We are grateful to Anna Lisa Varri and Antonio Sollima for interesting discussions. We are grateful to Sverre Aarseth and Keigo Nitadori for making \textsc{nbody6} publicly available, and to Dan Foreman-Mackey for providing the \textsc{emcee} software and for maintaining the online documentation; we also thank Mr. Dave Munro of the University of Surrey for hardware and software support. MG acknowledges financial support from the Royal Society (University Research Fellowship) and AZ acknowledges financial support from the Royal Society (Newton International Fellowship). MP, MG and AZ acknowledge the European Research Council (ERC-StG-335936, CLUSTERS). VHB acknowledges support from the Radboud Excellence Initiative Fellowship.

\bibliographystyle{mnras}

\appendix

\section{$N$-body data and MCMC results}

\begin{table*}
\caption{Properties of the snapshots from the $N$-body model N1 with 100\% initial BH and NS retention. We list the age in units of the initial half-mass relaxation time, the total bound mass in $\msun$, the half-mass radius in pc, the number of BHs in the cluster, the number of NSs in the cluster and the Jacobi radius $\rj$ in pc calculated as in Section~\ref{subsec:BoundSel}. For the bound mass and the number of BHs and NSs, we also give the percentage relative to the initial values in brackets.}
\label{tab:NBodyN1}
\begin{tabular}{lccccc}
\hline
Age      & $M_{\rm CL}$         & Half-mass radius & Number of BHs    & Number of NSs    & $\rj$    \\
($\trhz$)& ($\msun$)            & (pc)             &                  &                  &  (pc)      \\
\hline
$0$     & $37353$ ($100\%$) & $2.06$           & $121$ ($100\%$) & $632$ ($100\%$) & $30.12$  \\
$2.4$   & $30408$ ($81\%$)  & $5.03$           & $119$ ($56\%$)  & $551$ ($87\%$)  & $27.98$  \\
$4.9$   & $25737$ ($69\%$)  & $6.03$           & $85$ ($40\%$)   & $518$ ($82\%$)  & $26.51$  \\
$7.3$   & $22077$ ($59\%$)  & $6.56$           & $59$ ($28\%$)   & $503$ ($80\%$)  & $25.24$  \\
$9.7$   & $19109$ ($51\%$)  & $6.58$           & $55$ ($26\%$)   & $490$ ($78\%$)  & $24.08$  \\
$12.1$  & $15827$ ($42\%$)  & $6.54$           & $43$ ($20\%$)   & $474$ ($75\%$)  & $22.68$  \\
$14.6$  & $12618$ ($34\%$)  & $6.27$           & $35$ ($17\%$)   & $461$ ($73\%$)  & $21.10$  \\
$17.0$  & $9434$ ($25\%$)   & $5.71$           & $26$ ($12\%$)   & $439$ ($69\%$)  & $19.22$  \\
$19.4$  & $6602$ ($18\%$)   & $4.98$           & $18$ ($8.5\%$)  & $416$ ($66\%$)  & $17.11$  \\
$21.9$  & $4094$ ($11\%$)   & $4.09$           & $12$ ($5.7\%$)  & $368$ ($58\%$)  & $14.62$  \\
$24.3$  & $2046$ ($5.5\%$)  & $2.92$           & $6$ ($2.8\%$)   & $301$ ($48\%$)  & $11.68$  \\
$26.7$  & $497$ ($1.3\%$)   & $1.58$           & $0$ ($0.0\%$)   & $149$ ($24\%$)  & $7.52$   \\
\hline
\end{tabular}
\end{table*}

\begin{table*}
\caption{Properties of the snapshots from the $N$-body model N0.3 with 33\% initial BH and NS retention. We list the age in units of the initial half-mass relaxation time, the total bound mass in $\msun$, the half-mass radius in pc, the number of BHs in the cluster, the number of NSs in the cluster and the Jacobi radius $\rj$ in pc calculated as in Section~\ref{subsec:BoundSel}. For the bound mass and the number of BHs and NSs, we also give the percentage relative to the initial values in brackets.}
\label{tab:NBodyN0.3}
\begin{tabular}{lccccc}
\hline
Age     & $M_{\rm CL}$         & Half-mass radius & Number of BHs    & Number of NSs    & $\rj$    \\
($\trhz$) & ($\msun$)           & (pc)               &                 &                 &  (pc)      \\
\hline
$0$     & $34404$ ($100\%$) & $2.06$           & $66$ ($100\%$)  & $211$ ($100\%$) &  $29.33$  \\
$2.3$   & $31131$ ($90\%$)  & $3.07$           & $22$ ($33\%$)   & $199$ ($94\%$)  &  $28.14$  \\
$4.7$   & $29061$ ($84\%$)  & $3.13$           & $8$ ($12\%$)    & $189$ ($90\%$)  &  $27.50$  \\
$7.0$   & $26824$ ($78\%$)  & $3.12$           & $1$ ($1.5\%$)   & $179$ ($85\%$)  &  $26.78$  \\
$9.4$   & $24151$ ($70\%$)  & $3.27$           & $0$ ($0\%$)     & $150$ ($71\%$)  &  $25.88$  \\
$11.7$  & $21002$ ($61\%$)  & $3.57$           & $0$ ($0\%$)     & $119$ ($56\%$)  &  $24.72$  \\
$14.1$  & $17922$ ($52\%$)  & $3.57$           & $0$ ($0\%$)     & $94$ ($45\%$)   &  $23.45$  \\
$16.4$  & $14861$ ($43\%$)  & $3.61$           & $0$ ($0\%$)     & $75$ ($36\%$)   &  $22.06$  \\
$18.8$  & $11785$ ($34\%$)  & $3.43$           & $0$ ($0\%$)     & $59$ ($28\%$)   &  $20.45$  \\
$21.1$  & $8808$ ($26\%$)   & $3.28$           & $0$ ($0\%$)     & $46$ ($22\%$)   &  $18.59$  \\
$23.5$  & $6077$ ($18\%$)   & $2.88$           & $0$ ($0\%$)     & $37$ ($18\%$)   &  $16.49$  \\
$25.8$  & $3651$ ($11\%$)   & $2.61$           & $0$ ($0\%$)     & $26$ ($12\%$)   &  $13.99$  \\
$28.1$  & $1489$ ($4.3\%$)  & $1.89$           & $0$ ($0\%$)     & $15$ ($7.1\%$)  &  $10.53$  \\
$30.5$  & $268$ ($0.8\%$)   & $1.08$           & $0$ ($0\%$)     & $7$ ($3.3\%$)   &  $6.06$   \\
\hline
\end{tabular}
\end{table*}

\begin{table*}
\caption{Properties of the snapshots from the $N$-body model N0.1 with 10\% initial BH and NS retention. We list the age in units of the initial half-mass relaxation time, the total bound mass in $\msun$, the half-mass radius in pc, the number of BHs in the cluster, the number of NSs in the cluster and the Jacobi radius $\rj$ in pc calculated as in Section~\ref{subsec:BoundSel}. For the bound mass and the number of BHs and NSs, we also give the percentage relative to the initial values in brackets.}
\label{tab:NBodyN0.1}
\begin{tabular}{lccccc}
\hline
Age     & $M_{\rm CL}$         & Half-mass radius & Number of BHs    & Number of NSs    & $\rj$    \\
($\trhz$) & ($\msun$)           & (pc)               &                 &                 &  (pc)      \\
\hline
$0$     & $33476$ ($100\%$) & $2.04$           & $22$ ($100\%$)  & $56$ ($100\%$)  & $29.06$  \\
$2.3$   & $31156$ ($93\%$)  & $2.46$           & $5$ ($23\%$)    & $54$ ($96\%$)   & $28.15$  \\
$4.7$   & $28719$ ($86\%$)  & $2.76$           & $0$ ($0\%$)     & $50$ ($89\%$)   & $27.41$  \\
$7.0$   & $25388$ ($76\%$)  & $3.27$           & $0$ ($0\%$)     & $24$ ($43\%$)   & $26.33$  \\
$9.4$   & $22112$ ($66\%$)  & $3.52$           & $0$ ($0\%$)     & $19$ ($34\%$)   & $25.16$  \\
$11.7$  & $18956$ ($57\%$)  & $3.60$           & $0$ ($0\%$)     & $12$ ($21\%$)   & $23.91$  \\
$14.1$  & $15809$ ($47\%$)  & $3.65$           & $0$ ($0\%$)     & $7$ ($13\%$)    & $22.53$  \\
$16.4$  & $12625$ ($38\%$)  & $3.52$           & $0$ ($0\%$)     & $5$ ($8.9\%$)   & $20.94$  \\
$18.7$  & $9772$ ($29\%$)   & $3.42$           & $0$ ($0\%$)     & $3$ ($5.4\%$)   & $19.26$  \\
$21.1$  & $6926$ ($21\%$)   & $3.11$           & $0$ ($0\%$)     & $1$ ($1.8\%$)   & $17.23$  \\
$23.4$  & $4373$ ($13\%$)   & $2.75$           & $0$ ($0\%$)     & $0$ ($0\%$)     & $14.85$  \\
$25.8$  & $2172$ ($6.5\%$)  & $2.29$           & $0$ ($0\%$)     & $0$ ($0\%$)     & $11.89$  \\
$28.1$  & $6780$ ($2.0\%$)  & $1.48$           & $0$ ($0\%$)     & $0$ ($0\%$)     & $8.19$   \\
\hline
\end{tabular}
\end{table*}

\begin{table*}
\caption{Properties of the snapshots from the $N$-body model N0 with no initial BH and NS retention. We list the age in units of the initial half-mass relaxation time, the total bound mass in $\msun$, the half-mass radius in pc and the Jacobi radius $\rj$ in pc calculated as in Section~\ref{subsec:BoundSel}. For the bound mass we also give the percentage relative to the initial value in brackets.}
\label{tab:NBodyN0}
\begin{tabular}{lcccc}
\hline
Age     & $M_{\rm Cl}$        & Half-mass radius & $\rj$    \\
($\trhz$) & ($\msun$)           & (pc)               & (pc)      \\
\hline
$0$     & $33042$ ($100\%$) & $2.04$           & $28.93$    \\
$2.3$   & $30886$ ($93\%$)  & $2.26$           & $28.07$    \\
$4.7$   & $27539$ ($83\%$)  & $2.93$           & $27.03$    \\
$7.0$   & $24261$ ($73\%$)  & $3.25$           & $25.93$    \\
$9.3$   & $21223$ ($64\%$)  & $3.46$           & $24.80$    \\
$11.7$  & $18167$ ($55\%$)  & $3.53$           & $23.56$    \\
$14.0$  & $15053$ ($46\%$)  & $3.56$           & $22.15$    \\
$16.3$  & $12026$ ($36\%$)  & $3.47$           & $20.58$    \\
$18.7$  & $9183$ ($28\%$)   & $3.18$           & $18.84$    \\
$21.0$  & $6367$ ($19\%$)   & $2.89$           & $16.75$    \\
$23.3$  & $3955$ ($12\%$)   & $2.55$           & $14.33$    \\
$25.7$  & $1951$ ($5.9\%$)  & $2.12$           & $11.42$    \\
$28.0$  & $455$ ($1.4\%$)   & $1.06$           & $7.18$    \\
\hline
\end{tabular}
\end{table*}

\begin{landscape}

\begin{table}
\caption{Mass bins of the different snapshots of $N$-body model N1. We list the age in units of the initial half-mass relaxation time and for each mass bin the total mass $M_j$ and the mean mass $m_j$ in units of $\msun$. There are a total of 11 mass bins: five for the MSs, one for the ESs, three for the WDs and one each for the NS and BHs.}
\label{tab:MFN1}
\begin{tiny}
\begin{tabular}{lccccccccccccccccccccccc}
\hline
Age     & \multicolumn{2}{c|}{MS1}       & \multicolumn{2}{c|}{MS2}       & \multicolumn{2}{c|}{MS3}        & \multicolumn{2}{c|}{MS4}      & \multicolumn{2}{c|}{MS5}        & \multicolumn{2}{c|}{ES}       & \multicolumn{2}{c|}{WD1}      & \multicolumn{2}{c|}{WD2}       & \multicolumn{2}{c|}{WD3}       & \multicolumn{2}{c|}{NS}      & \multicolumn{2}{c|}{BH}      \\
        & $\Mj$     &  $\mj$ & $\Mj$     &  $\mj$ & $\Mj$ &  $\mj$      & $\Mj$ & $\mj$     &  $\Mj$ &  $\mj$     &  $\Mj$ &  $\mj$   & $\Mj$ & $\mj$     & $\Mj$ &  $\mj$     & $\Mj$ &  $\mj$     & $\Mj$ &  $\mj$   & $\Mj$ & $\mj$    \\
($\trhz$) & ($\msun$)       & ($\msun$)        & ($\msun$)       & ($\msun$)        & ($\msun$)       & ($\msun$)         & ($\msun$)       & ($\msun$)       & ($\msun$)        & ($\msun$)        & ($\msun$)       & ($\msun$)       & ($\msun$)       & ($\msun$)       & ($\msun$)       & ($\msun$)        & ($\msun$)       & ($\msun$)        & ($\msun$)      & ($\msun$)       & ($\msun$)      & ($\msun$)       \\
\hline
$2.4$   & $2498$        & $0.13$         & $3288$        & $0.2$          & $4792$        & $0.32$          & $6155$        & $0.5$         & $3630$         & $0.72$         & $185$         & $0.82$        & $3650$        & $0.6$        & $2346$        & $0.78$         & $1521$        & $1.08$         & $841$       & $1.53$         & $1504$       & $12.6$        \\
$4.9$   & $2024$        & $0.13$         & $2702$        & $0.2$          & $4004$        & $0.32$          & $5287$        & $0.5$         & $3194$         & $0.72$         & $164$         & $0.82$        & $3152$        & $0.6$        & $2075$        & $0.78$         & $1391$        & $1.08$         & $789$       & $1.52$         & $955$        & $11.2$        \\
$7.3$   & $1626$        & $0.13$         & $2182$        & $0.2$          & $3354$        & $0.32$          & $4572$        & $0.5$         & $2878$         & $0.72$         & $152$         & $0.82$        & $2775$        & $0.6$        & $1877$        & $0.78$         & $1299$        & $1.09$         & $766$       & $1.52$         & $596$        & $10.1$        \\
$9.7$   & $1252$        & $0.13$         & $1722$        & $0.2$          & $2773$        & $0.32$          & $3943$        & $0.5$         & $2585$         & $0.72$         & $138$         & $0.82$        & $2475$        & $0.6$        & $1704$        & $0.78$         & $1230$        & $1.09$         & $748$       & $1.53$         & $538$        & $9.79$        \\
$12.1$  & $892$         & $0.13$         & $1280$        & $0.2$          & $2149$        & $0.32$          & $3283$        & $0.5$         & $2235$         & $0.72$         & $118$         & $0.82$        & $2104$        & $0.6$        & $1507$        & $0.78$         & $1154$        & $1.09$         & $725$       & $1.53$         & $379$        & $8.81$        \\
$14.6$  & $568$         & $0.13$         & $874$         & $0.2$          & $1543$        & $0.32$          & $2549$        & $0.5$         & $1894$         & $0.72$         & $104$         & $0.83$        & $1728$        & $0.6$        & $1317$        & $0.79$         & $1055$        & $1.09$         & $706$       & $1.53$         & $280$        & $7.99$        \\
$17.0$  & $317$         & $0.13$         & $528$         & $0.2$          & $987$         & $0.32$          & $1818$        & $0.51$        & $1519$         & $0.72$         & $82$          & $0.83$        & $1299$        & $0.6$        & $1060$        & $0.79$         & $957$         & $1.1$          & $675$       & $1.54$         & $192$        & $7.38$        \\
$19.4$  & $144$         & $0.13$         & $269$         & $0.2$          & $544$         & $0.33$          & $1172$        & $0.51$        & $1119$         & $0.72$         & $62$          & $0.83$        & $881$         & $0.6$        & $817$         & $0.79$         & $831$         & $1.1$          & $642$       & $1.54$         & $119$        & $6.63$        \\
$21.9$  & $47.5$        & $0.13$         & $98$          & $0.2$          & $230$         & $0.33$          & $611$         & $0.51$        & $728$          & $0.73$         & $41.4$        & $0.83$        & $496$         & $0.61$       & $541$         & $0.79$         & $647$         & $1.11$         & $574$       & $1.56$         & $80$         & $6.68$        \\
$24.3$  & $8.99$        & $0.13$         & $22.7$        & $0.21$         & $63$          & $0.33$          & $213$         & $0.52$        & $325$          & $0.73$         & $15.7$        & $0.83$        & $204$         & $0.61$       & $272$         & $0.8$          & $418$         & $1.13$         & $474$       & $1.57$         & $30.7$       & $5.11$        \\
$26.7$  & $0.12$        & $0.12$         & $0.24$        & $0.24$         & $4.52$        & $0.35$          & $14.4$        & $0.53$        & $50$           & $0.74$         & $2.51$        & $0.84$        & $15.4$        & $0.62$       & $47.2$        & $0.81$         & $119$         & $1.18$         & $243$       & $1.63$         & $0$          & $0$           \\
\hline
\end{tabular}
\end{tiny}
\end{table}

\begin{table}
\caption{Mass bins of the different snapshots of $N$-body model N0.3. We list the age in units of the initial half-mass relaxation time and for each mass bin the total mass $M_j$ and the mean mass $m_j$ in units of $\msun$. There are a total of 11 mass bins: five for the MSs, one for the ESs, three for the WDs and one each for the NS and BHs.}
\label{tab:MFN0.3}
\begin{tiny}
\begin{tabular}{lccccccccccccccccccccccc}
\hline
Age     & \multicolumn{2}{c|}{MS1}       & \multicolumn{2}{c|}{MS2}       & \multicolumn{2}{c|}{MS3}        & \multicolumn{2}{c|}{MS4}      & \multicolumn{2}{c|}{MS5}        & \multicolumn{2}{c|}{ES}       & \multicolumn{2}{c|}{WD1}      & \multicolumn{2}{c|}{WD2}       & \multicolumn{2}{c|}{WD3}       & \multicolumn{2}{c|}{NS}      & \multicolumn{2}{c|}{BH}      \\
        & $\Mj$   &  $m_{\rm j}$   & $\Mj$     &  $m_{\rm j}$ & $\Mj$ &  $m_{\rm j}$      & $\Mj$ & $m_{\rm j}$     &  $\Mj$ &  $m_{\rm j}$     &  $\Mj$ &  $m_{\rm j}$   & $\Mj$ & $m_{\rm j}$     & $\Mj$ &  $m_{\rm j}$     & $\Mj$ &  $m_{\rm j}$     & $\Mj$ &  $m_{\rm j}$   & $\Mj$ & $m_{\rm j}$    \\
($\trhz$) & ($\msun$)       & ($\msun$)        & ($\msun$)       & ($\msun$)        & ($\msun$)       & ($\msun$)         & ($\msun$)       & ($\msun$)       & ($\msun$)        & ($\msun$)        & ($\msun$)       & ($\msun$)       & ($\msun$)       & ($\msun$)       & ($\msun$)       & ($\msun$)        & ($\msun$)       & ($\msun$)        & ($\msun$)      & ($\msun$)       & ($\msun$)      & ($\msun$)       \\
\hline
$2.3$   & $2746$        & $0.13$         & $3609$        & $0.2$          & $5268$        & $0.32$          & $6732$        & $0.5$         & $3935$         & $0.72$         & $200$         & $0.81$        & $3975$        & $0.6$        & $2516$        & $0.78$         & $1622$        & $1.08$         & $302$       & $1.52$         & $224$        & $10.2$        \\
$4.7$   & $2465$        & $0.13$         & $3276$        & $0.2$          & $4845$        & $0.32$          & $6340$        & $0.5$         & $3783$         & $0.72$         & $196$         & $0.82$        & $3805$        & $0.6$        & $2423$        & $0.78$         & $1588$        & $1.08$         & $285$       & $1.51$         & $53$         & $6.62$        \\
$7.0$   & $2113$        & $0.13$         & $2869$        & $0.2$          & $4365$        & $0.32$          & $5889$        & $0.5$         & $3637$         & $0.72$         & $184$         & $0.82$        & $3603$        & $0.6$        & $3603$        & $0.78$         & $1546$        & $1.08$         & $269$       & $1.50$         & $6.31$       & $6.31$        \\
$9.3$   & $1731$        & $0.13$         & $2386$        & $0.2$          & $3783$        & $0.32$          & $5356$        & $0.5$         & $3457$         & $0.72$         & $179$         & $0.82$        & $3329$        & $0.6$        & $2235$        & $0.78$         & $1478$        & $1.08$         & $218$       & $1.45$         & $0$          & $0$           \\
$11.7$  & $1337$        & $0.13$         & $1782$        & $0.2$          & $3118$        & $0.32$          & $4691$        & $0.5$         & $3205$         & $0.72$         & $164$         & $0.82$        & $3005$        & $0.6$        & $2087$        & $0.78$         & $1354$        & $1.07$         & $168$       & $1.41$         & $0$          & $0$           \\
$14.0$  & $964$         & $0.13$         & $1406$        & $0.2$          & $2471$        & $0.32$          & $4003$        & $0.5$         & $2941$         & $0.72$         & $155$         & $0.82$        & $2675$        & $0.6$        & $1933$        & $0.78$         & $1244$        & $1.06$         & $130$       & $1.38$         & $0$          & $0$           \\
$16.3$  & $635$         & $0.13$         & $990$         & $0.2$          & $1861$        & $0.32$          & $3308$        & $0.5$         & $2636$         & $0.72$         & $143$         & $0.82$        & $2274$        & $0.61$       & $1772$        & $0.79$         & $1140$        & $1.05$         & $102$       & $1.36$         & $0$          & $0$           \\
$18.7$  & $385$         & $0.13$         & $622$         & $0.2$          & $1274$        & $0.32$          & $2595$        & $0.51$        & $2263$         & $0.72$         & $126$         & $0.82$        & $1859$        & $0.61$       & $1559$        & $0.79$         & $1024$        & $1.05$         & $80$        & $1.35$         & $0$          & $0$           \\
$21.0$  & $189$         & $0.13$         & $341$         & $0.2$          & $779$         & $0.33$          & $1862$        & $0.51$        & $1847$         & $0.73$         & $104$         & $0.82$        & $1390$        & $0.61$       & $1329$        & $0.79$         & $904$         & $1.04$         & $61$        & $1.34$         & $0$          & $0$           \\
$23.3$  & $74$          & $0.13$         & $144$         & $0.2$          & $404$         & $0.33$          & $1169$        & $0.52$        & $1394$         & $0.73$         & $83$          & $0.82$        & $949$         & $0.61$       & $1055$        & $0.8$          & $756$         & $1.04$         & $49.1$      & $1.33$         & $0$          & $0$           \\
$25.7$  & $19.1$        & $0.13$         & $45.8$        & $0.2$          & $144$         & $0.33$          & $587$         & $0.52$        & $892$          & $0.73$         & $55$          & $0.81$        & $533$         & $0.61$       & $729$         & $0.8$          & $612$         & $1.04$         & $43.4$      & $1.32$         & $0$          & $0$           \\
$28.0$  & $2.92$        & $0.13$         & $5.93$        & $0.2$          & $21.5$        & $0.35$          & $148$         & $0.53$        & $367$          & $0.74$         & $20.7$        & $0.83$        & $161$         & $0.61$       & $349$         & $0.81$         & $393$         & $1.04$         & $19.7$      & $1.31$         & $0$          & $0$           \\
$30.3$  & $0.17$        & $0.17$         & $0.39$        & $0.39$         & $0$           & $0$             & $6.1$         & $0.55$        & $38.5$         & $0.74$         & $3.32$        & $0.83$        & $8.71$        & $0.62$       & $69$          & $0.85$         & $133$         & $1.06$         & $9.18$      & $1.31$         & $0$          & $0$           \\ 
\hline
\end{tabular}
\end{tiny}
\end{table}

\end{landscape}

\begin{landscape}

\begin{table}
\caption{Mass bins of the different snapshots of $N$-body model N0.1. We list the age in units of the initial half-mass relaxation time and for each mass bin the total mass $M_j$ and the mean mass $m_j$ in units of $\msun$. There are a total of 11 mass bins: five for the MSs, one for the ESs, three for the WDs and one each for the NS and BHs.}
\label{tab:MFN0.1}
\begin{tiny}
\begin{tabular}{lccccccccccccccccccccccc}
\hline
Age     & \multicolumn{2}{c|}{MS1}       & \multicolumn{2}{c|}{MS2}       & \multicolumn{2}{c|}{MS3}        & \multicolumn{2}{c|}{MS4}      & \multicolumn{2}{c|}{MS5}        & \multicolumn{2}{c|}{ES}       & \multicolumn{2}{c|}{WD1}      & \multicolumn{2}{c|}{WD2}       & \multicolumn{2}{c|}{WD3}       & \multicolumn{2}{c|}{NS}      & \multicolumn{2}{c|}{BH}      \\
        & $\Mj$     &  $\mj$ & $M_j$     &  $\mj$ & $\Mj$ &  $\mj$      & $\Mj$ & $\mj$     &  $\Mj$ &  $\mj$     &  $\Mj$ &  $\mj$   & $\Mj$ & $\mj$     & $\Mj$ &  $\mj$     & $\Mj$ &  $\mj$     & $\Mj$ &  $\mj$   & $\Mj$ & $\mj$    \\
($\trhz$) & ($\msun$)       & ($\msun$)        & ($\msun$)       & ($\msun$)        & ($\msun$)       & ($\msun$)         & ($\msun$)       & ($\msun$)       & ($\msun$)        & ($\msun$)        & ($\msun$)       & ($\msun$)       & ($\msun$)       & ($\msun$)       & ($\msun$)       & ($\msun$)        & ($\msun$)       & ($\msun$)        & ($\msun$)      & ($\msun$)       & ($\msun$)      & ($\msun$)       \\
\hline
$2.3$   & $2779$        & $0.13$         & $3658$        & $0.2$          & $5338$        & $0.32$          & $6844$        & $0.5$         & $3976$         & $0.72$         & $204$         & $0.81$        & $4039$        & $0.6$        & $2538$        & $0.78$         & $1644$        & $1.08$         & $82$        & $1.52$         & $53$         & $10.6$        \\
$4.7$   & $2394$        & $0.13$         & $3220$        & $0.2$          & $4800$        & $0.32$          & $6374$        & $0.5$         & $3801$         & $0.72$         & $200$         & $0.82$        & $3827$        & $0.6$        & $2451$        & $0.78$         & $1577$        & $1.08$         & $75$        & $1.50$         & $0$          & $0$           \\
$7.0$   & $1954$        & $0.13$         & $2683$        & $0.2$          & $4103$        & $0.32$          & $5703$        & $0.5$         & $3560$         & $0.72$         & $186$         & $0.81$        & $3515$        & $0.6$        & $2317$        & $0.78$         & $1334$        & $1.06$         & $33$        & $1.36$         & $0$          & $0$           \\
$9.4$   & $1516$        & $0.13$         & $2142$        & $0.2$          & $3405$        & $0.32$          & $5015$        & $0.5$         & $3290$         & $0.72$         & $180$         & $0.82$        & $3183$        & $0.6$        & $2171$        & $0.78$         & $1185$        & $1.05$         & $26$        & $1.35$         & $0$          & $0$           \\
$11.7$  & $1131$        & $0.13$         & $1652$        & $0.2$          & $2757$        & $0.32$          & $4323$        & $0.5$         & $3011$         & $0.72$         & $168$         & $0.82$        & $2824$        & $0.6$        & $2009$        & $0.78$         & $1065$        & $1.04$         & $16.3$      & $1.36$         & $0$          & $0$           \\
$14.1$  & $776$         & $0.13$         & $1207$        & $0.2$          & $2121$        & $0.32$          & $3606$        & $0.5$         & $2706$         & $0.72$         & $158$         & $0.82$        & $2455$        & $0.6$        & $1838$        & $0.78$         & $935$         & $1.03$         & $9.26$      & $1.33$         & $0$          & $0$           \\
$16.4$  & $480$         & $0.13$         & $787$         & $0.2$          & $1504$        & $0.32$          & $2845$        & $0.51$        & $2345$         & $0.72$         & $147$         & $0.82$        & $2064$        & $0.61$       & $1635$        & $0.79$         & $811$         & $1.02$         & $6.51$      & $1.3$          & $0$          & $0$           \\
$18.7$  & $262$         & $0.13$         & $456$         & $0.2$          & $971$         & $0.32$          & $2148$        & $0.51$        & $2025$         & $0.73$         & $130$         & $0.82$        & $1651$        & $0.61$       & $1414$        & $0.79$         & $711$         & $1.01$         & $3.91$      & $1.3$          & $0$          & $0$           \\
$21.1$  & $107$         & $0.13$         & $227$         & $0.2$          & $528$         & $0.33$          & $1411$        & $0.51$        & $1600$         & $0.73$         & $113$         & $0.82$        & $1179$        & $0.61$       & $1169$        & $0.79$         & $592$         & $1.01$         & $1.3$       & $1.3$          & $0$          & $0$           \\
$23.4$  & $33.4$        & $0.13$         & $81$          & $0.2$          & $227$         & $0.33$          & $745$         & $0.52$        & $1121$         & $0.73$         & $89$          & $0.82$        & $724$         & $0.61$       & $871$         & $0.79$         & $481$         & $1.0$          & $0$         & $0$            & $0$          & $0$           \\
$25.8$  & $5.55$        & $0.13$         & $14.1$        & $0.2$          & $57$          & $0.33$          & $291$         & $0.53$        & $596$          & $0.74$         & $55$          & $0.83$        & $305$         & $0.61$       & $511$         & $0.8$          & $337$         & $1.01$         & $0$         & $0$            & $0$          & $0$           \\
$28.1$  & $0.14$        & $0.14$         & $0.48$        & $0.24$         & $3.68$        & $0.33$          & $53$          & $0.55$        & $196$          & $0.75$         & $21.6$        & $0.83$        & $57$          & $0.61$       & $192$         & $0.82$         & $155$         & $1.01$         & $0$         & $0$            & $0$          & $0$           \\
\hline
\end{tabular}
\end{tiny}
\end{table}

\begin{table}
\caption{Mass bins of the different snapshots of $N$-body model N0. We list the age in units of the initial half-mass relaxation time and for each mass bin the total mass $M_j$ and the mean mass $m_j$ in units of $\msun$. There are a total of nine mass bins: five for the MSs, one for the ESs and three for the WDs.}
\label{tab:MFN0}
\begin{tiny}
\begin{tabular}{lccccccccccccccccccc}
\hline
Age     & \multicolumn{2}{c|}{MS1}       & \multicolumn{2}{c|}{MS2}       & \multicolumn{2}{c|}{MS3}        & \multicolumn{2}{c|}{MS4}      & \multicolumn{2}{c|}{MS5}        & \multicolumn{2}{c|}{ES}       & \multicolumn{2}{c|}{WD1}      & \multicolumn{2}{c|}{WD2}       & \multicolumn{2}{c|}{WD3}       \\
        & $\Mj$   &  $\mj$   & $\Mj$     &  $\mj$ & $\Mj$ &  $\mj$      & $\Mj$ & $\mj$     &  $\Mj$ &  $\mj$     &  $\Mj$ &  $\mj$   & $\Mj$ & $\mj$     & $\Mj$ &  $\mj$     & $\Mj$ &  $\mj$     \\
($\trhz$) & ($\msun$)       & ($\msun$)        & ($\msun$)       & ($\msun$)        & ($\msun$)       & ($\msun$)         & ($\msun$)       & ($\msun$)       & ($\msun$)        & ($\msun$)        & ($\msun$)       & ($\msun$)       & ($\msun$)       & ($\msun$)       & ($\msun$)       & ($\msun$)        & ($\msun$)       & ($\msun$)        \\
\hline
$2.3$   & $2743$        & $0.13$         & $3613$        & $0.2$          & $5300$        & $0.32$          & $6817$        & $0.5$         & $3987$         & $0.72$         & $210$         & $0.82$        & $4016$        & $0.6$        & $2541$        & $0.78$         & $1659$        & $1.08$         \\
$4.7$   & $2285$        & $0.13$         & $3091$        & $0.2$          & $4614$        & $0.32$          & $6174$        & $0.5$         & $3742$         & $0.72$         & $197$         & $0.81$        & $3703$        & $0.6$        & $2378$        & $0.78$         & $1356$        & $1.06$         \\
$7.0$   & $1847$        & $0.13$         & $2543$        & $0.2$          & $3944$        & $0.32$          & $5515$        & $0.5$         & $3489$         & $0.72$         & $185$         & $0.81$        & $3385$        & $0.6$        & $2218$        & $0.78$         & $1135$        & $1.04$         \\
$9.4$   & $1450$        & $0.13$         & $2055$        & $0.2$          & $3292$        & $0.32$          & $4874$        & $0.5$         & $3236$         & $0.72$         & $174$         & $0.81$        & $3074$        & $0.6$        & $2081$        & $0.78$         & $987$         & $1.03$         \\
$11.7$  & $1071$        & $0.13$         & $1589$        & $0.2$          & $2665$        & $0.32$          & $4210$        & $0.5$         & $2984$         & $0.72$         & $159$         & $0.81$        & $2708$        & $0.6$        & $1917$        & $0.78$         & $867$         & $1.02$         \\
$14.1$  & $730$         & $0.13$         & $1147$        & $0.2$          & $2053$        & $0.32$          & $3495$        & $0.51$        & $2667$         & $0.72$         & $140$         & $0.81$        & $2329$        & $0.6$        & $1747$        & $0.78$         & $745$         & $1.01$         \\
$16.4$  & $453$         & $0.13$         & $747$         & $0.2$          & $1466$        & $0.32$          & $2763$        & $0.51$        & $2336$         & $0.72$         & $123$         & $0.82$        & $1945$        & $0.61$       & $1557$        & $0.78$         & $636$         & $1.0$          \\
$18.8$  & $236$         & $0.13$         & $432$         & $0.2$          & $935$         & $0.32$          & $2045$        & $0.51$        & $1961$         & $0.73$         & $107$         & $0.82$        & $1550$        & $0.61$       & $1362$        & $0.79$         & $555$         & $0.99$         \\
$21.1$  & $98$          & $0.13$         & $196$         & $0.2$          & $506$         & $0.33$          & $1316$        & $0.51$        & $1518$         & $0.73$         & $87$          & $0.82$        & $1098$        & $0.61$       & $1086$        & $0.79$         & $463$         & $0.99$         \\
$23.5$  & $26.7$        & $0.13$         & $68$          & $0.2$          & $213$         & $0.33$          & $707$         & $0.52$        & $1045$         & $0.73$         & $66$          & $0.82$        & $681$         & $0.61$       & $795$         & $0.79$         & $353$         & $0.98$         \\
$25.8$  & $4.32$        & $0.13$         & $13$          & $0.21$         & $52$          & $0.34$          & $248$         & $0.53$        & $578$          & $0.74$         & $43$          & $0.83$        & $287$         & $0.61$       & $473$         & $0.8$          & $252$         & $0.98$         \\
$28.1$  & $0.11$        & $0.11$         & $0.61$        & $0.2$          & $2.31$        & $0.33$          & $25.4$        & $0.54$        & $139$          & $0.76$         & $10.9$        & $0.84$        & $29$          & $0.62$       & $138$         & $0.82$         & $111$         & $0.98$         \\
\hline
\end{tabular}
\end{tiny}
\end{table}

\end{landscape}

\begin{landscape}

\begin{table}
\caption{Results from the MCMC fitting process for the snapshots from the $N$-body model N1 with initial 100\% BH and NS retention. We list here the age of each snapshot in units of the initial half-mass relaxation time, the total cluster mass in $\msun$, the half-mass radius in pc, the dimensionless central concentration for the global mean mass, the central mean mass, and the ES stars, the truncation parameter $g$, the mass segregation parameter $\delta$, the anisotropy radius for the global mean mass and the central mean mass in pc, the anisotropy parameter $\eta$ and the truncation radius $\rt$ in pc. All parameters except the dimensionless central concentration for the global mean mass, and the ES stars, the anisotropy radius for the global mean mass and the truncation radius $\rt$, are fitting parameters; the other values were obtained from the multimass models of the 10 last walker positions of each MCMC chain. The median of the marginalized posterior distribution of each parameter is used to estimate its best-fitting value and the 16th and 84th percentiles as proxy for the $1 \sigma$ uncertainties. }
\label{tab:MCMCN1}
\begin{tabular}{lccccccccccc}
\hline
Age     & $M_{\rm Cl}$         & $\rh $                   & $W_{0}$ & $W_{0,\rm{CMM}}$ & $W_{0\rm ,ES}$          & $g$                      & $\delta$                    & $\ra$            & $r_{\rm a,CMM}$            & $\eta$                    & $\rt$               \\
($\trhz$) & ($\msun$)               & (pc)                       &                               &                                &                         &                          &                          & (pc)                     &  (pc)                        &              & (pc)                        \\
\hline
$2.4$   & $30455_{-14}^{+14}$   & $5.11_{-0.02}^{+0.02}$ & $1.7_{-0.1}^{+0.1}$    & $16.9_{-0.4}^{+0.4}$         & $2.8_{-0.1}^{+0.1}$   & $1.43_{-0.02}^{+0.02}$ & $0.350_{-0.008}^{+0.009}$ & $4.3_{-0.2}^{+0.15}$  & $7.5_{-0.6}^{+0.7}$    & $0.17_{-0.02}^{+0.02}$  & $31.2_{-0.2}^{+0.2}$    \\
$4.9$   & $25790_{-12}^{+13}$   & $6.13_{-0.02}^{+0.02}$ & $1.8_{-0.1}^{+0.1}$    & $20.4_{-0.2}^{+0.2}$         & $3.3_{-0.2}^{+0.03}$  & $1.25_{-0.02}^{+0.02}$ & $0.368_{-0.008}^{+0.008}$ & $8.4_{-0.4}^{+0.41}$  & $3.6_{-0.4}^{+0.4}$    & $-0.26_{-0.03}^{+0.04}$ & $28.37_{-0.09}^{+0.1}$  \\
$7.3$   & $22140_{-11}^{+11}$   & $6.59_{-0.02}^{+0.02}$ & $2.40_{-0.07}^{+0.07}$ & $23.6_{-1.0}^{+0.7}$         & $4.1_{-0.1}^{+0.1}$   & $1.05_{-0.01}^{+0.01}$ & $0.362_{-0.005}^{+0.005}$ & $20_{-2.3}^{+2.6}$    & $6.2_{-0.9}^{+0.8}$    & $-0.39_{-0.06}^{+0.06}$ & $26.73_{-0.06}^{+0.06}$ \\
$9.7$   & $19175_{-9.9}^{+9.3}$ & $6.61_{-0.02}^{+0.01}$ & $2.68_{-0.08}^{+0.07}$ & $18_{-1.2}^{+1.2}$           & $4.3_{-0.10}^{+0.08}$ & $0.91_{-0.01}^{+0.01}$ & $0.331_{-0.007}^{+0.007}$ & $61_{-31}^{+\infty}$  & $29_{-17}^{+4770}$     & $-0.4_{-3.6}^{+0.2}$    & $25.56_{-0.05}^{+0.05}$ \\
$12.1$  & $15878_{-9.1}^{+8.5}$ & $6.51_{-0.02}^{+0.02}$ & $2.74_{-0.06}^{+0.06}$ & $20_{-1.5}^{+1.2}$           & $4.4_{-0.08}^{+0.08}$ & $0.8_{-0.01}^{+0.02}$  & $0.360_{-0.007}^{+0.006}$ & $102_{-43}^{+\infty}$ & $39_{-18}^{+208}$      & $-0.4_{-0.3}^{+0.2}$    & $23.96_{-0.05}^{+0.09}$ \\
$14.6$  & $12659_{-8.2}^{+7.7}$ & $6.23_{-0.02}^{+0.02}$ & $2.85_{-0.06}^{+0.06}$ & $19_{-1.2}^{+1.6}$           & $4.4_{-0.08}^{+0.08}$ & $0.72_{-0.02}^{+0.02}$ & $0.377_{-0.007}^{+0.007}$ & --              & $4416_{-2895}^{+2814}$ & $-4_{-4}^{+4}$          & $22.39_{-0.04}^{+0.06}$ \\
$17.0$  & $9464_{-7.3}^{+6.8}$  & $5.64_{-0.02}^{+0.02}$ & $3.09_{-0.06}^{+0.07}$ & $20_{-1.2}^{+1.3}$           & $4.6_{-0.08}^{+0.08}$ & $0.64_{-0.02}^{+0.02}$ & $0.390_{-0.007}^{+0.007}$ & --              & $3740_{-2428}^{+2365}$ & $-4_{-4}^{+4}$          & $20.27_{-0.05}^{+0.07}$ \\
$19.4$  & $6623_{-6.9}^{+6.2}$  & $4.96_{-0.03}^{+0.03}$ & $3.20_{-0.09}^{+0.08}$ & $18_{-1.7}^{+2.9}$           & $4.4_{-0.1}^{+0.1}$   & $0.64_{-0.02}^{+0.02}$ & $0.42_{-0.01}^{+0.01}$    & --              & $1017_{-741}^{+2149}$  & $-4_{-4}^{+4}$          & $18.23_{-0.05}^{+0.07}$ \\
$21.9$  & $4107_{-5.7}^{+5.0}$  & $4.03_{-0.04}^{+0.04}$ & $3.5_{-0.13}^{+0.13}$  & $17_{-2.1}^{+3.0}$           & $4.3_{-0.1}^{+0.1}$   & $0.71_{-0.04}^{+0.04}$ & $0.42_{-0.02}^{+0.02}$    & --              & $3041_{-1941}^{+1843}$ & $-4_{-4}^{+4}$          & $15.9_{-0.2}^{+0.3}$    \\
$24.3$  & $2050_{-5.0}^{+4.2}$  & $2.92_{-0.04}^{+0.04}$ & $4.3_{-0.23}^{+0.04}$  & $23_{-5.4}^{+4.0}$           & $4.5_{-0.2}^{+0.4}$   & $0.7_{-0.04}^{+0.04}$  & $0.50_{-0.03}^{+0.02}$    & --              & $874_{-576}^{+600}$    & $-4_{-4}^{+4}$          & $12.9_{-0.1}^{+0.1}$    \\
$26.7$  & $495_{-3.9}^{+2.7}$   & $1.7_{-0.1}^{+0.08}$   & $5_{-1.2}^{+1.1}$      & $8.9_{-0.7}^{+1.0}$          & $2.7_{-1.1}^{+1.2}$   & $0.7_{-0.1}^{+0.2}$    & $0.9_{-0.3}^{+0.4}$       & $72_{-54}^{+266}$     & $21_{-11}^{+12}$       & $-4_{-4}^{+4}$          & $8.7_{-0.2}^{+0.4}$     \\
\hline
\end{tabular}
\end{table}

\begin{table}
\caption{Results from the MCMC fitting process for the snapshots from the $N$-body model N0.3 with initial 33\% BH and NS retention. We list here the age of each snapshot in units of the initial half-mass relaxation time, the total cluster mass in $\msun$, the half-mass radius in pc, the dimensionless central concentration for the global mean mass, the central mean mass, and the ES stars, the truncation parameter $g$, the mass segregation parameter $\delta$, the anisotropy radius for the global mean mass and the central mean mass in pc, the anisotropy parameter $\eta$ and the truncation radius $\rt$ in pc. All parameters except the dimensionless central concentration for the global mean mass, and the ES stars, the anisotropy radius for the global mean mass and the truncation radius $\rt$, are fitting parameters; the other values were obtained from the multimass models of the 10 last walker positions of each MCMC chain. The median of the marginalized posterior distribution of each parameter is used to estimate its best-fitting value and the 16th and 84th percentiles as proxy for the $1 \sigma$ uncertainties.}
\label{tab:MCMCN0.3}
\begin{tabular}{lccccccccccc}
\hline
Age     & $M_{\rm Cl}$         & $\rh $                   & $W_{0}$ & $W_{0,\rm{CMM}}$ & $W_{0\rm ,ES}$          & $g$                      & $\delta$                    & $\ra$            & $r_{\rm a,CMM}$            & $\eta$                    & $\rt$               \\
($\trhz$) & ($\msun$)               & (pc)                       &                               &                                &                         &                          &                          & (pc)                     &  (pc)                        &              & (pc)                        \\
\hline
$2.3$   & $31166_{-14}^{+13}$    & $3.14_{-0.01}^{+0.01}$    & $4.22_{-0.06}^{+0.06}$      & $47.8_{-0.7}^{+0.2}$    & $7.76_{-0.08}^{+0.07}$  & $1.52_{-0.02}^{+0.02}$    & $0.370_{-0.003}^{+0.003}$ & $5.9_{-0.1}^{+0.2}$     & $25.0_{-2.8}^{+2.6}$ & $0.44_{-0.02}^{+0.03}$  & $32.0_{-0.4}^{+0.4}$    \\
$4.7$   & $29103_{-13}^{+14}$    & $3.18_{-0.01}^{+0.01}$    & $4.65_{-0.03}^{+0.02}$      & $13.26_{-0.05}^{+0.05}$ & $8.30_{-0.03}^{+0.02}$  & $1.22_{-0.01}^{+0.01}$    & $0.356_{-0.001}^{+0.002}$ & $7.8_{-0.1}^{+0.1}$     & $13.0_{-0.3}^{+0.3}$ & $0.35_{-0.02}^{+0.02}$  & $28.6_{-0.2}^{+0.1}$    \\
$7.0$   & $26860_{-13}^{+14}$    & $3.13_{-0.01}^{+0.01}$    & $4.69_{-0.04}^{+0.04}$      & $11.23_{-0.04}^{+0.04}$ & $9.03_{-0.04}^{+0.04}$  & $1.11_{-0.01}^{+0.01}$    & $0.414_{-0.003}^{+0.003}$ & $6.7_{-0.2}^{+0.2}$     & $6.5_{-0.3}^{+0.3}$  & $-0.03_{-0.04}^{+0.03}$ & $27.55_{-0.1}^{+0.1}$  \\
$9.3$   & $24196_{-12}^{+12}$    & $3.38_{-0.01}^{+0.01}$    & $5.38_{-0.08}^{+0.08}$      & $13.5_{-0.1}^{+0.1}$    & $10.75_{-0.07}^{+0.08}$ & $1.102_{-0.009}^{+0.009}$ & $0.458_{-0.006}^{+0.006}$ & $6.3_{-0.2}^{+0.2}$     & $4.0_{-0.2}^{+0.2}$  & $-0.45_{-0.04}^{+0.05}$ & $26.8_{-0.1}^{+0.1}$    \\
$11.7$  & $21047_{-11}^{+11}$    & $3.64_{-0.01}^{+0.01}$    & $5.67_{-0.09}^{+0.1}$       & $13.8_{-0.2}^{+0.2}$    & $10.97_{-0.09}^{+0.08}$ & $1.03_{-0.01}^{+0.01}$    & $0.460_{-0.009}^{+0.009}$ & $7.5_{-0.2}^{+0.3}$     & $4.0_{-0.2}^{+0.3}$  & $-0.67_{-0.05}^{+0.05}$ & $25.56_{-0.09}^{+0.09}$ \\
$14.0$  & $17972_{-10}^{+9}$     & $3.68_{-0.01}^{+0.009}$   & $5.7_{-0.1}^{+0.1}$         & $13.9_{-0.3}^{+0.2}$    & $11.0_{-0.1}^{+0.1}$    & $0.97_{-0.01}^{+0.01}$    & $0.49_{-0.01}^{+0.01}$    & $7.9_{-0.3}^{+0.3}$     & $3.8_{-0.3}^{+0.3}$  & $-0.81_{-0.06}^{+0.06}$ & $24.34_{-0.08}^{+0.09}$ \\
$16.3$  & $14905_{-9}^{+9}$      & $3.60_{-0.01}^{+0.008}$   & $5.5_{-0.09}^{+0.1}$        & $12.0_{-0.1}^{+0.1}$    & $10.10_{-0.07}^{+0.07}$ & $0.89_{-0.01}^{+0.01}$    & $0.50_{-0.01}^{+0.01}$    & $9.6_{-0.5}^{+0.6}$     & $4.5_{-0.2}^{+0.2}$  & $-0.97_{-0.07}^{+0.08}$ & $22.92_{-0.07}^{+0.07}$ \\
$18.7$  & $11820_{-9}^{+8}$      & $3.47_{-0.01}^{+0.01}$    & $6.1_{-0.1}^{+0.1}$         & $11.9_{-0.2}^{+0.2}$    & $10.16_{-0.09}^{+0.08}$ & $0.85_{-0.01}^{+0.01}$    & $0.48_{-0.01}^{+0.01}$    & $10.4_{-0.7}^{+0.9}$    & $4.7_{-0.3}^{+0.3}$  & $-1.1_{-0.1}^{+0.1}$    & $21.33_{-0.07}^{+0.08}$ \\
$21.0$  & $8832_{-8}^{+7}$       & $3.32_{-0.04}^{+0.02}$    & $6.0_{-0.1}^{+0.1}$         & $11.6_{-0.2}^{+0.2}$    & $9.7_{-0.1}^{+0.1}$     & $0.82_{-0.02}^{+0.03}$    & $0.53_{-0.02}^{+0.02}$    & $12_{-1}^{+2}$          & $6.0_{-0.5}^{+0.6}$  & $-1.2_{-0.2}^{+0.2}$    & $19.6_{-0.1}^{+0.2}$    \\
$23.3$  & $6089_{-8}^{+7}$       & $2.94_{-0.02}^{+0.02}$    & $6.7_{-0.2}^{+0.2}$         & $11.2_{-0.2}^{+0.2}$    & $9.4_{-0.1}^{+0.1}$     & $0.8_{-0.03}^{+0.03}$     & $0.51_{-0.02}^{+0.02}$    & $14_{-2}^{+4}$          & $5.7_{-0.7}^{+0.9}$  & $-1.8_{-0.4}^{+0.4}$    & $17.3_{-0.2}^{+0.2}$    \\
$25.7$  & $3658_{-6}^{+6}$       & $2.57_{-0.03}^{+0.02}$    & $7.3_{-0.2}^{+0.2}$         & $10.9_{-0.2}^{+0.2}$    & $9.1_{-0.2}^{+0.2}$     & $0.75_{-0.03}^{+0.03}$    & $0.48_{-0.03}^{+0.03}$    & $14_{-4}^{+7}$          & $5.5_{-0.8}^{+1.4}$  & $-2.3_{-0.8}^{+0.6}$    & $14.91_{-0.1}^{+0.1}$  \\
$28.0$  & $1494_{-4}^{+3}$       & $1.92_{-0.04}^{+0.04}$    & $6.4_{-0.3}^{+0.3}$         & $9.4_{-0.3}^{+0.4}$     & $7.4_{-0.2}^{+0.2}$     & $0.76_{-0.05}^{+0.05}$    & $0.65_{-0.07}^{+0.07}$    & $2621_{-1921}^{+8346}$  & $1034_{-689}^{+696}$ & $-4._{-4.1}^{+4.1}$     & $11.6_{-0.2}^{+0.2}$    \\
$30.3$  & $268_{-3}^{+2}$        & $1.10_{-0.06}^{+0.06}$    & $7.1_{-0.6}^{+0.4}$         & $8.4_{-0.5}^{+0.7}$     & $4.7_{-1.1}^{+1.0}$     & $0.6_{-0.1}^{+0.2}$       & $0.6_{-0.2}^{+0.2}$       & $1140_{-782}^{+1444}$   & $737_{-497}^{+504}$  & $-4._{-4.1}^{+4.1}$     & $6.5_{-0.2}^{+0.5}$     \\
\hline
\end{tabular}
\end{table}

\end{landscape}

\begin{landscape}

\begin{table}
\caption{Results from the MCMC fitting process for the snapshots from the $N$-body model N0.1 with initial 10\% BH and NS retention. We list here the age of each snapshot in units of the initial half-mass relaxation time, the total cluster mass in $\msun$, the half-mass radius in pc, the dimensionless central concentration for the global mean mass, the central mean mass, and the ES stars, the truncation parameter $g$, the mass segregation parameter $\delta$, the anisotropy radius for the global mean mass and the central mean mass in pc, the anisotropy parameter $\eta$ and the truncation radius $\rt$ in pc. All parameters except the dimensionless central concentration for the global mean mass, and the ES stars, the anisotropy radius for the global mean mass and the truncation radius $\rt$, are fitting parameters; the other values were obtained from the multimass models of the 10 last walker positions of each MCMC chain. The median of the marginalized posterior distribution of each parameter is used to estimate its best-fitting value and the 16th and 84th percentiles as proxy for the $1 \sigma$ uncertainties.}
\label{tab:MCMCN0.1}
\begin{tabular}{lccccccccccc}
\hline
Age     & $M_{\rm Cl}$         & $\rh $                   & $W_{0}$ & $W_{0,\rm{CMM}}$ & $W_{0\rm ,ES}$          & $g$                      & $\delta$                    & $\ra$            & $r_{\rm a,CMM}$            & $\eta$                    & $\rt$               \\
($\trhz$) & ($\msun$)               & (pc)                       &                               &                                &                         &                          &                          & (pc)                     &  (pc)                        &              & (pc)                        \\
\hline
$2.3$   & $31177_{-15}^{+15}$   & $2.51_{-0.01}^{+0.01}$ & $5.41_{-0.03}^{+0.04}$ & $16.2_{-0.1}^{+0.2}$ & $8.84_{-0.03}^{+0.03}$  & $1.47_{-0.01}^{+0.01}$ & $0.295_{-0.002}^{+0.002}$ & $11.7_{-0.4}^{+0.3}$    & $41_{-1}^{+1}$      & $0.67_{-0.02}^{+0.02}$  & $31.1_{-0.2}^{+0.3}$    \\
$4.7$   & $28763_{-14}^{+14}$   & $2.82_{-0.01}^{+0.01}$ & $6.18_{-0.06}^{+0.07}$ & $12.9_{-0.1}^{+0.1}$ & $11.56_{-0.05}^{+0.05}$ & $1.30_{-0.01}^{+0.01}$ & $0.387_{-0.006}^{+0.006}$ & $4.33_{-0.08}^{+0.07}$  & $4.6_{-0.2}^{+0.2}$ & $0.05_{-0.02}^{+0.02}$  & $30.1_{-0.2}^{+0.2}$    \\
$7.0$   & $25430_{-12}^{+11}$   & $3.36_{-0.02}^{+0.01}$ & $5.62_{-0.08}^{+0.09}$ & $12.7_{-0.1}^{+0.1}$ & $11.46_{-0.07}^{+0.08}$ & $1.12_{-0.01}^{+0.01}$ & $0.453_{-0.007}^{+0.008}$ & $5.4_{-0.1}^{+0.1}$     & $4.9_{-0.2}^{+0.2}$ & $-0.11_{-0.03}^{+0.03}$ & $27.7_{-0.1}^{+0.1}$    \\
$9.4$   & $22153_{-12}^{+11}$   & $3.56_{-0.01}^{+0.01}$ & $5.9_{-0.1}^{+0.1}$    & $12.9_{-0.1}^{+0.1}$ & $11.60_{-0.08}^{+0.09}$ & $1.04_{-0.01}^{+0.01}$ & $0.46_{-0.01}^{+0.01}$    & $6.6_{-0.2}^{+0.2}$     & $4.5_{-0.2}^{+0.2}$ & $-0.45_{-0.05}^{+0.05}$ & $26.2_{-0.1}^{+0.1}$    \\
$11.7$  & $18998_{-10}^{+10}$   & $3.67_{-0.01}^{+0.01}$ & $5.8_{-0.1}^{+0.1}$    & $12.2_{-0.1}^{+0.1}$ & $11.22_{-0.07}^{+0.08}$ & $0.94_{-0.01}^{+0.01}$ & $0.46_{-0.01}^{+0.01}$    & $8.0_{-0.3}^{+0.3}$     & $4.6_{-0.2}^{+0.2}$ & $-0.68_{-0.06}^{+0.05}$ & $24.76_{-0.08}^{+0.08}$ \\
$14.1$  & $15846_{-10}^{+9}$    & $3.69_{-0.02}^{+0.02}$ & $6.1_{-0.1}^{+0.1}$    & $11.7_{-0.1}^{+0.1}$ & $11.00_{-0.07}^{+0.08}$ & $0.85_{-0.01}^{+0.01}$ & $0.46_{-0.01}^{+0.01}$    & $8.7_{-0.4}^{+0.4}$     & $4.4_{-0.2}^{+0.2}$ & $-0.97_{-0.07}^{+0.08}$ & $23.40_{-0.1}^{+0.06}$  \\
$16.4$  & $12658_{-8}^{+8}$     & $3.58_{-0.02}^{+0.02}$ & $6.1_{-0.2}^{+0.1}$    & $11.8_{-0.2}^{+0.2}$ & $10.96_{-0.09}^{+0.1}$  & $0.84_{-0.01}^{+0.01}$ & $0.50_{-0.02}^{+0.02}$    & $9.3_{-0.5}^{+0.6}$     & $4.5_{-0.3}^{+0.3}$ & $-1.1_{-0.1}^{+0.1}$    & $21.90_{-0.08}^{+0.09}$ \\
$18.7$  & $9799_{-8}^{+8}$      & $3.45_{-0.01}^{+0.02}$ & $6.8_{-0.2}^{+0.2}$    & $12.1_{-0.2}^{+0.2}$ & $11.1_{-0.1}^{+0.1}$    & $0.80_{-0.01}^{+0.01}$ & $0.49_{-0.02}^{+0.02}$    & $10.1_{-0.7}^{+0.9}$    & $4.5_{-0.3}^{+0.4}$ & $-1.4_{-0.2}^{+0.1}$    & $20.2_{-0.1}^{+0.1}$ \\
$21.1$  & $6945_{-8}^{+7}$      & $3.06_{-0.03}^{+0.03}$ & $6.8_{-0.2}^{+0.2}$    & $11.1_{-0.2}^{+0.2}$ & $10.4_{-0.1}^{+0.1}$    & $0.84_{-0.02}^{+0.02}$ & $0.53_{-0.02}^{+0.02}$    & $14_{-2}^{+3}$          & $6.6_{-0.7}^{+0.8}$ & $-1.7_{-0.4}^{+0.3}$    & $18.5_{-0.1}^{+0.2}$    \\
$23.4$  & $4384_{-7}^{+6}$      & $2.77_{-0.03}^{+0.03}$ & $7.4_{-0.1}^{+0.2}$    & $10.8_{-0.2}^{+0.2}$ & $10.1_{-0.1}^{+0.1}$    & $0.80_{-0.02}^{+0.03}$ & $0.53_{-0.03}^{+0.03}$    & $1594_{-1418}^{+8138}$  & $504_{-471}^{+491}$ & $-4._{-4}^{+4}$         & $15.9_{-0.2}^{+0.2}$    \\
$25.8$  & $2177_{-5}^{+4}$      & $2.29_{-0.05}^{+0.05}$ & $8.2_{-0.2}^{+0.3}$    & $10.7_{-0.3}^{+0.3}$ & $9.9_{-0.2}^{+0.2}$     & $0.79_{-0.04}^{+0.05}$ & $0.49_{-0.05}^{+0.05}$    & $1099_{-772}^{+2822}$   & $466_{-304}^{+305}$ & $-4._{-4}^{+4}$         & $13.2_{-0.2}^{+0.2}$    \\
$28.1$  & $681_{-3}^{+2}$       & $1.42_{-0.06}^{+0.04}$ & $7.5_{-0.3}^{+0.3}$    & $8.6_{-0.3}^{+0.3}$  & $8.0_{-0.2}^{+0.2}$     & $0.68_{-0.06}^{+0.09}$ & $0.5_{-0.1}^{+0.1}$       & $1338_{-874}^{+1719}$   & $858_{-584}^{+589}$ & $-4._{-4}^{+4}$         & $9.2_{-0.2}^{+0.3}$     \\
\hline
\end{tabular}
\end{table}

\begin{table}
\caption{Results from the MCMC fitting process for the snapshots from the $N$-body model N0 with no initial BH and NS retention. We list here the age of each snapshot in units of the initial half-mass relaxation time, the total cluster mass in $\msun$, the half-mass radius in pc, the dimensionless central concentration for the global mean mass, the central mean mass, and the ES stars, the truncation parameter $g$, the mass segregation parameter $\delta$, the anisotropy radius for the global mean mass and the central mean mass in pc, the anisotropy parameter $\eta$ and the truncation radius $\rt$ in pc. All parameters except the dimensionless central concentration for the global mean mass, and the ES stars, the anisotropy radius for the global mean mass and the truncation radius $\rt$, are fitting parameters; the other values were obtained from the multimass models of the 10 last walker positions of each MCMC chain. The median of the marginalized posterior distribution of each parameter is used to estimate its best-fitting value and the 16th and 84th percentiles as proxy for the $1 \sigma$ uncertainties.}
\label{tab:MCMCN0}
\begin{tabular}{lccccccccccc}
\hline
Age     & $M_{\rm Cl}$         & $\rh $                   & $W_{0}$ & $W_{0,\rm{CMM}}$ & $W_{0\rm ,ES}$          & $g$                      & $\delta$                    & $\ra$            & $r_{\rm a,CMM}$            & $\eta$                    & $\rt$               \\
($\trhz$) & ($\msun$)               & (pc)                       &                               &                                &                         &                          &                          & (pc)                     &  (pc)                        &              & (pc)                        \\
\hline
$2.3$   & $30921_{-15}^{+13}$   & $2.30_{-0.01}^{+0.01}$ & $5.65_{-0.07}^{+0.07}$ & $11.9_{-0.1}^{+0.1}$ & $11.29_{-0.05}^{+0.05}$ & $1.57_{-0.01}^{+0.01}$ & $0.415_{-0.007}^{+0.007}$ & $4.7_{-0.1}^{+0.1}$    & $7.0_{-0.3}^{+0.4}$ & $0.45_{-0.02}^{+0.02}$  & $33.0_{-0.2}^{+0.3}$    \\
$4.7$   & $27575_{-13}^{+13}$   & $3.01_{-0.01}^{+0.01}$ & $5.9_{-0.1}^{+0.1}$    & $13.1_{-0.1}^{+0.1}$ & $12.09_{-0.08}^{+0.08}$ & $1.28_{-0.02}^{+0.01}$ & $0.439_{-0.008}^{+0.008}$ & $4.60_{-0.09}^{+0.08}$ & $5.4_{-0.3}^{+0.3}$ & $0.17_{-0.02}^{+0.02}$  & $29.4_{-0.3}^{+0.1}$    \\
$7.0$   & $24303_{-13}^{+12}$   & $3.32_{-0.01}^{+0.01}$ & $5.7_{-0.1}^{+0.1}$    & $12.8_{-0.1}^{+0.1}$ & $11.93_{-0.07}^{+0.08}$ & $1.16_{-0.01}^{+0.01}$ & $0.466_{-0.009}^{+0.009}$ & $5.8_{-0.2}^{+0.2}$    & $5.1_{-0.2}^{+0.3}$ & $-0.14_{-0.04}^{+0.04}$ & $27.12_{-0.07}^{+0.09}$ \\
$9.4$   & $21266_{-11}^{+11}$   & $3.55_{-0.02}^{+0.01}$ & $6.1_{-0.1}^{+0.1}$    & $13.1_{-0.1}^{+0.1}$ & $12.17_{-0.09}^{+0.1}$  & $1.08_{-0.01}^{+0.01}$ & $0.47_{-0.01}^{+0.01}$    & $7.4_{-0.3}^{+0.3}$    & $4.8_{-0.2}^{+0.3}$ & $-0.51_{-0.05}^{+0.05}$ & $25.74_{-0.1}^{+0.09}$ \\
$11.7$  & $18207_{-10}^{+10}$   & $3.59_{-0.02}^{+0.01}$ & $6.2_{-0.1}^{+0.1}$    & $12.7_{-0.2}^{+0.2}$ & $11.93_{-0.1}^{+0.09}$  & $0.99_{-0.01}^{+0.01}$ & $0.47_{-0.01}^{+0.01}$    & $8.1_{-0.3}^{+0.4}$    & $4.6_{-0.3}^{+0.3}$ & $-0.75_{-0.07}^{+0.06}$ & $24.28_{-0.08}^{+0.09}$ \\
$14.1$  & $15089_{-9}^{+9}$     & $3.65_{-0.01}^{+0.01}$ & $6.6_{-0.2}^{+0.2}$    & $13.3_{-0.2}^{+0.2}$ & $12.3_{-0.1}^{+0.1}$    & $0.94_{-0.01}^{+0.01}$ & $0.48_{-0.01}^{+0.01}$    & $9.0_{-0.4}^{+0.5}$    & $4.5_{-0.3}^{+0.3}$ & $-0.95_{-0.08}^{+0.08}$ & $23.05_{-0.1}^{+0.07}$ \\
$16.4$  & $12058_{-9}^{+8}$     & $3.50_{-0.01}^{+0.01}$ & $7.0_{-0.1}^{+0.2}$    & $12.3_{-0.1}^{+0.1}$ & $11.86_{-0.1}^{+0.09}$  & $0.86_{-0.01}^{+0.01}$ & $0.45_{-0.02}^{+0.02}$    & $9.7_{-0.6}^{+0.8}$    & $4.4_{-0.2}^{+0.2}$ & $-1.3_{-0.1}^{+0.1}$    & $21.40_{-0.06}^{+0.07}$ \\
$18.8$  & $9209_{-8}^{+7}$      & $3.22_{-0.02}^{+0.02}$ & $6.4_{-0.1}^{+0.1}$    & $11.0_{-0.2}^{+0.2}$ & $10.73_{-0.09}^{+0.1}$  & $0.81_{-0.02}^{+0.02}$ & $0.52_{-0.02}^{+0.02}$    & $11_{-1}^{+1}$         & $5.5_{-0.4}^{+0.4}$ & $-1.4_{-0.2}^{+0.2}$    & $19.62_{-0.07}^{+0.1}$ \\
$21.1$  & $6384_{-7}^{+6}$      & $2.95_{-0.03}^{+0.03}$ & $7.5_{-0.2}^{+0.2}$    & $11.7_{-0.2}^{+0.2}$ & $11.2_{-0.1}^{+0.2}$    & $0.86_{-0.03}^{+0.02}$ & $0.50_{-0.02}^{+0.02}$    & $12_{-1}^{+2}$         & $4.7_{-0.4}^{+0.5}$ & $-2.0_{-0.3}^{+0.3}$    & $17.8_{-0.1}^{+0.2}$    \\
$23.5$  & $3966_{-6}^{+5}$      & $2.54_{-0.03}^{+0.03}$ & $7.3_{-0.2}^{+0.2}$    & $10.6_{-0.2}^{+0.2}$ & $10.1_{-0.1}^{+0.2}$    & $0.79_{-0.03}^{+0.03}$ & $0.55_{-0.04}^{+0.04}$    & $16_{-4}^{+11}$        & $7_{-1}^{+2}$       & $-2.5_{-1.0}^{+0.7}$    & $15.2_{-0.1}^{+0.2}$    \\
$25.8$  & $1955_{-5}^{+4}$      & $2.10_{-0.04}^{+0.04}$ & $7.4_{-0.3}^{+0.3}$    & $10.3_{-0.3}^{+0.3}$ & $9.5_{-0.2}^{+0.1}$     & $0.81_{-0.04}^{+0.04}$ & $0.66_{-0.06}^{+0.06}$    & $1369_{-983}^{+3098}$  & $592_{-391}^{+390}$ & $-4_{-4}^{+4}$          & $12.6_{-0.3}^{+0.3}$    \\
$28.1$  & $455_{-3}^{+2}$       & $1.01_{-0.05}^{+0.04}$ & $7.6_{-0.4}^{+0.3}$    & $8.7_{-0.3}^{+0.4}$  & $8.2_{-0.3}^{+0.3}$     & $1.0_{-0.1}^{+0.1}$    & $0.6_{-0.2}^{+0.2}$       & $19_{-10}^{+17}$       & $12_{-6}^{+6}$      & $-4_{-4}^{+4}$          & $8.4_{-0.5}^{+0.7}$     \\
\hline
\end{tabular}
\end{table}

\label{lastpage}

\end{landscape}

\end{document}